\documentclass[preprint]{aastex}

\shorttitle{OH$^+$, H$_2$O$^+$, and H$_3$O$^+$}
\shortauthors{D. Hollenbach}

\newcommand{\be}{\begin{equation}}

\newcommand{\htho}{H$_3$O$^+$}
\newcommand{\hto}{H$_2$O$^+$}
\newcommand{\oh}{OH$^+$}
\newcommand{\crt}{$\zeta _{crp}$}

\def\ltsimeq{\,\raise 0.3 ex\hbox{$ < $}\kern -0.75 em
 \lower 0.7 ex\hbox{$\sim$}\,}
\def\gtsimeq{\,\raise 0.3 ex\hbox{$ > $}\kern -0.75 em
 \lower 0.7 ex\hbox{$\sim$}\,}

\let\gta=\gtsimeq
\let\lta=\ltsimeq
\newcommand{\beq}{\begin{equation}}
\newcommand{\eeq}{\end{equation}}

\newcommand{\cms}{\rm {cm^{3}\;\;s^{-1}}}
\begin{document}

\title{The Chemistry of Interstellar OH$^+$, H$_2$O$^+$, and H$_3$O$^+$: Inferring the Cosmic Ray Ionization Rates from Observations
of Molecular Ions}

\author{David Hollenbach$^1$, M. J. Kaufman$^{2}$, D. Neufeld$^3$, M. Wolfire$^4$, and J. R. Goicoechea$^5$}

\affil{$^1$SETI Institute, Mountain View, CA, 94043-5203}

\affil{$^2$Department of Physics \& Astronomy, San Jose State University, San Jose, CA 
95192-0106}

\affil{$^3$Department of Physics and Astronomy, The Johns Hopkins University, Baltimore, MD, 21218}

\affil{$^4$Department of Astronomy, University of Maryland, College Park, MD, 20742 }

\affil{$^5$Departamento de Astrof\'{\i}sica, Centro de Astrobiolog\'ia
(CSIC-INTA), 28850, Madrid, Spain }

\begin{abstract}
We model the production of  \oh, \hto, and \htho\   in interstellar clouds, using a steady state photodissociation region
code that treats the freeze-out of gas species, grain surface chemistry, and desorption of ices from grains.  The
code includes PAHs, which have important effects on the chemistry.    All three ions generally have two peaks in abundance
as a function of depth into the cloud, one at $A_V \lta 1$ and one at $A_V \sim 3-8$, the exact values depending
on the ratio of incident ultraviolet flux to gas density.   For relatively low values of the incident 
far ultraviolet flux on the cloud ($\chi \lta 1000$; $\chi= 1$= local interstellar value), the columns 
of \oh\  and \hto\  scale roughly as the cosmic ray primary ionization rate \crt \ divided by the hydrogen nucleus density $n$. 
The \htho\ column is dominated
by the second peak, and we show that if PAHs are present, $N$(\htho)$\sim 4\times10^{13}$ cm$^{-2}$ independent
of \crt\  or $n$.  If there are no PAHs or very small grains at the second peak, $N$(\htho) can attain such columns only if 
low ionization potential metals
are heavily depleted.
We also model diffuse and translucent clouds in the interstellar medium,
and show how observations of $N$(\oh)/$N$(H) and $N$(\oh)/$N$(\hto) can be used to estimate \crt/$n$, $\chi/n$ and $A_V$ in them.
We compare our models to {\it Herschel} observations of these two ions, and estimate
\crt$\sim 4-6\times 10^{-16} (n/100\ {\rm cm^{-3}})$ s$^{-1}$ and $\chi/n=0.03$ cm$^3$ for diffuse foreground clouds towards W49N.

\end{abstract}


\section{INTRODUCTION}

The \oh, \hto, and \htho\ ions form the backbone of interstellar chemistry and are important probes of the 
cosmic ray ionization rates in diffuse clouds, on the surfaces ($A_V < 1$)  of molecular clouds, and (with less reliability
as we will discuss in this paper) in the interiors ($A_V \sim 4-8$) of molecular clouds.  They are the backbone
of chemistry because once H$_2$ forms,  cosmic ray ionization of H or H$_2$ leads to the formation
of these ions, and \htho\ recombination with electrons  leads to OH and H$_2$O.  Reaction of OH with
C or C$^+$ leads to CO.\footnote {CO also has another route initiated by the production of the CH$^+$ ion, which we
do not discuss in detail in this paper although it is included in our models.}    Once these basic molecules are formed, 
many of the other polyatomic and rare species follow.

The pathway of cosmic ray ionization of hydrogen  to these molecular ions follows two routes 
(see Figure 1).   In gas with significant H atoms the ionization of H leads to H$^+$ that then proceeds via a series of
reactions (see Figure 1 top) to \oh,  \hto\ and \htho\ ions.  We note that the charge exchange of O with H$^+$ is slightly endothermic, so
the reaction rate is proportional to exp(-230 K/$T$); this means that this reaction slows down at cooler temperatures, 
and a greater fraction of the cosmic ray ionizations of H are followed by recombination of H$^+$ with neutral or negatively charged polycyclic
aromatic  hydrocarbons (PAHs or PAH$^-$) or  electrons
rather than proceeding to form O$^+$ and then \oh.  It is this atomic route to \oh\ which is primarily important in
diffuse clouds and in the $A_V \lta 2$ surfaces of molecular clouds.  A second route dominates deeper in the opaque
interiors of molecular clouds.   Here the ionization of H$_2$ leads to H$_2^+$, which then proceeds via a series of 
reactions (see Figure 1 bottom) to
 \oh,  \hto\ and \htho.  A key competitor here to the formation of \oh \ is the dissociative recombination of H$_3^+$ with electrons,
which has a high rate coefficient compared to the reaction of H$_3^+$ with O.   In addition, the reaction of 
H$_3^+$ with CO  dominates that with O  when the CO abundance exceeds that of O.   Therefore, low electron abundances and
high O abundances are needed to ensure that a large fraction of cosmic ray ionizations of H$_2$ 
eventually produces  \oh.

As we shall show in this paper, these two routes lead generally to two peaks in the abundances of \oh, \hto, and
\htho\ as a function of depth or $A_V$ into a cloud.    The first peak (at $A_V \sim 0.01-1$ depending on
the ratio of the incident FUV flux to the gas density) 
occurs in atomic gas where the  cosmic ray ionization of H
begins the chemical chain.\footnote{FUV is defined as photons in the range 6eV$< h\nu < 13.6$ eV.}   
The second peak (at $A_V \sim 3-8$, again dependent on the FUV flux/gas density
ratio) occurs in molecular gas where the cosmic ray ionization of H$_2$ begins the chemical chain.   Deeper 
in the cloud the gas phase oxygen freezes out as water ice on grain surfaces (e.g., Hollenbach et al 2009, hereafter H09), and the gas phase abundances
of the three ions drop.   Whereas significant columns of \oh\ and \hto\ are produced in the first peak, most of the
\htho\ column arises in the second peak.

In warm ($T>300$ K) neutral gas with significant abundances of H$_2$ there are other
 dominant pathways to form \oh, \hto, and \htho\ ions than the two paths initiated by cosmic rays and
shown in Figure 1.  
One of the goals of this paper is to show under what conditions (e.g., FUV flux, gas density,
cosmic ray ionization rate)  cosmic rays initiate the formation of the \oh, \hto, and \htho\ ions, and under what conditions other
chemical routes dominate.

If cosmic rays dominate, then the observed columns of \oh, \hto, and \htho\ serve as a probe of the cosmic ray ionization rate.
The major goal of this paper is to show how these columns depend on the cosmic ray ionization rate, the gas density, 
the FUV flux, the total column or  optical extinction $A_{Vt}$ through the cloud, and the molecular hydrogen abundance.    Comparison with our models allow the cosmic ray 
ionization rate to be estimated from these ion columns and the atomic H column.    

Our method of estimating cosmic ray ionization rates from the abundance of eventual products of this ionization 
dates back to
the 1970's, when Black \& Dalgarno (1973) pointed out the sensitivity of the OH and HD abundances to cosmic ray
ionization rates.  A later comprehensive paper by van Dishoeck \& Black (1986) showed how the abundance of OH in diffuse or
translucent clouds could be used to estimate the cosmic ray rate.  One advantage of using \oh\ or \hto\ to probe cosmic
ray ionization rates over using OH is that OH is produced after a series of reactions following the formation of \oh\ (see Figure 1). 
This makes the inference of cosmic ray ionization rate dependent on knowledge of the rate coefficients of these additional reactions.  

Recent observations strongly motivate the theoretical study of the $\rm OH^+$, $\rm H_2O^+$, and $\rm H_3O^+$ molecular ions.  Prior to 2010, the only observations of these three ions within the interstellar medium\footnote{Note, however, that the $\rm H_2O^+$ ion has long been observed in cometary comae (e.g. Wehinger et al. 1974), where it is produced by the photoionization of water by solar ultraviolet radiation.} consisted of a few detections of $\rm H_3O^+$ in rich interstellar sources of submillimeter line emission and absorption (Wootten et al 1991; Phillips, van Dishoeck \& Keene 1992, hereafter PvDK92; Goicoechea \& Cernicharo 2001; van der Tak et al 2006).  Thanks largely to absorption-line spectroscopy with the HIFI instrument on the {\it Herschel Space Observatory}, together with additional ground-based detections of OH$^+$ obtained with the APEX telescope (submillimeter rotational transition) and the ESO Paranal observatory (near-UV electronic transition), the observational picture has improved radically over the past two years.  All three molecular ions have now been detected in both the diffuse and dense Galactic interstellar medium (Gerin et al.\ 2010; Ossenkopf et al.\ 2010; Wyrowski et al.\ 2010; Neufeld et al.\ 2010; Schilke et al.\ 2010; Gupta et al.\ 2010; Bruderer et al.\ 2010; Benz et al.\ 2010; Krelowski, Beletsky, \& Galazutdinov 2010) and in external galaxies (van der Tak et al 2008; Weiss et al. 2010; van der Werf et al.\ 2010; Gonz\'alez-Alfonso et al.\ 2010; Aalto et al 2011).  The most reliable column density determinations are obtained for foreground molecular clouds lying along the sight-lines to bright submillimeter continuum sources in the Galactic disk. Here, the absorbing material typically covers a wide range of line-of-sight velocities, arising in widely distributed material along the sight-line.  The total inferred column densities of the ions lie in the range $\rm few \times 10^{13} - few \times 10^{14} \rm \, cm^{-2}$, the largest values being attained for $\rm OH^+$ and the smallest for $\rm H_3O^+$ (Gerin et al.\ 2010; Neufeld et al.\ 2010).  

In their study of OH$^+$ and H$_2$O$^+$ absorption along the sight-line to the luminous star-forming region W49N, Neufeld et al.\ (2010, hereafter N10) measured an average OH$^+$/H$_2$O$^+$ abundance ratio of $\sim 10$, with variations over the range $\sim 3 - 15$.  These observed ratios are considerably larger than the value $\sim 1$ expected in fully molecular gas.\footnote{In fully molecular gas the formation rates of these ions per unit volume are equal because every \oh\ formation is followed by \hto\ formation  and the destruction rate per ion is nearly the same
(due to reaction with H$_2$), leading to similar abundances for both ions (N10).}  By means of a simple analytical treatment of the chemistry of OH$^+$ and H$_2$O$^+$, confirmed by more detailed pure gas-phase  models performed using the Meudon PDR model (Le Petit et al 2006; Goicoechea \& Le Bourlot 2007), N10 concluded that the molecular fraction in the absorbing material lies in the range 2 -- 8 $\%$.  This conclusion was supported by the observed distribution in velocity space of the OH$^+$ and $\rm H_2O^+$ absorption, which proved similar to that of the atomic gas probed by 21 cm absorption studies, and quite dissimilar from that of the molecular gas traced by HF or CH absorption.  The analytic treatment introduced by N10 also allowed the cosmic ray ionization rate to be inferred from the abundances of OH$^+$ and $\rm H_2O^+$ relative to atomic hydrogen.  The resulting estimate of the cosmic ray ionization rate in the range \crt= 0.6 -- 2.4 $\times 10^{-16}\  \rm s^{-1}$ (primary ionization rate per H atom) was broadly consistent with earlier values inferred quite independently from observations of the H$_3^+$ molecular ion toward different sight lines (Indriolo et al. 2007).  One of the goals of the present study is to refine the N10 analytic treatment of diffuse cloud chemistry through detailed modeling, as an aid to interpreting the growing body of observational data now available.

This paper is organized as follows.  In \S 2 we describe the 
chemical/thermal model of  both an opaque molecular cloud illuminated
by FUV radiation, or a diffuse cloud illuminated by the interstellar radiation field.  
In \S 3 we show the model columns and column ratios of \oh, \hto, and \htho\ 
as functions of the cloud gas density $n$, the incident FUV flux $\chi$ (in units of the local
average interstellar field, see below),
and the primary cosmic ray ionization rate \crt \ per H atom for the case of opaque molecular clouds.
We also show the same results for diffuse clouds, but with the total column or $A_{Vt}$ through the cloud as an additional
parameter.    In \S 4 we
compare our model results with previous \htho\ observations and recent \oh, \hto, and \htho\ observations by {\it Herschel}, and show how observations compared
to models can constrain the cosmic ray ionization rates.   We summarize our results in \S 5. Appendix A presents tables of key reaction rate coefficients and
adopted abundances and grain parameters.   Appendix B includes analytic expressions that explain the variation of \oh\  with $A_V$ over the first peak, and
the variation of \htho\ with $A_V$ over the second peak.  Appendix C assesses the sensitivity of our results to certain chemical rate coefficients, the
gas phase abundance of elemental oxygen and low ionization potential metals, and the freeze-out of species.

\section{THE CHEMICAL AND THERMAL MODEL OF A CLOUD}

\subsection{Summary of Prior PDR Model and Modifications}

The numerical code we have developed to model the chemical
and thermal structure of an opaque cloud externally illuminated by FUV
flux is based on our previous PDR model described in H09.
This 1D code models a 
constant density $n$ (the hydrogen nucleus density) slab of
gas, illuminated from one side by an FUV flux of $2.7\times 10^{-3}\chi$ erg 
cm$^{-2}$ s$^{-1}$ incident perpendicular to the slab. The unitless parameter $\chi$ is defined above in
such a way that $\chi \sim 1$ corresponds to the average local interstellar radiation
field in the FUV band (Draine 1978).\footnote{ Note that $\chi=1$ corresponds to $G_0=1.7$
in the Tielens \& Hollenbach (1985) units based on the Habing (1968) local interstellar
radiation field.  The shape of the FUV spectrum for $\chi>1$ is implicitly assumed to mimic that of the Draine field,
which is approximately that of a $T_{eff}\sim 30,000$ K star. }  The code calculates the steady state chemical abundances and
the gas temperature from thermal balance as a function of depth into the
cloud.  It incorporates 63 chemical species, $\sim 300$ chemical 
reactions, and a 
large number of heating mechanisms and cooling processes.  The chemical
reactions include FUV photoionization and photodissociation; cosmic ray
ionization; neutral-neutral, ion-neutral, and electronic recombination
reactions.  H$_2$ self-shielding is included as described in H09, and CO self-shielding and the partial shielding of CO by H$_2$ is included as described in Visser et al (2009).
The code includes the photodissociation of molecules by ``secondary" FUV photons produced
(ultimately) by cosmic rays (Prasad \& Tarafdar 1983).
We also include reactions with charged dust grains and PAHs; and the 
formation of H$_2$, OH, H$_2$O, CH, CH$_2$, CH$_3$, and CH$_4$ on grain surfaces. 
The code treats the freezing of all condensable species to grain surfaces and three desorption processes:
thermal desorption, photodesorption, and cosmic ray desorption.  The only significant difference
in the desorption code used here versus the H09 code is the
inclusion of the new (higher by factor $\sim 3$) rate of photodesorption 
of CO (Oberg et al 2009).  The code does not include photodesorption by the secondary FUV photons;
this process is negligible at the peaks in the \oh, \hto, and \htho\ abundances.  The  code has been used
to model regions which lie at hydrogen nucleus column densities 
$N \lta 4 \times 10^{22}$ cm$^{-2}$ (or $A_V \lta 20$) from the surface
of a cloud.  Therefore, it applies not only to the                   
photodissociated  surface region, where gas phase hydrogen and oxygen
are nearly entirely atomic and where gas phase carbon is mostly C$^+$,
but also to regions deeper into the molecular cloud where hydrogen is in
H$_2$ molecules and carbon is in CO molecules.  Even in these molecular regions,
the attenuated FUV field can play a significant role in photodissociating
H$_2$O and O$_2$, in photodesorbing species adsorbed on grain surfaces,
 and in heating the gas.  However, the code is now sufficiently general that it finds
 the steady state solutions for abundances and temperature in any region of a molecular cloud,
 even where FUV is insignificant.

 We emphasize that we present a steady-state model of chemical abundances
 as a function of depth into the cloud.   The chemical timescales can be quite long, which might
 suggest that time dependent models are more appropriate.   For example, the timescale
 to convert atomic H gas to fully molecular H$_2$ gas is $t_{H_2} \sim 10^9/n$ years, where
 $n$ is the hydrogen nucleus density in units of cm$^{-3}$.   However, as we show below, the
 \oh\ and  \hto \ ions peak in abundance when $x($H$_2) \sim 0.03 - 0.1$, which occur typically
 at $A_V \sim 0.1$.  [Note, abundances in
 this paper are defined relative to hydrogen nuclei, so that $x($H$_2) \equiv n($H$_2)/n$ and
 $n\simeq n($H$) + 2n($H$_2$)].  Therefore, the timescales for
 even diffuse clouds of density $n\sim 100$ cm$^{-3}$ to reach these abundances are less than
 $\sim 10^6$ years, which is shorter than the typical lifetime of a diffuse cloud (Wolfire et al 2003).   Steady state solutions
 therefore generally apply, at least for computing the columns of these ions. The steady state models may somewhat 
 underestimate the ion columns for low density diffuse clouds with $A_V > 0.1$, since the steady state
 solutions can lead to lower abundances of these two ions than time dependent solutions  in the high $A_V$ regions.
 This arises because the steady state abundances of H$_2$ are higher than time dependent models which start with fully atomic gas.   Higher H$_2$ abundances lead to more destruction of \oh\ and \hto\ 
 in the high $A_V$ regions where H$_2$ and not electrons dominate the destruction of these ions.
   Liszt (2007) provides a detailed analysis of the time dependent formation of
 H$_2$ and \oh\ as a function of $A_V$ in  diffuse clouds with initial atomic conditions.
 
 We also emphasize that our model does not include turbulent dissipation and heating of small 
 pockets of gas along the line of sight (e.g., Godard, Falgarone, \& Pineau des Forets 2009 and references therein).   We do, however, run PDR models with small fractions of the line of sight having either enhanced temperature or enhanced
 rates of ion-neutral drift
 to test the possible effects of turbulence, and find the effects on \oh, \hto, \htho\ column densities are likely 
 small. 
  
 For simplicity we assume constant H nucleus density $n$ in our models.    At low $A_V$ and low
 $x({\rm H_2}) < 0.1$, where much of the \oh\  and \hto\ columns often arise, the temperature is quite constant so
 that constant density implies constant thermal pressure.    If thermal pressure (and not turbulence) dominates
 deeper into a high $A_V$ cloud, then the density will rise as one moves from its warmer PDR surface to
 its CO-cooled molecular interior.  In addition, self gravity can raise the density of the interior regions at higher $A_V$.  Still another effect is
 the transition of atomic hydrogen to molecular hydrogen which can raise $n$ by a factor of 2 if thermal pressure is conserved.  We ignore the possible rise in density, which mainly affects the second peaks of the ions  deep 
 ($A_V>4$) in the cloud.  If such a density enhancement occurs,
  it tends to depress the  abundances of \oh\ and \hto\ in the second peaks, but the abundance of \htho\ at the second peak 
  is not sensitive to hydrogen density if PAHs are present(see \S 3).\footnote{If a reader is interested in the second peak and knows the density $n$  of a given source there, then
  our models with that $n$ will give a good prediction of the behavior of the second peaks.}
  The second peak is dependent on the abundance of PAHs, very small grains and low ionization potential metal ions there, which can control the electron density, $n_e$.   The 
  abundance of PAHs and very small grains at high $A_V$ is uncertain due to their possible coagulation on larger grain surfaces.
     
  One significant difference between the chemical code used in this paper and that used in H09
   is the inclusion of PAHs and very small grains (VSGs, radius $\lta 50$ \AA), which affect the ionization balance by enhancing the recombination   of 
  positive atomic ions.  In the rest of this paper we shall often use the term ``PAH" to denote both PAHs and VSGs.  PAHs also affect the second peaks of the ions by destroying electrons in the reaction
  e + PAH $\rightarrow$ PAH$^-$.  (The effects of PAHs on ionization balance has been treated extensively
  before in the literature: e.g., Lepp \& Dalgarno 1988, Bakes \& Tielens 1998, Weingartner \& Draine 2001a, Flower \& Pineau des Forets 2003, Liszt 2003, Wolfire et al 2003, 2008).   H09
   focused on the H$_2$O and O$_2$ peaks, which occur at relatively
  high $A_V\sim 3-7$.    Although the PAH abundances are not known deep in molecular clouds, H09
   assumed that their abundances are low, due to coagulation of PAHs on the surfaces of larger dust
  grains.  Here, however, we are mostly interested in the chemistry at low $A_V < 1$, and in particular in the
  chemistry of diffuse clouds.  These are the regions where PAHs are observed to be present, and we use
  PAH parameters derived from the literature.   For simplicity, we adopt a single size PAH for the PAH distribution,
  and assume that the standard PAH has 100 C atoms, and a number abundance of  $x_{{\rm PAH}}=2\times10^{-7}$ with respect to H nuclei (see Wolfire et al 2008 for a discussion of the amount
  of carbon in PAHs; here we adopt 100 C atoms and not the 35 C atoms that
  Wolfire et al adopted because of the result of  Draine \& Li 2007, which suggested that the distribution in mass peaks
  at 100 C atoms).  One of the key impacts of PAHs is in the recombination of H$^+$.
  As shown in Figure 1 (top), cosmic ray ionization of H can lead to \oh, but a competing route is
  the neutralization of  H$^+$ by an electron, PAH, or PAH$^-$.   PAHs therefore lower the production 
  of OH$^+$ and the columns of all the ions in the first peak.  Conversely, in the second peaks,  the
  reduction in electron abundance caused by PAHs cause fewer H$_3^+$ ions to recombine with electrons, and
  lead to greater production rates of the ions as well as smaller destruction rates for \htho.  Therefore, the columns
  of the ions increase with the presence of PAHs in the second peaks.  We also discuss results with no PAHs or very small grains at
  high $A_V$.
 
 In Appendix A we present Table 1 which lists the rate coefficients of key reactions
 in the pathways to the \oh, \hto, and \htho\ ions, as well as reactions that are either new
 or changed since H09.   Of particular note here are the
 photoionization of OH and H$_2$O, the photodissociation of OH$^+$ and H$_2$O$^+$, the treatment of PAHs--especially the photodetachment reaction rate for 
 PAH$^-$, the fine structure level population dependence in the charge exchange reaction of O with
 H$^+$, and some minor modifications in 
  reaction rates key to determining the abundances of the 
  \oh, \hto, and \htho\ ions, such as their dissociative recombination rates with electrons and their
 reactions rates with H$_2$.

We have also included in Table 1 rates that are important in producing H$^+$ by chemical means rather 
than by cosmic ray ionization of H in PDRs.    There are two main chemical routes.  The first is initiated 
by the FUV photoionization of C to C$^+$.    The C$^+$ reacts with OH to form CO$^+$.   The 
 CO$^+$ charge exchanges with H to form H$^+$.     The second is also initiated by FUV photoionization
 of C to C$^+$.    The C$^+$ reacts with H$_2$ to form CH$^+$.   The CH$^+$ is photodissociated by 
 FUV photons to produce H$^+$.    Both these reaction chains are very much enhanced by high
 ($\gta 300$ K)  temperatures in gas with appreciable H$_2$.  The first chain is enhanced because high gas temperatures lead to 
 significant amounts of OH being formed by the reaction H$_2$ + O $\rightarrow$ OH + H.   This reaction has
 an activation barrier of $\Delta E/k = 3160$ K and so is insignificant for low gas temperatures.
 The second chain similarly is enhanced because the reaction C$^+$ + H$_2$ $\rightarrow$ CH$^+$ + H has 
 an activation barrier of $\Delta E/k = 4640$ K (see Table 1).\footnote{As discussed in Tielens \& Hollenbach (1985), our code also
 includes the reaction of vibrationally excited H$_2$ with C$^+$ to form CH$^+$.   This reaction has no
 activation barrier and can be moderately important at low $A_V \lta 0.6$.} In regions where either of these two 
 chains dominate the production of H$^+$ over cosmic ray ionization, the ions \oh, \hto, and \htho\ 
 only provide upper limits to the cosmic ray ionization rates.

Other alternate routes to the production of \oh\ and \hto\ ions include three routes which produce OH or H$_2$O
without the ions:  the production of water on grain surfaces
followed by photodesorption of the water to produce gas phase OH and H$_2$O, the radiative association of O
with H to form OH, and the reaction of FUV-pumped H$_2$ in excited vibrational states with O  to form OH.    The gas phase OH and H$_2$O
can be photoionized by FUV photons to produce \oh\ and \hto \ (see Table 1).  
Unlike the routes described in the preceeding paragraphs,
these routes are never dominant in producing the ion column densities. 

An analogous route to the formation of H$_2$O on grain surfaces followed by photodesorption, but one not treated in this work, is the time-dependent evaporation of water ice which occurs around newly
formed stars.   The rapid rise in embedded luminosity heats the dust grains above about 100 K, and the 
icy mantles on grains are then thermally evaporated.   This sudden injection of high abundances of water vapor into the
gas is followed by reaction of the gas phase water with HCO$^+$ and H$_3^+$ to form \htho.    Eventually, the
system relaxes to the steady state chemistry described in this paper, but for a short time, there might be a large enhancement
in \htho\ (Millar, Herbst, \& Charnley 1991, PvDK92).  If this release of 
H$_2$O from the icy grain surface to the gas occurs in regions with elevated FUV fields,
the photoionization of H$_2$O could also result in enhanced \hto\ abundances (Gupta et al 2010)

We also present in Appendix A Table 2, which lists the gas phase elemental abundances, the PAH properties, 
 and the grain surface area per H nucleus adopted in our code.

\subsection{Interstellar Cloud Models}   

As noted above, our opaque molecular cloud models invoke a constant density $n$
slab, illuminated from one side by a 1D normal FUV flux $\chi$.  We solve for the gas temperature and
the gas phase and ice abundances of the various species as a function of depth in the slab.   
Note that depth is synonymous with hydrogen nucleus column $N$ [=$N$(H)+2$N$(H$_2$)]  into the cloud or $A_V$ into the cloud.
We take $N= 2 \times 10^{21}A_V$ cm$^{-2}$.   We 
follow the chemistry to $A_V \sim 20$, beyond which there is little contribution to the columns of
\oh, \hto, and \htho\ ions.

We use the primary cosmic ray ionization rate per H atom as an input parameter, since our code calculates the
secondary ionizations caused by cosmic rays and these depend on the H$_2$ and electron abundances.
In order to probe the range of cosmic ray ionization rates suggested in the literature, we include
cases with  primary
ionization rates of \crt $= 2\times 10^{-17}$ s$^{-1}$ and $2\times10^{-16}$ s$^{-1}$ per H atom, which correspond to total
rates (including secondary ionizations)  of about  $\zeta _{crt} \sim 3- 5\times 10^{-17}$ s$^{-1}$ and $3-5\times 10^{-16}$ s$^{-1}$ respectively.  Note that the primary rate per H$_2$ molecule is 2\crt.   We also
adopt \crt\ as the primary rate for He atoms; however, He has insignificant secondary ionizations.
Although there is 
evidence that the cosmic ray ionization rate may decrease with depth (e.g., Rimmer et al 2011, Indriolo \& McCall 2012, and discussion later in this paper), for simplicity we assume the primary
cosmic ray ionization rate does not vary with depth into a cloud in a given model.   The main effect of this assumption is that we may
have overestimated the abundances and columns of the \oh\ and \hto\ ions in the second 
(deeper) peak relative to the first peak.  However, since we vary \crt\ in our parameter study, the reader
can use lower \crt\ for the columns in the second peak if so desired.   The abundance of \htho\ in the second peak is not sensitive
to \crt\ if PAHs are present, as we will show below.    

The main parameters that we explore for our one-sided, opaque molecular cloud models are
the gas density $n$, the incident FUV flux $\chi$, and the primary cosmic ray ionization rate \crt.   We study 
the parameter space $ 10$ cm$^{-3}$$<  n < 10^7$ cm$^{-3}$, 1$< \chi < 10^6$, and
$2\times 10^{-17}$ s$^{-1} < $\crt$ < 2\times 10^{-16}$ s$^{-1}$.   We  explore the sensitivity of the columns of
\oh, \hto, and \htho\ ions to assumptions about the PAH chemistry, the elemental gas phase abundances of low ionization
potential metals, and the rate coefficient for the formation of H$_2$ on grain surfaces. 

The diffuse cloud models treat a constant density slab of total thickness $A_{Vt}$  illuminated on both sides by a 1D normal FUV flux $\chi/2$.    
We explore the parameter space $10^{-17.5}$ s$^{-1} <$ \crt$/n_2  < 10^{-14.5}$ s$^{-1}$, $0.01 < A_{Vt} < 3$,  $ 30$ cm$^{-3}$$< n < 300$ cm$^{-3}$, 1$< \chi\ < 10$,  and $1 \leq  \chi/n_2 \leq 10$ 
where $n_2= n/100$ cm$^{-3}$.   The models of Wolfire et al (2003) provide a good estimate of $\chi/n_2$ in the Galaxy 
by estimating $\chi$ from the star formation rate and the dust opacity as a function of galactocentric radius $R$, and setting the thermal
pressure to provide two phases, a cold diffuse cloud phase and a warm intercloud medium that fills most of the
volume.     Assuming $R=8.5$ kpc as the solar location, Wolfire et al find $\chi=1.0$, $n=33$ cm$^{-3}$, and $\chi/n_2=3.0$ for diffuse clouds at $R=8.5$ kpc; $\chi=2.35$, $n=49$ cm$^{-3}$, and $\chi/n_2=4.8$ at $R=5$ kpc; $\chi=3.0$, $n=54$ cm$^{-3}$, and $\chi/n_2= 5.5$ at $R=4$ kpc; and
$\chi=3.8$, $n=60$ cm$^{-3}$, $\chi/n_2=6.3$ at $R=3$ kpc.     Therefore, our prime region for study is $\chi/n_2= 3-6$.
We are especially interested
in how the ratio of the columns $N$(\oh)/$N$(\hto)  and $N$(\oh)/$N$(H) vary as  functions of $A_{Vt}$, $\chi/n$ and
\crt/$n$.  

We note that the columns our models predict are columns perpendicular to the 1-sided (molecular cloud) slab or
2-sided (diffuse cloud) slab.   If the slabs are viewed at an angle $\theta$ with respect to the normal, the observed
columns will increase by ($\cos \theta)^{-1}$.  Another effect that will obviously raise the columns is if there are
more than one diffuse cloud (in a given velocity range) along the line of sight (los), or if the molecular cloud is clumpy
and FUV scattering then introduces several ``surfaces" along the los.  However, as we will show, \crt/$n$ in diffuse clouds 
can be estimated from our models from the ratios $N$(\oh)/$N$(\hto)  and $N$(\oh)/$N$(H), and
the ratios are independent of the geometric effects.  Nevertheless, there is some degeneracy in the solution for \crt/$n$, 
depending on the combination of the $A_{Vt}$ of a single cloud, $\chi/n$,  and the 
enhancement in columns created by the geometrical effects.

\section{MODEL RESULTS}

\subsection {PDR surfaces of opaque molecular clouds}

\subsubsection {The chemical and thermal structure of individual clouds}

In order to understand the columns of \oh, \hto, and \htho\ ions produced as a function of $n$, $\chi$,
and \crt, we first study the detailed chemical and thermal structure of a few specific (standard) cases.  
 Figure 2 shows the chemical abundances as a function of depth 
$A_V$ into the cloud for the case $n= 10^2$ cm$^{-3}$, $\chi = 1$, and \crt = $2 \times 10^{-17}$ 
s$^{-1}$.  This case is chosen not only because it may be appropriate for the ambient interstellar radiation
field (ISRF) incident on a relatively low density GMC, but also because the surface ($A_V \lta 3$) structure
(i.e., $T$ and chemical abundances as a function of depth or $A_V$) is illustrative of the depth structure of a diffuse or translucent cloud.
In addition, the cosmic ray rate may be appropriate to the interior of molecular clouds.

The main chemical result is that the \oh, \hto, and \htho\ ions all have a peak
at $0.03 < A_V < 0.3$, and then have a second peak at $A_V \sim 6$ for this combination of
$n$ and $\chi$.  We first discuss the surface peaks at low $A_V$.   \oh\  peaks at $A_V \sim 0.03$, 
where the molecular hydrogen abundance
is $x($H$_2) \sim 0.01$.  The \hto\  peaks slightly deeper, at $A_V \sim 0.1$, where $x($H$_2) \sim 0.1$.
Finally, \htho\ peaks at $A_V \sim 0.3$, where $x($H$_2)\sim 0.25$.  Appendix B provides an 
approximate analytic solution to the chemistry that explains the \oh\ first peak, and its relation to
$x($H$_2)$.   
 From $A_V\sim 0.01$ to either $A_V\sim 0.03$ (\oh) or $A_V\sim 0.1$ (\hto) or $A_V\sim 0.3$ (\htho) the three ions 
rise in abundance with increasing
$A_V$ because of the rise in the H$_2$ abundance, which drives the cosmic ray-produced
O$^+$ to the molecular ions.  At larger $A_V$, but for $x$(H$_2$)  significantly less than its maximum value
of $0.5$, the abundance of \oh\ and \hto\ 
plateau at the peak value because here both their formation and destruction rates are proportional to $x$(H$_2$).
\hto\ tends to peak at somewhat higher $A_V$ than \oh\ because, even at the peak of \oh\ abundance, not all cosmic ray
ionizations lead to \hto\ and so its abundance continues to rise as the H$_2$ abundance rises with increasing $A_V$.
Finally, as $x$(H$_2$) approaches 0.5, two effects lead to a drop in the \oh\ and \hto\ abundances with increasing $A_V$.  One is
that the formation rates of these ions saturate as nearly every cosmic ray ionization leads to their production, whereas
the destruction rates still scale as $x$(H$_2$), which increases with increasing $A_V$.    The other dominant effect
is that the H abundance drops so that the \oh\ formation rate via the top chain of reactions in Figure 1 drops.   The bottom chain at
relatively low $A_V$ is not as efficient at producing \oh, as the electrons are relatively abundant ($x_e \sim 10^{-4}$)  in these surface 
regions, and H$_3^+$
recombines with electrons rather than forming \oh.\footnote{As a result of the inefficiency of the bottom chain when the electron abundance is 
relatively high, the first peak is always dominated by the top chain of reactions in Figure 1, even as $x$(H$_2$) approaches 0.5.   The bottom
chain dominates in the second peak, however, because the H$_2$ abundance is high and the electron abundance is low.}
  The \htho\ is destroyed not by 
H$_2$, but by electrons, whose abundance stays quite constant (supplied mostly by  C$^+$ but with possible contribution by H$+$ at high \crt/$n$) at
the cloud surface.   The \htho\ starts to drop in abundance once the gas becomes predominantly H$_2$, due to the second effect described above.

We next discuss the second deeper peak in the  \oh, \hto, and \htho\ ions.   As one moves deeper 
into the cloud (typically, $A_V \gta 2$), the electron abundance starts to drop.   As this happens, a greater fraction of
the cosmic ray ionizations of H$_2$ leads to the production of the three ions, and their formation rates
rise.    The destruction  of \oh\ and \hto\ is by H$_2$, which now has constant abundance
(the gas is fully H$_2$), so the destruction rates hold constant.   
Therefore, these two ions rise in abundance.   Finally, they peak and
fall in abundance for $A_V > 6$ because gas phase oxygen freezes out as water ice, and again
the oxygen reaction with H$_3^+$  cannot compete with H$_3^+$ electronic recombination or its reaction
with CO.  Therefore,
the formation rates of all three ions drop.    \htho\ behaves somewhat more dramatically, because
its destruction is mainly by dissociative recombination with electrons.   Thus, as the electron
abundance drops, not only is the formation rate of \htho\ enhanced, but the destruction rate is
suppressed.    Therefore,  \htho\  rises to much higher abundances than  \oh\ and  \hto\ 
in the second peak.   It also drops at very high $A_V$ because gas phase elemental oxygen from which \htho\ 
forms freezes out as
water ice.

In this particular case, there is a column of $N$(\oh)=$2\times 10^{11}$ cm$^{-2}$ in the first surface peak
and $4\times 10^{11}$ cm$^{-2}$  in the deeper peak; for \hto\ the columns are $1.4\times 10^{11}$ cm$^{-2}$
and $2\times 10^{11}$  cm$^{-2}$; and for \htho\ the columns are $1.1\times 10^{11}$ cm$^{-2}$ and
$9.4\times 10^{13}$ cm$^{-2}$.    The columns of \oh\  and \hto\ are not detectable.   Typically,
columns of $\gta 10^{12}$ cm$^{-2}$ are needed for detection via absorption spectroscopy.    In particular, 
observations of \oh\ often imply columns $\gta 10^{13}$ cm$^{-2}$, which suggests that higher cosmic ray rates are 
required.   Therefore, for the rest of our standard cases we use \crt $= 2\times 10^{-16}$ s$^{-1}$.
Most of the \htho\ column is produced in the second deeper peak, and in our model the predicted
columns are detectable.  For an absorption measurement a background submillimeter source behind  or in a cloud with $A_V>6$ from
observer to submillimeter source is required.

Figure 3 shows the case $n=10^2$ cm$^{-3}$, $\chi= 1$, and \crt = $2\times 10^{-16}$ s$^{-1}$: in other
words, the same as Figure 2 but with ten times the cosmic ray ionization rate.   The abundances 
of all three ions in the first peak rise in proportion to \crt\ .   The \oh\ and \hto\ ion abundances in the second peak 
also scale roughly with \crt.   The column of $N$(\oh)=$2.2\times 10^{12}$ cm$^{-2}$ in the first  peak
and $3.2\times 10^{12}$ cm$^{-2}$  in the deeper peak; for \hto\ the columns are $1.5\times 10^{12}$ cm$^{-2}$
and $6.9\times 10^{12}$  cm$^{-2}$; and for \htho\ the columns are $9.2\times 10^{11}$ cm$^{-2}$ and
$7.3\times 10^{13}$ cm$^{-2}$.     Because the electron abundance is low in the second peak, a significant fraction ($.1-0.3$) of
 cosmic ray ionizations of H$_2$
lead to \oh\ and \hto, and their destruction is by H$_2$, which does not change in abundance with varying \crt.
Thus, the scaling of \oh\ and \hto\ abundances and columns with \crt.\footnote{Even at the second peak, both electrons and CO
compete with O in reacting with H$_3^+$.}
However, the \htho\ abundance in the second peak does not rise linearly with \crt, but stays fairly constant, because although the formation rate
scales with \crt, the destruction rate also increases as \crt\  is raised, due to the higher electron abundances produced
by the enhanced cosmic ray flux. Unlike \oh\ and \hto, which are destroyed by H$_2$, \htho\ is destroyed by
dissociative recombinations with electrons. 
In fact, to first order, we would expect the abundance of \htho\
to scale as \crt/$n_e=$\crt$/(x_en)$ at the second peak.  

The abundance of electrons $x_e$ deep in the cloud depends on the uncertain PAH abundance deep in the cloud.
 If the PAH abundance remains as high as is indicated in diffuse clouds and cloud surfaces (as we assume
 in our standard models),
then we obtain the following result.   Electrons are formed by cosmic ray ionization of H$_2$.  Electrons are 
mainly destroyed by collisional attachment to neutral PAHs, and the PAHs are mostly neutral.    Therefore,
the abundance of electrons $x_e \propto $\crt$ /n$.  
As a result, if PAHs are abundant in this deep peak, we predict that $x$(\htho) $\propto$ \crt$/(x_en)$
 will be independent of both $n$ and \crt!   Note that Figure 3 compared to Figure 2 shows that the electron abundance in the second
peak does scale roughly as \crt, and that the abundance of \htho\  in the second peak does not change significantly as we increase the cosmic ray ionization rate by 10. Appendix B presents an analytic solution for $x$(\htho) near the second peak if PAHs are present.  Since cosmic ray ionization of H$_2$ is similar to X-ray ionization of H$_2$, this
result implies that, {\it if PAHs are present}, regions of enhanced X-ray ionization will not show enhanced \htho\  columns.
We discuss in \S 3.1.3 the case of no PAHs at high $A_V$.

Figure 4 shows the same case as Figure 3, but plots a parameter $\epsilon$, first discussed by
N10.   Here, the parameter $\epsilon$ is defined as the rate per unit volume of formation of \oh\ 
divided by the total (not primary) rate per unit volume of cosmic ray ionization of H and H$_2$.  In effect, $\epsilon$ is
an efficiency parameter in determining the formation of \oh\ from cosmic rays.    If PAH, PAH$^-$ and electron abundances 
are relatively low, and H$_2$ and gas phase O abundances are high, then $\epsilon$ is near unity.  Essentially,
one needs O$^+$ to react with H$_2$ before H$^+$ reacts with e, PAH, or PAH$^-$.  Although this
qualitative limit is clear from the chemical pathways shown in Figure 1, we derive in Appendix B
an analytic formula for $\epsilon$ which makes this statement more quantitative.   Roughly, the condition
for $\epsilon$ to be of order unity is:
 
 \begin{equation}
 \left({ x({\rm H_2})\over 0.5}\right)\left( {x({\rm O})\over 10^{-4}}\right) \gta 0.028e^{230/T} \left[\left({x_e \over 3\times 10^{-4}}\right)
 +4.4\left({x({\rm PAH^-})\over 1.5\times 10^{-8}}\right) + 2.7\left( {x({\rm PAH})\over 1.85\times 10^{-7}}\right) \right]
\end{equation}
However, if the reverse is true,
then H$^+$  can recombine with PAH, PAH$^-$, or electrons and disrupt the chain of reactions that
lead to \oh, leading to low ($<1$) values of $\epsilon$.  Figure 4 also
plots the temperature and repeats the plots of the abundances of H$_2$ and electrons, since they help determine the value of $\epsilon$, as well as the abundance of \oh\ (see Appendix B).    Finally, we add the abundance of H$_3^+$ to Figure
4 since it is also used to estimate cosmic ray rates (e.g., Indriolo et al 2007, 2011).    Because H$_3^+$ is formed by
the reaction of H$_2$ with H$_2^+$ and often destroyed by electrons (see Figure 1), we see the H$_3^+$ abundance rise with
$A_V$ as the H$_2$ abundance rises and the electron abundance falls.    In general, the H$_3^+$ probes the
cosmic ray ionization rates at higher $A_V$ than the first peak in \oh.

  Figures 5  and 6 show  the case $n= 10^4$ cm$^{-3}$, $\chi = 100$, and \crt = $2 \times 10^{-16}$ 
s$^{-1}$.  This case is nearly identical to the standard case in H09, and is
representative of a Giant Molecular Cloud (GMC) surface illuminated by an FUV field somewhat higher than
the interstellar radiation field because of the presence of nearby O and B stars.    The total cosmic
ray ionization rate ($\sim 3-5\times 10^{-16}$ s$^{-1}$ including secondary ionizations) may represent values
on the surfaces of GMCs, but may be somewhat high for the interior.   We see from Figures 4 and 6 that the gas
temperatures of the molecular interiors of these clouds at $A_V > 5$ is of order 30 K for $n=10^4$ cm$^{-3}$
and 70 K for $n= 100$ cm$^{-3}$, due to cosmic ray heating when \crt = $2\times 10^{-16}$ s$^{-1}$.   
 Typically, temperatures in molecular cloud interiors
are observed to be $\lta 30$ K, suggesting that such high cosmic ray ionization rates may not be appropriate
for molecular cloud interiors.   However, as we shall see, such high cosmic ray rates are required to explain observations
of diffuse clouds, which should have the same cosmic ray rates as the surfaces ($A_V \lta 2$) of GMCs.  This suggests
that \crt\ may be higher on the surface of a molecular cloud than deep in its interior.

If chemistry is driven by FUV photoreactions and particle-particle reactions, such as H/H$_2$
chemistry, then the chemical abundances mainly depend on the ratio $\chi / n$.   Therefore,
one expects and sees that the H$_2$ abundance of Figure 5  closely matches that of Figures 2 and 3, which have the same
$\chi / n$ ratio.   The H$_2$ abundance in the higher $\chi$ case is
a bit lower because of FUV photodissociation of H$_2$ in FUV-pumped excited vibrational states of
H$_2$.    

However, the molecular ion abundances in the first peak depend to first order on
the ratio \crt /$n$ (see Appendix B) and are not too sensitive  to $\chi$.   Thus, in this $n= 10^4$ cm$^{-3}$ case, their abundances in
the first peak drop by a factor of nearly 10-30 compared to Figure 3 (which has the same value of \crt)  as the density rises by 100 from  the $n= 10^2$ cm$^{-3}$ assumed in Figure 3.  Slight differences in electron abundances, H$_2$ abundances, and $T$ explain the divergence from the expected $n^{-1}$
dependence.   

The second peaks of \oh\ and \hto\  nearly scale as $n^{-1}$, as expected.  However, the second
peak of the \htho\ abundance ($\sim 10^{-8}$) is independent of $n$, as predicted above by the scalings 
of the electron density with $n$ and \crt\ if PAHs are present. 

Figures 7  shows the case $n=10^6$ cm$^{-3}$, $\chi=10^5$, and \crt = $2\times 10^{-16}$ s$^{-1}$. This case
may represent strongly illuminated PDRs such as may occur around embedded compact or ultracompact HII regions, or
possibly embedded protostars illuminating the opaque walls of outflow cones.
This high density and high FUV flux case was chosen because most of the column of all three ions
is produced not by
cosmic ray ionization, but by other chemical reactions.  In Figure 7 we see an enormous enhancement of \oh\
abundance at $A_V \sim 1.6$.  Here, $T \sim 1000$ K and at the same time
the abundance of H$_2$ is moderately high, $\sim 10^{-2}$.   At these elevated temperatures, as discussed in \S 2.1, the H$_2$ can
 react rapidly with O  to form OH or with C$^+$ to form CH$^+$, leading to reaction chains that make \oh, \hto, and \htho.  
 One of the key heating mechanisms providing this high $T$ is the FUV pumping and the H$_2$ formation pumping of
 excited vibrational levels of H$_2$, followed by collisional de-excitation of these levels which leads to gas heating (e.g.,
 Tielens \& Hollenbach 1985).
 The essential point is that the \oh, \hto, and \htho\ 
columns are not provided by cosmic ray ionization, and therefore cannot diagnose the cosmic ray
ionization rate, except to give an upper limit.

\subsubsection{Contour plots of integrated columns of ions}

Figure 8 shows the contours of the integrated (from $A_V=0$ to $A_V=20$) columns
of \oh\ ions for a primary cosmic ray ionization rate of \crt=$2\times 10^{-16}$ s$^{-1}$ per H atom, and plotted as functions of $n$ and $\chi$.
  Note that the \oh\  columns include
both the first and second peaks.  We emphasize that
these are columns perpendicular to the face of the PDR slab.   If clouds are observed
obliquely, proportionately more column will be in the line of sight.  Similarly, if we are observing a slab
illuminated on both sides, the columns will be raised by a factor of 2 if the slab is quite optically thick ($A_V >>1$)
so that first and second peaks occur on both sides. The upper right hand portion of the figure is blacked out
because radiation pressure, photoelectric emission, and photodesorption forces on dust grains, when $\chi/n_2 \gta 300$,
 drives dust rapidly through the PDR
(Weingartner \& Draine 2001b).   Such high ratios of $\chi/n$ rarely occur in nature, and require a much more
detailed PDR code.
 
 Figure 8 shows two main features.   First, at low $\chi \lta 1000$ or low $n \lta 10^4$
 cm$^{-3}$, the columns of \oh\
 are roughly proportional to $n^{-1}$ and almost independent of $\chi$, as discussed above.  There is a weak
 dependence on $\chi$ for $\chi \lta 1000$ because higher $\chi$ leads to higher gas $T$, and the
 O$^+$ abundance rises with exp(-230 K/$T$).  From this figure it is clear that since \oh\ columns of at
 least $10^{12}$ cm$^{-2}$ are needed to make absorption observations feasible, low density ($n \lta 300$
 cm$^{-3}$)  clouds should be targeted if  \crt= $2\times 10^{-16}$ s$^{-1}$.  It should also be noted, 
 as is obvious in Figures 2, 3 and 5, that 
 both the first and second peaks contribute to the \oh\  column so that there is substantial column at both
 $A_V <1$ and $A_V > 1$.   The \oh\ in this case is quite cold, $T \lta 200$ K.   Second, these relations 
 completely break down in the upper right portion 
 ($n \gta 10^4$ cm$^{-3}$ and $\chi \gta 10^3$) of this contour plot.   Here, as discussed above, the high $\chi$
 and $n$ (high density brings the H/H$_2$ interface closer to the surface, where the gas is warm) lead to
 very warm ($T\sim 1000$ K) H$_2$ near the surface ($A_V \lta 2$) of the cloud.   This warm H$_2$ drives reactions
 which produce \oh\ without the need for cosmic ray ionization.   In this way, observable columns
 of \oh\ can be formed at these elevated values of $\chi$ and $n$, and the columns occur at moderate
 $A_V \sim 2$.  The \oh\ in this case is quite warm, $T \gta 300$ K, and its column is independent of \crt, but depends
 on $n$ and $\chi$.

Figure 9 shows the contours of the integrated (from $A_V=0$ to $A_V=20$) columns
of \hto\ ions for a primary cosmic ray ionization rate of \crt=$2\times 10^{-16}$ s$^{-1}$ per H atom.   This figure can be compared
with Figure 8, which treats the same cosmic ray case for \oh.   We focus on regions where these ions might
be detectable, that is, for columns $> 10^{12}$ cm$^{-2}$, and in the cosmic ray dominated zones of $n$ and $\chi$.   
In these low density regions, comparison of Figure 9 with 8 
reveals that the column ratios are of order unity, as might be expected.   The formation rates of these two ions
are nearly the same (stemming from the cosmic ray ionization rate) and the destruction rates of these two molecules
are nearly the same (via H$_2$ in the regions where most of the columns are generated).  
Note that  in Figures 2, 3,  and 5 the local \oh\ abundance peaks at lower values of $A_V$ than the \hto\  abundance,
 and that Figures 8 and 9 present integrated columns through a high $A_V$ slab.   Therefore, we can obtain
 higher column ratios of these two molecular ions if we truncate our slabs to small diffuse clouds with $A_V     <1$
 (see below, \S 3.2.2).
 
 We do not provide a contour plot similar to Figures 8 and 9 but with different \crt\ because the results are simple to
 describe.   For $\chi \lta 1000$ the \oh\ and \hto\ columns scale as \crt.   For $\chi \gta 1000$ the columns are independent
 of \crt\ because cosmic rays do not produce the \oh\ and \hto.
 Forcing a fit with the expected (to first order)
\crt/$n$ dependence in the cosmic ray-dominated region $\chi \lta 100$, we find roughly (factor of 2) the simple
expressions
 \begin{equation}
 N(\rm{OH^+}) \sim 5\times 10^{12}  \chi^{1/3} \left({{100\ \rm{cm^{-3}}}\over n}\right)\left({\zeta_{crp} \over {2\times 10^{-16} 
 \ \rm{s^{-1}}}}\right)
 \ \rm{cm^{-2}}.
 \end{equation}
 and
  \begin{equation}
 N(\rm{H_2O^+}) \sim 10^{13}\chi^{1/4}  \left({{100\ \rm{cm^{-3}}}\over n}\right) \left({\zeta_{crp} \over{ 2\times 10^{-16} \ \rm{s^{-1}}}}\right)
 \ \rm{cm^{-2}}.
 \end{equation}
We emphasize that these columns are the total \oh\ and \hto\ columns in both the first and second peaks.

We do not present integrated  (from $A_V=0$ to $A_V=20$) columns of \htho\  ions because, as discussed above, most
of their column arises at high $A_V\sim 6$ and this second peak column is quite constant as a function of $n$, $\chi$, and \crt\ 
if PAHs are present at the same abundances as in diffuse clouds.
With PAHs present in the second peak, we find that the \htho\ column is slightly dependent on the gas density, varying from $10^{14}$ cm$^{-2}$ at $n \sim 100$
cm$^{-3}$ to $10^{13.5}$ cm$^{-2}$ at $n\sim 10^6$ cm$^{-3}$.  At densities of $n= 100$ cm$^{-3}$ and with $\chi=1$
and \crt=$2\times 10^{-16}$ s$^{-1}$, we remind the reader (see discussion of Figure 3) that $N$(\htho)$= 9.2\times 10^{11}$
cm$^{-2}$ in the first peak; like \oh\ and \hto\ the column of \htho\ in the first peak scales with \crt/$n$ so that even lower densities or higher \crt \ are needed to make \htho\ detectable in the first peak (e.g., in a diffuse foreground cloud).


\subsubsection{Sensitivity to PAH abundance and other parameters}

PAHs may not have a high abundance deep in the high $A_V$ regions of a molecular cloud, perhaps due to coagulation of
PAHs on larger grains (see, for example, discussion in H09).    We have run a 
number of cases with the PAH abundance
set to zero to see the effect on the second peaks.  In these runs, we also reduce the number of small grains by setting
the minimum grain size to 100 \AA, but we still refer below to this case as the "no PAH" case for high $A_V$ cloud interiors .  

The results depend on the gas phase abundance of metals, like S, Si, Mg, and Fe; see Table 2 of Appendix A for our 
assumed gas phase abundances of these species in the interstellar radiation field.    Our code follows the freezeout of
these species as a function of depth into the cloud, as photodesorption no longer is able to keep the metal atoms off the grain
surfaces.    The photodesorption yields of these species are not known.   We find that for yields greater than 10$^{-6}$, there
are still sufficient gas phase metals present at the second peak to supply electrons and suppress the \htho \ abundance by factors
of $\gta 10$ relative to the case with PAHs.    However, if the photodesorption yields are very low so that the gas phase 
abundances of these metals at the second peak are $\lta 3\times 10^{-8}$, then we obtain \htho \ columns in the second
peak comparable to the case with PAHs.    For example, assuming $n= 10^{4}$ and $\chi=100$, we obtain $N$(\htho)$=4\times 10^{13}$ cm$^{-2}$ if \crt/$n_2=2\times 10^{-16}$ s$^{-1}$ per H atom, and $N$(\htho)$=8\times 10^{13}$ cm$^{-2}$ if \crt/$n_2=2\times 10^{-15}$ s$^{-1}$ per H atom. The main difference between the PAH and no PAH
case lies in the electron loss mechanisms: with PAHs the electrons are lost by encountering PAHs; with no PAHs the
electrons are lost by recombining with ions.    The atomic ions have slower rates of recombination than the molecular ions,
so the presence of metal atomic ions leads to higher electron abundances and lower \htho\ columns.   The \oh\ and \hto\ abundances and columns also decrease in the second peak, but by smaller factors ($\lta 3$),
because their destruction is not by electrons but by H$_2$, whose abundance does not change. 

In Appendix C we discuss the lack of sensitivity of the results on certain key reaction rate coefficients, the assumed abundance
of gas phase oxygen,  and on
the rate coefficient for H$_2$ formation on grain surfaces.   We discuss the effect of {\it raising} the gas phase abundances
of metals.   We also treat the increase in the \htho\ columns in the second peak if elemental O does not freeze (as water ice) at high $A_V$.

\subsection{Diffuse and Translucent Clouds}

   Diffuse clouds have relatively small columns, $A_V \lta 1$, whereas translucent clouds
   have columns intermediate between diffuse clouds and GMCs, $1 \lta A_V \lta 5$.   Although
   GMCs have sufficient column to incorporate both peaks of the \oh, \hto, and \htho\ abundances,
   diffuse and translucent clouds typically (with the possible exception of the highest $A_V$ translucent
   clouds) only contain the surface peak.   In fact, a diffuse cloud may have such small $A_V$ that
   the cloud may truncate the full peak of one or all of the ions.   Therefore, the parameter
   $A_{Vt}$, the total $A_V$ through the cloud becomes an important new parameter in
   determining, for example, ratios of the columns of the ions.  
   
   Both diffuse and translucent clouds have insufficient column to be gravitationally bound, so generally they
   are in thermal pressure equilibrium with the interstellar medium.    The local typical value of
   the thermal pressure is $nT \sim 3-4\times 10^3$ cm$^{-3}$ K, and this pressure may rise
   by factors of 2-3 for clouds in the molecular ring of the Galaxy (Wolfire et al 2003).  As discussed above (\S 2.2), depending on
   the galactocentric radius $R$,  they
   have densities 30 cm$^{-3} < n < 100$ cm$^{-3}$, temperatures $T \sim 50-100$ K,  and incident FUV fluxes characterized
   by 1$ < \chi < 4$.   The ratio $\chi/n _2$, critical to the photochemistry, varies typically from about 3-6, although we explore
   a somewhat larger range 3-10 here.
   
   In the low density regime, $n< 10^3$ cm$^{-3}$, appropriate for diffuse and translucent clouds, there is a scaling of
   the results of thermochemical models.    The parameters which control the thermochemical structure of a cloud
   are just $\chi/n$, \crt/$n$, and $A_{Vt}$.    We note that this scaling does not apply to the denser 
    PDR models discussed in the previous section.
   At high density, $n>10^3$ cm$^{-3}$, the cooling of the gas is suppressed by collisional de-excitation of the
   fine structure states of C$^+$ and O.   Therefore, higher density gas will typically give higher gas temperatures,
   even when $\chi/n$ is held constant.   However, given that this scaling  applies for diffuse and translucent clouds,
   we generally in this section plot our results as functions of $\chi/n$, \crt/$n$, and $A_{Vt}$.  We note that this implies that
   the observations may give us a measure of \crt/$n$, but to get an estimate of \crt \ we will then have to estimate the
   gas density $n$ in the region observed.
   

  \subsubsection{The total column of \oh\ through a diffuse cloud of size $A_{Vt}$}
      
   Figure 10 plots the column of \oh, $N$(\oh), as a function of $A_{Vt}$ for four values of \crt/$n_2$
   (recall, $n_2 \equiv n/ 100$ cm$^{-3}$) and for $\chi/n_2= 3.16$.   The  dotted lines plot $x_c($H$_2)$,
   the abundance of H$_2$ at cloud center, and these values appear on the right of the figure.  Recall $A_{Vt}$ is the total
   $A_V$ through the diffuse slab, measured perpendicular to the slab and the ion columns plotted are perpendicular
   to the slab.
   
   Figure 10 shows that $N$(\oh) increases with \crt/$n$ and with $A_{Vt}$, although
   one sees a saturation of the column with $A_{Vt}$ at $A_{Vt} \gta 0.3$ for the lower three cases of \crt/$n$.  This saturation
   is due to the peaking of the local \oh\ abundance at lower values of $A_V$.  Figure 10 shows 
   that $N$(\oh) correlates with $x_c($H$_2)$, and that moderate molecular hydrogen abundances
    $x_c($H$_2) \gta 0.03$ are required to obtain substantial columns of
   \oh.    Observed \oh\ columns in broad ($\Delta v \sim 20$ km s$^{-1}$, see \S 4.1.1) velocity components in the ISM are of
   order $3\times 10^{13}-3\times 10^{14}$ cm$^{-2}$.  With this value of $\chi/n_2=3.16$, we see that it requires very high cosmic ray
   ionization rates, \crt/$n_2 \gta 3\times 10^{-16}$ s$^{-1}$ and high $A_{Vt}\gta 0.3$ for a single cloud, seen face on,
   to produce these columns.   Therefore, it appears that geometrical effects
   such as seeing several clouds along the line of sight, or viewing the cloud obliquely, may be required to match
   observation.   In fact, as we discuss in \S 4.1.1, many clouds may to required to cover the broad ($\Delta v \sim 20$ km s$^{-1}$) absorption
   components since single clouds would have much narrower velocity widths.   Because of geometrical effects and variation in $A_{Vt} $
   and $\chi/n$, the observation of the column of \oh\ is not sufficient to tightly restrict \crt/$n$.
   
    Figure 11 plots  $N$(\oh) as a function of $A_{Vt}$ for four values of \crt/$n_2$
   and for $\chi/n_2= 10$: in other words, the same as Figure 10 but
   with a ratio of $\chi/n$ that is 3.16 times higher.       The main effect of raising $\chi/n$
   is to push the H/H$_2$ transition deeper into the cloud to higher $A_V$.   This is seen in the plot of $x_c($H$_2)$ which
   rises to 0.01 at $A_{Vt} \sim 0.3$ in this case, compared to $A_{Vt}\sim 0.1$ in the previous case of $\chi/n_2= 3.16$.
   Since H$_2$ is required
   to make \oh, this moves the peak of \oh\  to higher values of $A_V$, and thus pushes the 
   rise of $N$(\oh) to higher values of $A_{Vt}$ than in Figure 10.
   However, this first peak has more column of \oh\ compared with the case $\chi/n_2= 3.16$.   Therefore, to produce
   the observed columns of \oh\ now requires only \crt/$n_2 \gta 3\times 10^{-17}$ s$^{-1}$ and high $A_{Vt}\gta 1$ for
    a single  cloud, seen face on.
   
  \subsubsection{$N$({\rm \oh})/$N$({\rm \hto}) and  $N$({\rm \oh})$/N$({\rm H}) through a diffuse cloud of size $A_{Vt}$}
 
    The results on the total column of \oh\ suggests that other observations must be brought into play to further restrict
    the cosmic ray ionization rate.    Since \hto\ is often observed as well, we first consider the ratio  $N$(\oh)/$N$(\hto).
    As seen in Figures 2, 3, and 5, the \hto\ abundance peaks at higher values of $A_V$ than the \oh\ abundance.
    Therefore, the ratio of the columns of these two ions may give us some measure of $A_{Vt}$.
    
    Figure 12 plots the ratio  $N$(\oh)/$N$(\hto) as a function of $A_{Vt}$ for four values of \crt/$n_2$
    and for $\chi/n_2=3.16$.     As expected, at low
   values of $A_{Vt}\lta 0.1$, the ratio  $N$(\oh)/$N$(\hto) is high, $\sim 10-30$, as we have not yet reached
   high enough values of $A_V$ to include the \hto\ peak.  However, as $A_{Vt}$ increases, the ratio of the columns
   continues to drop until at $A_{Vt} \gta 1$, the ratio is approximately 2-3, except in the case of the highest
   cosmic ray rate \crt/$n_2 = 3.16\times 10^{-15}$ s$^{-1}$.\footnote{The \crt/$n_2= 3.16\times 10^{-15}$ s$^{-1}$ case is 
   extreme;  cosmic ray ionization of H
   produces an electron abundance of $x_e\simeq 3\times10^{-3}$.   In this case, electrons and not H$_2$ dominate
   the destruction of \oh\ and \hto\ even when $x($H$_2)\sim 0.5$.   Therefore, the fraction of cosmic ray ionizations that produce \oh \ is higher than the
   fraction that produce \hto, even at the peaks of their abundance.   Hence we have much more \oh \ than \hto as seen in this figure, and,
    once
   we reach high enough $A_{Vt}$ where the ion chemistry shown in Figure 1(top) dominates, the ratio rises as a function of $A_{Vt}$. This holds as well
   in X-ray ionized regions with high ionization rates. }  Once we have incorporated both the \oh\ and the \hto\ 
   abundance peaks, the column ratios are near unity, as noted in \S 3.1.

   Figure 13 plots the ratio  $N$(\oh)/$N$(\hto) as a function of $A_{Vt}$ for four values of \crt/$n_2$ 
   and for $\chi/n_2= 10$: in other words, the same cases as
   Figure 12 except the ratio $\chi/n$ is raised by a factor of 3.16.  
   The rise in $\chi/n$ pushes the H/H$_2$ transition and the peaks of \oh\ and \hto\ to higher values of $A_V$.
   Therefore, compared to Figure 12, we see the drop in the  $N$(\oh)/$N$(\hto) ratio at higher values of $A_{Vt}$.
   The observed ratios of  $N$(\oh)/$N$(\hto) range from 3-15.   If $\chi/n_2= 3.16$ and if we assume that
   \crt/$n_2 \lta 3\times 10^{-16}$ s$^{-1}$, Figure 12 suggests that $A_{Vt} \sim 0.1 -0.3$ to  obtain these ratios.
   However, if this is true, Figure 10 implies that even with as high a value of \crt/$n_2 = 3\times 10^{-16}$ s$^{-1}$,
   we will need a ``geometrical factor" of $\sim 10$ in order to obtain $N$(\oh)$\sim 10^{14}$ cm$^{-2}$.
   This geometrical factor is the combination of many clouds along the line of sight, along with the enhancement in
   column due to viewing angle of the cloud.   The situation changes with $\chi/n_2= 10$, as seen in Figure 13.  Here,
   again assuming that \crt/$n_2 \lta 3\times 10^{-16}$ s$^{-1}$, the observed ratios can be obtained with $A_{Vt} \sim 1-3$.
   In this case, using Figure 11, the geometrical factor needs to be only $\sim 1-3$ to produce \oh\ columns of $10^{14}$
   cm$^{-2}$ when \crt/$n_2 \sim 3\times 10^{-16}$ s$^{-1}$.
   
   Observations of HI 21 cm are often available along the same lines of sight as the  \oh\ and \hto\ measurements.   In this case,
   the column of HI, $N$(H), along the line of sight associated with the velocity feature of the ions can be estimated.  To be precise,
   the observations directly give $N$(H)/$T$, where $T$ is an average temperature along the line of sight.
    In Figures 14 and 15 we plot the ratio $N$(\oh)/[$N$(H)/$T_2$]  on the vertical axis  ($T_2=T/100$ K) and the ratio  $N$(\oh)/$N$(\hto)
   on the horizontal axis for two fixed values of $\chi/n_2=3.16$ and $10$.  The results of our 2 sided diffuse cloud models are shown as contours
   on this figure.    As noted above, higher average abundances of \oh\ (or $N$(\oh)/$N$(H))
   requires higher \crt/$n$.  On the other hand, higher $N$(\oh)/$N$(\hto) requires either low $A_{Vt}$ or high $\chi/n$.  In the next
   section we apply Figures 10, 11, 14, and 15 to the observational data to constrain \crt/$n$ along sightlines to W49N.

  \subsubsection{The possible effects of turbulence}
   We have run additional models to estimate how turbulence might affect the production of \oh, \hto, and \htho.   
It has been recognized for several decades that standard astrochemical models for diffuse molecular clouds greatly underpredict the observed column densities ($\sim 10^{13}$ cm$^{-2}$) of CH$^+$ (e.g., Falgarone et al 2010a,b; and references therein.)  The models presented here are no exception; 
in our models with $\chi \sim 1$, $n \sim 100$ cm$^{-3}$, and $A_{Vt} >1$, for example, we obtain a predicted CH$^+$ column density
$ \sim 3\times 10^{10}$ cm$^{-2}$, more than two orders of magnitude lower than that typically observed.
Both shock waves with large scale order (Elitzur \& Watson 1978, Draine \& Katz 1986, Flower \& Pineau des Forets 1998), and interstellar turbulence (e.g.\ Joulain et al.\ 1998; Lesaffre et al 2007, Godard et al 2009, Falgarone et al 2010a,b) have been proposed as sources of heating that could enhance the CH$^+$ abundance within some small fraction of the cloud volume; in these models, CH$^+$ is formed by the endothermic reaction of C$^+$ with H$_2$.
It has also been long recognized (e.g. Draine \& Katz 1986) that models invoking ion-neutral drift -- either in C-type shock waves or in turbulent dissipation regions -- are most successful in simultaneously matching the observed column densities of CH$^+$ and OH.  In particular, while models in which the gas temperature is merely elevated without significant ion-neutral drift tend to overpredict the OH/CH$^+$ ratio  (the neutral-neutral endothermic reaction $\rm O + H_2 \rightarrow OH + H$ being enhanced along with the reaction of C$^+$ with H$_2$), models incorporating ion-neutral drift preferentially enhance the endothermic ion-molecule reactions that produce CH$^+$ (and SH$^+$; Godard et al. 2012), {\it without} producing more OH than the observations permit.
We have crudely estimated the effects of turbulence by positing an enhanced
rate of endothermic ion-neutral reactions in a small fraction of the cloud volume.  Motivated by recent calculations performed by Myers \& McKee (private communication), we adopted enhanced ion-neutral reaction rates - equal to the thermal rate coefficients at 1000 K - in 3\% of the cloud volume: this leads to a predicted CH$^+$ column in accord with what is typically observed.\footnote{In particular, we add 97\% of the columns from our standard model to 3\% of our columns from the same model but
with the enhanced ion-neutral rates to obtain the total columns of, for example, CH$^+$.} We found that turbulence does not greatly enhance the \oh, \hto, and \htho\ column densities through such a cloud.  There was a modest enhancement of these oxygen hydride ions within the region of ion-neutral drift, due to the faster reaction of H$^+$ with O, but since that region occupies only a small fraction of the total volume, this effect did not affect the total column densities significantly.  Therefore, we conclude 
that turbulent dissipation or shocks in small regions of the clouds  are unlikely to affect our conclusions concerning OH$^+$, \hto\ or \htho\ in the  models presented here.


  \section{OBSERVATIONS AND COMPARISON WITH MODEL PREDICTIONS}

\subsection{OH$^+$, H$_2$O$^+$, and H$_3$O$^+$ Absorption in Diffuse Galactic Molecular Clouds}

As discussed in \S1, recent {\it Herschel} observations of bright Galactic continuum sources have revealed OH$^+$ and H$_2$O$^+$ absorption features arising in multiple foreground clouds along the targeted sight-lines (Gerin et al.\ 2010; N10; Neufeld et al.\ 2011).   The three crosses in Figure 14  denote the $N({\rm OH^+})/N({\rm H_2O^+})$ and $N({\rm OH}^+)/[N({\rm H})/T_2]$ ratios for the two such sources for which results have been reported to date: G10.6--0.4 (a.k.a.\ W31C) (the lower limits) and W49N (2 points to the right).
Those given for G10.6--0.4 apply to all the material along the sight-line, while those given for W49N apply to two separate velocity intervals (30-50 km s$^{-1}$ and 50-78 km s$^{-1}$).  The OH$^+$ absorption in G10.6--0.4 is largely saturated, and thus the plotted values are lower limits. For W49N, the columns of H derived assumed $T=100$ K or $T_2=1$; the quoted
values of $N$(H)/$T_2$ are $6.95\times 10^{21}$ cm$^{-2}$ for the $30-50$ km s$^{-1}$ feature and $7.23\times 10^{21}$  cm$^{-2}$ for the
$50-78$ km s$^{-1}$ feature.   The total columns $N$(\oh) observed in
these two velocity intervals are $3.6\times 10^{14}$ and $2.1\times 10^{14}$ cm$^{-2}$, respectively.  

The $N({\rm OH^+})/N({\rm H_2O^+})$ ratios measured toward W49N ($\sim 10$) and the wide velocity range  of the absorbing clouds suggest that the absorbing material is comprised of multiple clouds of small extinction.  The line of sight to W49N is approximately 11 kpc long, and passes
as close as $R=5$ kpc to the Galactic center.   Assuming $\chi/n_2=3.16$, models with $A_{Vt} = 0.32$ and $0.25$  best account for all the data (Figure 14), with cosmic-ray ionization rates of $\sim 6$ and $ 4 \times 10^{-16}\, n_2 \, \rm s^{-1}$, respectively, for the $30 - 50$ km s$^{-1}$ and $50 -
78$ km s$^{-1}$ velocity intervals.   As seen in Figure 10 for $\chi/n_2=3.16$,
the column of \oh\ in a single cloud of size $A_{Vt}=0.32$ is $N$(\oh)$ \simeq 2\times 10^{13}$ cm$^{-2}$ and for $A_{Vt}=0.25$ is
$N$(\oh)$ \simeq 1.5\times 10^{13}$ cm$^{-2}$.    Therefore, we either need $\sim 15$ clouds along the line of sight, or a smaller number but with geometrical enhancement effects.    Note that a single cloud might have an \oh\ absorption
linewidth of only $\sim 1-3$ km s$^{-1}$.   Therefore, to cover the broad absorption features we are modeling ($\Delta v = 20$ or 28 km s$^{-1}$),
we need 10-30 clouds spread out along the line of sight so that their galactic rotational velocities coupled with their individual line widths cover this velocity range.   Therefore, this $\chi/n_2=3.16$ model has barely enough clouds to produce the relatively smooth absorption feature observed. Assuming $\chi/n_2=10$,  models with $A_{Vt}\simeq 0.75$  and $A_{Vt}\simeq 0.63$ and with \crt/$n_2 \sim  2\times 10^{-16}$ and
$1.2\times 10^{-16}$ s$^{-1}$, respectively, best account for the data (see Figure 15).   With these values of  $A_{Vt}$ and \crt/$n_2$, inspection of Figure 11 shows that we require 
$\sim 12$  clouds along the line of sight.    This value of $\chi/n_2$ has even fewer clouds to cover the broad absorption line observed. Moreover,
the temperatures in clouds with $\chi/n_2=10$ are higher ($\sim 200$ K) than is typically observed near the solar neighborhood (Heiles \&
Troland 2003).    And finally, the Wolfire et al (2003) results suggest that along the line of sight to W49N,
the average $\chi/n_2 \sim 4$.   A similar modeling of the case $\chi/n_2=1$, not shown in the figures, gives \crt/$n_2=2.5\times 10^{-15}$ and
$2.0\times 10^{-15}$ s$^{-1}$ with $A_{Vt}= 0.16$ and 0.08, respectively.    Again, because of the high cosmic ray rates, the gas temperatures
are high in these clouds ($\sim 150$ K).   In addition, these cosmic ray rates seem improbably high.    This model is driven to high cosmic ray
rates in part because otherwise (with low $\chi/n_2$) the gas is cold, which reduces the rate that H$^+$ can charge exchange with O.    Therefore, we finally conclude that our models suggest typical $A_{Vt} \sim 0.3$, $\chi/n_2\sim 3$,  and \crt/$n_2 \sim
6\times 10^{-16}$ s$^{-1}$ for the $30-50$ km s$^{-1}$ component, and \crt/$n_2\sim 4\times 10^{-16}$ s$^{-1}$ for the $50-78$
km s$^{-1}$ component.

A better approach is to ask,  ``What distribution of cloud $A_{Vt}$ or hydrogen nucleus column $N$ through the cloud will we encounter in traversing
the $\sim 11$ kpc to W49N?"   HI 21 cm observational data suggest that $dn_{cl}/dN \propto N^{-2}$, where $n_{cl}$ is the number of
clouds for $N\gta 2.6 \times 10^{20}$ cm$^{-2}$, and proportional to $N^{-1}$ for $N\lta 2.6 \times 10^{20}$ cm$^{-2}$  (Heiles \& Troland 2005).\footnote{Recall that $N$ is the column of H nuclei; however, in the above range, generally $N\simeq N$(H).}
We have crudely integrated a distribution of clouds with this dependence, with the constant of proportionality derived by requiring the integral
to obtain a total column of $N$(H)$=7\times 10^{21}$ cm$^{-2}$, the observed atomic H column in each velocity component of W49N.   We use
$N=6\times 10^{21}$ cm$^{-2}$ as an upper limit for the integration; above this, we enter small number statistics since the total column is of this order.   The lower
limit does not enter the integration significantly, as long as it is $<<2.6\times 10^{20}$ cm$^{-2}$, since these clouds contain very little of the total column
of any species.
The integration removes
$A_{Vt}$ as a variable, and leaves only the parameters \crt/$n_2$ and $\chi/n_2$ as free parameters to match the observed \oh/\hto\ ratio
and the total column of \oh.   We find that with $\chi/n_2=3.16$ and  \crt/$n_2=3.7 \times 10^{-16}$ s$^{-1}$, we obtain $N$(\oh)$\simeq 2.1\times 10^{14}$
cm$^{-2}$ and $N$(\hto)$\simeq 2.4\times 10^{13}$ cm$^{-2}$.    These values match the velocity interval $50-78$ km s$^{-1}$ extremely well.
 Increasing the cosmic ray rate to \crt/$N_2=6.3\times 10^{-16}$ s$^{-1}$, we obtain $N$(\oh)$\simeq 3.6\times 10^{14}$
cm$^{-2}$ and $N$(\hto)$\simeq 4.0\times 10^{13}$ cm$^{-2}$, a very good fit to the lower velocity component.  Considering the small number statistics, especially of the higher $A_{Vt}$ clouds, this is very good agreement.   Interestingly, the main contribution to $N$(\oh) comes from near
$A_{Vt}=0.2-0.6$.   This is because of the dependence of $N$(\oh) on $A_{Vt}$ (see Figure 10).   Note that  $N$(\oh) rises sharply with 
$A_{Vt}$ until about $A_{Vt}=0.3$, and then it levels off.    Therefore, the integral is weighted to that size cloud in the range of the turnover
($A_{Vt} \sim 0.3-0.6$), which has the peak \oh\ 
abundance (averaged through the cloud). In addition, the cloud distribution steepens when $A_{Vt}>0.13$, thereby decreasing the contribution
from higher $A_{Vt}$ clouds.  The total number of clouds in the range $A_{Vt}\simeq 0.1-1$ is approximately 20.  We note that the \hto\ columns
have greater contribution from higher $A_{Vt}$ clouds, which are less numerous, and therefore we predict the \hto\ absorption feature
to have greater fluctuations.  

We next examine the cases $\chi/n_2=1$ and 10.   Assuming $\chi/n_2 = 10$, we find that we need  cosmic ray rates of $1.1\times 10^{-15}$ s$^{-1}$  and $7\times 10^{-16}$
s$^{-1}$ to match the \oh\ columns in the two velocity components, respectively.   However, the \oh/\hto \ column ratio is 12.1, somewhat higher
than observed, and again, the clouds are too warm ($\sim 200 $ K).   The high field pushes the peaks of
\oh\ and \hto\  deeper into the cloud, and our clouds tend to truncate the \hto\ column, leading to the high \oh/\hto\ ratio.  We need much higher
cosmic ray rates than our solution above for $\chi/n_2=10$ with clouds of single $A_{Vt} \sim 0.7$ because the cloud distribution leads
to a significant column of H from clouds of low $A_{Vt}$ which have very little \oh.
Assuming $\chi/n_2 = 1$, we find that we need  cosmic ray rates of $2\times 10^{-15}$ s$^{-1}$  and $1.2\times 10^{-15}$
s$^{-1}$ to match the \oh\ columns in the two velocity components, respectively.   However, the \oh/\hto column ratio is 5.6, lower
than observed.    With low $\chi/n_2$, the clouds tend to contain both \oh\ and \hto \ peaks, driving the ratio down.  We conclude that
the best fit is with $\chi/n_2=3.16$, giving \crt/$n_2\sim  6\times 10^{-16}$ s$^{-1}$  and $4\times 10^{-16}$ s$^{-1}$ for the two velocity components, in good agreement
with our analysis above that did not use the cloud distribution.    However, the cloud distribution gives added weight to the conclusion
that $\chi/n_2 \sim 3$.

The cosmic ray rates derived from our models are roughly 2-4 times larger than those inferred by N10, who used a simple analytic treatment to infer cosmic ray ionization rates of $0.6$ and $1.2 \times 10^{-16}\, (n_2 / \epsilon) \, \rm s^{-1}$, where $\epsilon$ is the fraction of cosmic-ray ionizations that lead to OH$^+$. 
Using pure gas-phase model results from the Meudon PDR code (Le~Petit et al.\ 2006), N10 found that $\epsilon$ lay in the range 0.5 -- 1.0 for a wide variety of cloud conditions; this would imply cosmic ray ionization rates in the range $\sim 0.6 - 2.4 \times 10^{-16}\,n_2 \, \rm s^{-1}$ for the material along the sight-line.  The main source of the difference between our models and those of N10 is our inclusion of PAH$^-$ -- H$^+$ recombination and the charge exchange of H$^+$ with
neutral PAH in the present work,  processes -- previously omitted -- that reduce the fraction of cosmic-ray ionizations that lead to OH$^+$ production (in our
models $\epsilon\sim 0.1-0.3$ in the regions where most of the \oh\ lies).  

Figures 16 and 17 repeat Figures 14 and 15 except that the rates of PAH and PAH$^-$ neutralization of H$^+$  have been reduced by $>4$, so that they are negligible compared to the radiative recombination of H$^+$ with electrons (see Appendix B).  Note that this reduction could be achieved
by either lower PAH abundances or lower PAH rate coefficients with H$^+$, or a combination of both.   Another possibility is increased
photoionization rates of PAH$^-$ and PAH.   Figure 16 shows that the best fit to the data for $\chi/n = 3.16$  is $A_{Vt} \sim
0.316$ and \crt/$n_2 \sim 1-2 \times 10^{-16}$ s$^{-1}$, in complete agreement with N10.   Therefore, the PAH chemistry is extremely important
in deriving cosmic ray rates from the \oh\ and \hto\ ions.   We consider the rates from Figures 16 and 17  lower limits to the cosmic ray rates, whereas the rates from Figures 14 and 15 could be considered upper limits, although they represent our best estimates of PAH  chemistry at this time.

The cosmic-ray ionization rates suggested by the predictions shown in Figure 14 are also a factor $\sim 5-10$ larger than those typically inferred from observations of H$_3^+$ in the Galactic disk (Indriolo et al.\ 2007, Indriolo \& McCall 2012).  Indriolo \& McCall conclude, however, that cosmic ray rates vary by more
than an order of magnitude over various sight lines, from \crt$\sim  0.7 \times 10^{-16}$ s$^{-1}$ to $4.6 \times 10^{-16}$ s$^{-1}$, with a mean
of $\sim 1.5\times 10^{-16}$ s$^{-1}$ (note that we have converted their total ionization rate per H$_2$ to our primary ionization rate per H
nucleus by dividing their values by $2.3$; we also note that their typical density was $n\sim 200$ cm$^{-3}$ so their results imply \crt/$n_2$ that
are two times smaller than these values.  
  Their highest values of \crt/$n_2$ are still a factor of $\sim 2$ below what we find for W49N, so there appears to be a small discrepancy.   We point out that the
cosmic ray rate derived from H$_3^+$ observations is directly proportional to the assumed electron abundance.    Indriolo \& McCall assumed that the electrons were supplied by C$^+$, with an abundance of  $1.5\times 10^{-4}$.  However, in our models of diffuse clouds, where typically the gas density is $\sim 100$ cm$^{-3}$, the cosmic ray ionization of H can produce comparable or greater abundances of H$^+$ 
than C$^+$.  Therefore, we find for \crt/$n_2 = 2\times 10^{-16}$ s$^{-1}$, for example, that $x_e\simeq 3 \times 10^{-4}$ when $x$(H$_2$)$\simeq 0.1$ and $x_e\simeq 2 \times 10^{-4}$ when $x$(H$_2$)$\simeq 0.4$ .  This implies that the Indriolo et al rates may need to be increased, by factors as much as $\sim 2$, depending on \crt/$n_2$.     Given the 
small number of sources toward which $N({\rm OH}^+)/N({\rm H_2O^+})$ and $N({\rm OH}^+)/N({\rm H})$ ratios have so far been reported, the proven 
variation in cosmic ray rates along different
lines of sight, the lack of a common sight line to compare the cosmic ray rates derived by H$_3^+$ versus by \oh and \hto,  the possibility of higher electron abundances than assumed by Indriolo and coworkers, and the lack of certainty in PAH chemistry, it is not yet possible to draw firm conclusions about whether our results differ significantly from that of Indriolo et al.  What is clear is that cosmic ray rates in the diffuse ISM are higher than was previously thought.

{\it Herschel} has also detected absorption by $\rm H_3O^+$ in foreground material along the sight-line to G10.6 -- 0.4 (Gerin et al.\ 2010).  The {\it average} $\rm H_3O^+$/$\rm H_2O^+$ ratio along the sight-line is 0.7, but the distribution of $\rm H_3O^+$ is clearly different from that of H$_2$O$^+$ and OH$^+$, with the $\rm H_3O^+$ concentrated within a single narrow velocity component where the H$_2$ fraction is presumably largest.  This cloud is unassociated with the source, but is probably the foreground cloud with the largest $A_V$ along the line of
sight.   It appears in the HF spectrum (Neufeld et al. 2010) and CO spectrum (C. Vastel, private communication; see also Corbel \& Eikenberry
2004), suggesting a high
molecular fraction and therefore substantial $A_V$.   The \htho\ column in this narrow component is $\sim 1-2\times 10^{13}$ cm$^{-2}$, consistent with our predictions for the second peak in a high $A_V$ cloud.

\subsection{H$_3$O$^+$ Emission from Dense Galactic Molecular Clouds}

 PvDK92 searched for the 396, 364, and 307 GHz inversion-rotation lines of \htho\ in 20 Galactic targets -- including GMCs, star forming regions, 
Galactic center sources  and a few evolved stars -- and reported definitive detections of \htho\ emission 
from the G34.3+0.15 hot core, two positions in Sgr B2 (a Galactic center GMC),
and the W3 IRS5, W3(OH), W51M and Orion KL high-mass star forming regions.
In the case of Orion and Sgr B2, \htho\ emissions had been previously discovered by Wootten et al.\ (1991).
The \htho\  column densities derived by PvDK92 for the high-mass cores W3 IRS5, W3(OH), W51M  are fairly similar,
ranging from a few $\times 10^{13}$ to a few $\times 10^{14}$ cm$^{-2}$, and arguing for a similar environment and formation mechanism.  Given the large critical densities for the observed transitions, PvDK92 inferred relatively high densities ($n>10^6$ cm$^{-3}$) and temperatures  ($T>50$ K) for the \htho\-emitting gas.
The \htho\ column densities observed in these high-mass cores are in good agreement with the predictions of our models for either the case in which PAHs and/or very small grains are assumed to be present or the case where PAHs and low ionization potential metals are highly depleted\footnote{If PAHs and very small grains coagulate into larger aggregates deeper inside the cloud (e.g., Boulanger et al. 1990; Rapacioli et al. 2006) or onto larger grain surfaces, 
our models would predict \htho\  column densities that are roughly an order of magnitude lower than the values inferred from
the above observations, unless low ionization potential metal atoms are highly depleted at the second peak of \htho.}); the PAH case predicts \htho\ column densities of $3\times 10^{13}-10^{14}$ cm$^{-2}$, regardless of density $n$ or $\zeta_{crp}$.  

Herschel's HIFI instrument has allowed the detection of far-infrared and submillimeter \htho\  lines that are 
not accessible to ground-based observatories.  Several \htho\  emission lines in the THz domain have been reported towards high-mass YSOs in the W3 IRS5 region (Benz et al. 2010). 
A rotation diagram combining HIFI data with lower frequency data from PvDK92 indicates a rotational temperature $T_{rot}\sim 239$ K, suggesting that the observed emission may arise either from hot and dense gas or is radiatively excited by continuum photons from hot dust grains. Benz et al. (2010) inferred $N($\htho$)=8.5 \pm2 \times 10^{13}$ cm$^{-2}$ and interpreted the \htho\  emission as arising from the outflow walls heated and irradiated by the FUV radiation field from massive YSOs.
Their inferred \htho\  column densities and rotational temperatures are consistent with  our models of dense gas illuminated by a relatively strong FUV field (where large kinetic temperatures are achieved at relatively low $A_V \sim 1-2$).

\subsection{Combined H$_3$O$^+$ Emission and Absorption in Strong Far-infrared Continuum Sources}

For sources where a strong infrared radiation field is present, submillimeter \htho\ line emission can be accompanied
 by far-infrared absorption within an extended envelope.   The \htho\ {J$_K$=$2_{1}^{-}-1_{1}^{+}$} and
{$2_{0}^{-}-1_{0}^{+}$}   inversion-rotation and {$1_{1}^{-}-1_{1}^{+}$} pure-inversion FIR absorption lines at 100.577, 100.869 and 181.054 $\mu$m  were detected by ISO towards the optically thick FIR continuum emission of Sgr B2 (Goicoechea \& Cernicharo 2001). 
An \htho\  column density of $\sim 1.6\times 10^{14}$ cm$^{-2}$ was inferred in the warm and low density extended envelope of Sgr B2 ($n=10^3 - 10^4$ cm$^{-3}$, $\chi=10^3 - 10^4$ from [OI] and [CII] observations; Goicoechea et al. 2004).  However, the p-\htho\ {$3_{2}^{+}-2_{2}^{-}$} 364 GHz emission line was subsequently mapped around the Sgr B2(M) core at 18'' angular resolution with APEX by van der Tak et al. (2006), who inferred 
column densities $\sim 10^{15}-10^{16}$ cm$^{-2}$ on the basis of a detailed excitation model.  
This value is much larger than the estimates obtained 
from FIR absorption lines, presumably because the latter only probe the outer envelope of the source whereas the submillimeter observations probe material to a much larger depth.  The \htho\ column densities inferred by van der Tak et al.\ (2006) are much larger than our model predictions for externally-illuminated clouds; this discrepancy likely reflects the presence of a strong internal luminosity source in Sgr B2 that heats the dust grains within the interior and prevents the freeze-out of oxygen nuclei (see Appendix C).

\subsection{Extragalactic, OH$^+$, \hto\ and \htho\ }

Thanks to the much improved sensitivity of space- and ground-based receivers and detectors, molecular
ions can now be detected outside the Milky Way. 
Extragalactic \htho\  was first detected through JCMT observations of the p-\htho\  364 GHz emission line towards the prototypical ultraluminous infrared galaxy Arp 220, and towards M82, an
evolved starburst (van der Tak et al. 2008).   \htho\ was subsequently detected towards 
IC342, NGC253, NGC1068, NGC4418 and NGC 6240, and upper limits obtained towards IRAS 15250 and Arp 299 galaxies (Aalto et al. 2011).
Except for IC342 and M82, 
the typical  \htho\  column densities ($\sim ~10^{15}-10^{16}$ cm$^{-2}$)  derived from extragalactic \htho\ detections are much larger than the predictions of our models for single clouds.  These observations are of \htho\ in emission and the authors assume $T_{ex}\sim T_{rad}$ to obtain these columns with error bars of factor of 2.  
As in the case of Sgr B2, the discrepancy between our models and the observations may reflect the effect of enhanced dust temperatures that prevent the freeze-out of elemental oxygen.  Alternatively, there may be a large number of clouds along the line of sight,

For the starburst galaxies M82 and IC342 , however, where \htho\ column densities 
($\sim 10^{14}$ cm$^{-2}$) are inferred from the observations, enhanced dust temperatures or ionization rates are not apparently required.  
For example, van der Tak et al (2008) inferred $N$(\htho)$\simeq 1.1\times 10^{14}$ cm$^{-2}$ in M82 from an emission line which contained a combination of a broad ($\Delta v \sim260$ 
km s$^{-1}$) and a narrow ($\Delta v \sim 40$ km s$^{-1}$) component.    In addition,  a recent {\it Herschel}/HIFI detection (Weiss et al. 2010) of the o-\hto\ $1_{11} - 0_{00}$ ground-state line in absorption towards M82 revealed $N$(\hto)$\simeq 2.9\times 10^{14}$ cm$^{-2}$,
arising from a $\Delta v \simeq 77$ km s$^{-1}$ feature.   The large width of the lines suggest numerous clouds in the beam.   
Overall, these numbers are consistent with our models with the \hto\ arising from the first (and possibly second) peak of numerous relatively
low density clouds, whereas the \htho\ arises from the second peaks of many molecular clouds in the beam.

\section{Summary and Conclusions}

We model the production of  \oh, \hto, and \htho\   in interstellar clouds, using a steady state photodissociation region
code that treats the freeze-out of gas species, grain surface chemistry, and desorption of ices from grains.  The
code includes PAHs, which have an important effect on the chemistry.   

As a function of depth or $A_V$ into a molecular cloud, the ions tend to have two peaks, one at low $A_V \lta 1$ and one at 
high $A_V\sim 6$, the exact value depending on $\chi/n$.  In the first peak the ions are created by the cosmic ray ionization of
H to H$^+$, followed by reactions with O and H$_2$ (see Figure 1 top).   These peaks appear on the surfaces of molecular clouds
as well as in diffuse clouds.  PAHs can lower the production of the ions here, by neutralizing H$^+$ and interrupting the reaction
chain (Figure 1 top).  At most, they depress the ion abundances and columns by a factor of $\sim 3$.

In molecular clouds
a significant fraction of the column density of \oh and \hto \ is found in the first peak at the surface ($A_V <1$) of the cloud.   For relatively low values of the incident 
far ultraviolet flux on the cloud ($\chi \lta 1000$), the columns of \oh\  and \hto\ 
scale as the cosmic ray ionization rate divided by the gas density. Roughly, the columns
of \oh\ , \hto, and \htho\ in the first peak are $2.2\times 10^{12} \chi^{1/3} $(\crt/$2\times 10^{-16}$ s$^{-1}$) (100 cm$^{-3}$/$n$)
cm$^{-2}$,   $1.5\times 10^{12} \chi^{1/4}$ (\crt/$2\times 10^{-16}$ s$^{-1}$) (100 cm$^{-3}$/$n$)
cm$^{-2}$, and $9\times 10^{11} \chi^{1/3} $(\crt/$2\times 10^{-16}$ s$^{-1}$) (100 cm$^{-3}$/$n$)
cm$^{-2}$.

There is a second peak in \oh, \hto, and \htho \ abundances at larger depths ($A_V \sim 6$) in molecular clouds,  when the second route to \oh\ formation, initiated by the cosmic
ray ionization of H$_2$ becomes dominant (Figure 1 bottom).   Here, lower electron abundances enhance the abundances of the
ions by lowering the electronic recombination rate of H$_3^+$ (which raises the abundance of H$_3^+$ and the formation rates of the three ions), and by lowering
the electronic recombination of \htho \ (its main destruction path).
  However, even deeper in the cloud, the oxygen freezes out as water ice on the grain surfaces,
and the ion abundances fall as their formation rates fall, being 
starved for gas phase elemental oxygen.  
If PAHs or VSGs  are present
at these high values of $A_V\sim 6$, the electron abundance at the second peak is controlled by association with neutral PAH or VSG.  
 In this case, rather surprisingly, the column 
of \htho\  ($\sim 4\times 10^{13}$ cm$^{-3}$) in the second peak and the
peak abundance ($\sim 10^{-8}$) of \htho\  is independent
of both \crt\ and $n$.      Raising the cosmic ray ionization rate increases the production rate of \htho, but it also increases the destruction
rate by electrons by the same amount.    If PAHs  and VSGs are not
present at high $A_V$, the column of  \htho\  depends mainly on the gas phase abundance of elemental S, Si, Mg, and Fe
at the second peak.   If these abundances are low, $\lta 3\times 10^{-8}$, then columns comparable to the PAH case are obtained.
However, for higher abundances of the metals, the \htho\ column is reduced by a factor of roughly 10.
The columns of \oh\ and \hto\ in the second peaks are usually of order of  the columns in the first peaks.

The observed \htho \ columns of $\sim 4\times 10^{13}$ cm$^{-2}$ seen in many dense Galactic molecular clouds  therefore imply 
that either PAHs or VSGs are present deep in molecular  clouds or that the depletion of PAHs and VSGs are accompanied by a very
significant depletion of gas phase metals.    There are a few observations that imply much higher column ($\sim 10^{15}-10^{16}$ cm$^{-2}$) of \htho, and these can only be explained in the context of our models as arising in very high $A_V$ clouds where high gas
phase elemental O abundances are present throughout due to either desorption processes or time dependent effects.  We suspect
that the grains in these sources may be so warm, $\gta 100$ K, that thermal desorption keeps water ice from depleting the oxygen reservoir.
In the case of extragalactic sources with very broad velocity features, the columns may be increased by the presence of a large number
of clouds along the line of sight.

 For high values of the incident far ultraviolet flux ($\chi \gta 1000$)
and high gas densities ($\gta 10^4$ cm$^{-3}$), producing warm ($T>300$ K) gas with significant H$_2$ abundance
at $A_V \sim 1-2$, chemical reactions initiated by the photoionization of carbon 
in the mainly atomic surface can
form ionized hydrogen, which then leads to the formation of  \oh, \hto, and \htho.    In this case, their columns (typically,
of order $3\times 10^{13}$ cm$^{-2}$)
are not related to the cosmic ray ionization rates.    \htho \ emission from W3 IRS5 may be an example of such conditions.

We also model diffuse and translucent clouds in the interstellar medium,
and show how observations of $N$(\oh)/$N$(H)  (typically $10^{-8} - 10^{-7}$) and $N$(\oh)/$N$(\hto) 
(typically 3-15) can be used to estimate \crt/$n$. The ratio  $N$(\oh)/$N$(H), which is essentially the average abundance of \oh\ 
in all the clouds along the line of sight (within the same absorption velocity component), is mainly a measure of \crt/$n$. 
 The ratio of the \oh\  column
to the \hto\  column in diffuse clouds is mainly dependent on the ratio $\chi/n$ and  $A_{Vt}$ (i.e, the total hydrogen column
through a single typical cloud).  If $\chi/n$ is known, or at least constrained to a narrow range such as $\chi/n_2 \sim 3-6$
typical of diffuse clouds in random locations along the line of sight in the Galaxy, then observations of $N$(\oh)/$N$(H),
 $N$(\oh)/$N$(\hto), and $N$(\oh) can determine \crt/$n$, $A_{Vt}$ , and a geometrical factor which is a combination of the
 number of clouds along the line of sight and the typical angle that the line of sight makes as it passes through each diffuse cloud slab.
 Using the observed distribution of $A_{Vt}$ in diffuse clouds, the models can provide an estimate of $\chi/n_2$.

We discuss the relation of our models to recent observations 
of \oh\ and \hto\  by the Herschel Space Observatory, and the ability to constrain the cosmic ray ionization rate 
through comparison of observations with these models. We conclude that the cosmic ray primary ionization rates
\crt\ in the observed foreground diffuse clouds towards W49N have values
of approximately \crt$\sim 4-6\times 10^{-16}(n/100\ {\rm cm^{-3}})$ s$^{-1}$,  if our adopted PAH chemistry is correct.\footnote{We emphasize the need for further theoretical and laboratory work to investigate PAH reaction rates.}   We find a hard lower limit
of \crt/$n_2 \gta 1-2 \times 10^{-16}$ s$^{-1}$ by neglecting PAH chemistry in W49N.  Our best fit models suggest that $\chi/n_2 \sim 3$ in the diffuse
clouds towards W49N.    Our W49 models suggest that, in terms of producing \oh\ and \hto\ column, the typical $A_{Vt} \sim 0.3$ through
a single cloud towards W49N. and that a diffuse cloud $A_{Vt}$ distribution measured by Heiles \& Troland (2005) fits the data very well.   To produce the total column of \oh\  and H observed in the two velocity components requires $\sim 20$ clouds in each component along the line of sight.  We discuss differences between our estimates
and those of N10 and Indriolo et al (2007), pointing out the former neglected PAH chemistry and the latter may have slightly underestimated
the electron abundance in the diffuse foreground clouds with the highest cosmic ray ionization rates.     We look forward to further observations of
\oh\ and \hto\  along many sight lines to probe the cosmic ray ionization rates throughout the Galaxy.

\bigskip
\centerline{\bf Acknowledgements}
\medskip

We would like to thank A. Benz,  J.H. Black, E. Falgarone,   B. Godard, C. Heiles, V. Ossenkopf, 
and E. van Dishoeck  for many useful
discussions.  We especially thank M. Gerin, N. Indriolo, and the anonymous referee for  careful readings of the manuscript and many useful
suggestions.   We gratefully acknowledge the support of a grant from the NASA Herschel Science Center's Theoretical
Research/Laboratory Astrophysics Program.  Partial support for MGW, MJK and DJH was provided by a NASA Long Term Space Astrophysics Grant NNG05G46G.
 JRG is supported by a Ram\'on y Cajal research contract and thanks the Spanish
MICINN for funding support through grants AYA2009-07304 and CSD2009-00038.
\clearpage

\appendix{Appendix A: Tables of rate coefficients and adopted abundances}

Table 1 presents some of the key reactions and/or some of the reaction rates that have changed since H09.  We emphasize that this is
not a complete listing of the $\sim 300$ chemical reactions in the PDR code described in H09.   The PAH rates are scaled by the factor 
$\Phi_{\rm PAH}$, a scaling factor introduced in Wolfire et al (2008) that incorporates  the uncertainties in PAH reaction rates, PAH sizes,
and PAH abundances.   Wolfire et al (2008) found from a semi-empirical fit of our PDR models to C, C$^+$, H, and H$_2$ column densities in
diffuse clouds that $\Phi_{\rm PAH} \sim 0.5$, and we have adopted that value in all of our PDR models.

Table 2 presents the standard gas phase abundances and grain properties that we have adopted in most of our PDR models.   As discussed 
in the text, we have also run models with these values changed to test the sensitivity of the results to the standard values.  In particular, we have run 
models with $x$(PAH)= 0;  with $x$(Si),  $x$(Fe), $x$(S), and $x$(Mg) all set to $10^{-8}$ or $10^{-5}$; and with $x$(O) = $4.5\times 10^{-4}$.                     
\begin{deluxetable}{ll}
\tablecaption{Reaction Rates\label{tbl:reactions}}
\tablehead{
\colhead{Reaction} &
\colhead{Rate Coefficient} 
}
\startdata
${\rm PAH^-} + {\rm H^+} \rightarrow {\rm PAH^0} + {\rm H}$
     & $8.1\times 10^{-7} \Phi_{\rm PAH} (T/300\; {\rm K})^{-0.50}$ 
      $\cms$ \tablenotemark{a} \\
${\rm PAH^0} + {\rm H^+} \rightarrow {\rm PAH^+} + {\rm H}$
     & $7.0\times 10^{-8} \Phi_{\rm PAH}$ $\cms$ \tablenotemark{a}\\
${\rm PAH^+} + e \rightarrow {\rm PAH^0}$
     & $3.4\times 10^{-5} \Phi_{\rm PAH} (T/300\; {\rm K})^{-0.50}$ 
      $\cms$ \tablenotemark{a}\\
${\rm PAH^0} + e \rightarrow {\rm PAH^-} $
     & $3.0\times 10^{-6} \Phi_{\rm PAH}$ $\cms$ \tablenotemark{a}\\
${\rm PAH^-} +{\rm C^+} \rightarrow {\rm PAH^0} + {\rm C}$ 
     & $2.3\times 10^{-7}\Phi_{\rm PAH} (T/300\; {\rm K})^{-0.50}$ 
     $\cms$ \tablenotemark{a,b} \\
${\rm PAH^0} + {\rm C^+}\rightarrow {\rm PAH^+} + {\rm C}$ 
    & $2.0\times 10^{-8}\Phi_{\rm PAH}$  $\cms$ \tablenotemark{a,b} \\
${\rm PAH^0} + h\nu \rightarrow {\rm PAH^+} + e$ &
         $2.8\times 10^{-8}\chi \exp (-2.34 A_{\rm V})$ ${\rm s^{-1}}$ 
          \tablenotemark{c, d}\\
${\rm PAH^-} + h\nu \rightarrow {\rm PAH^0} + e$ &
         $5.7\times 10^{-7}\chi \exp (-1.09 A_{\rm V})$ ${\rm s^{-1}}$ 
         \tablenotemark{c, e}\\
${\rm PAH^0} + h\nu \rightarrow {\rm PAH^+} + e$ &
         $3.5\times 10^{-8}\chi \exp (-2.45 A_{\rm V})$ ${\rm s^{-1}}$ 
          \tablenotemark{f, d}\\         
${\rm PAH^-} + h\nu \rightarrow {\rm PAH^0} + e$ &
         $1.7\times 10^{-7}\chi \exp (-1.77 A_{\rm V})$ ${\rm s^{-1}}$ 
         \tablenotemark{f, e}\\
${\rm C} + h\nu \rightarrow {\rm C^+} + e$ &
         $3.1\times 10^{-10}\chi \exp (-3.33 A_{\rm V})$ ${\rm s^{-1}}$ 
         \tablenotemark{g}\\
${\rm H_2O} + h\nu \rightarrow {\rm H_2O^+} + e$ &
         $3.1\times 10^{-11}\chi \exp (-3.90 A_{\rm V})$   ${\rm s^{-1}}$ 
         \tablenotemark{g}\\
${\rm H_2O} + h\nu \rightarrow {\rm OH} + {\rm H}$ &
         $7.5\times 10^{-10}\chi \exp (-1.70 A_{\rm V})$   ${\rm s^{-1}}$ 
         \tablenotemark{g, h}\\
${\rm H_2O} + h\nu \rightarrow {\rm O} + {\rm H_2}$ &
         $4.8\times 10^{-11}\chi \exp (-2.20 A_{\rm V})$   ${\rm s^{-1}}$ 
         \tablenotemark{g, h}\\
${\rm OH} + h\nu \rightarrow {\rm O} + {\rm H}$ &
         $3.9\times 10^{-10}\chi \exp (-1.70 A_{\rm V})$  ${\rm s^{-1}}$ 
         \tablenotemark{g,h}\\
${\rm OH} + h\nu \rightarrow {\rm OH^+} + e$ &
         $2.2\times 10^{-11}\chi \exp (-4.05 A_{\rm V})$  ${\rm s^{-1}}$ 
         \tablenotemark{i}\\
${\rm OH^+} + h\nu \rightarrow {\rm O^+} + {\rm H}$ &
         $1.1\times 10^{-11}\chi \exp (-3.50 A_{\rm V})$ ${\rm s^{-1}}$   
         \tablenotemark{g}\\
${\rm CH^+} + h\nu \rightarrow {\rm H^+} + {\rm C}$ &
         $3.3\times 10^{-10}\chi \exp (-2.94 A_{\rm V})$ ${\rm s^{-1}}$   
         \tablenotemark{g}\\
${\rm H}  + {\rm CR} \rightarrow {\rm H^+} + e$    &
          $\zeta_{crp}$ ${\rm s^{-1}}$ \tablenotemark{j}\\
${\rm H_2}  + {\rm CR} \rightarrow {\rm H_2^+} + e$    &
          $2\zeta_{crp}$ ${\rm s^{-1}}$ \tablenotemark{j}\\
${\rm H^+} + e \rightarrow {\rm H}$  &
         $3.5\times 10^{-12} (T/300\; {\rm K})^{-0.75} $ $\cms$  
         \tablenotemark{k}\\
${\rm H_2^+} + e \rightarrow {\rm H} + {\rm H}$  &
         $1.6\times 10^{-8} (T/300\; {\rm K})^{-0.43} $ $\cms$  
         \tablenotemark{k,l}\\
${\rm H_3^+} + e \rightarrow {\rm H_2} + {\rm H}$  &
         $3.4\times 10^{-8} (T/300\; {\rm K})^{-0.50} $ $\cms$  
         \tablenotemark{m}\\
${\rm H_3^+} + e \rightarrow {\rm H}  +{\rm H} + {\rm H}$  &
         $3.4\times 10^{-8} (T/300\; {\rm K})^{-0.50} $ $\cms$  
         \tablenotemark{m}\\
${\rm OH^+} + e \rightarrow {\rm O} + {\rm H}$  &
         $3.8\times 10^{-8} (T/300\; {\rm K})^{-0.50} $ $\cms$  
         \tablenotemark{k,l}\\
${\rm H_2O^+} + e \rightarrow {\rm H_2} + {\rm O}$  &
         $3.9\times 10^{-8} (T/300\; {\rm K})^{-0.50} $ $\cms$  
         \tablenotemark{k}\\
${\rm H_2O^+} + e \rightarrow {\rm OH} + {\rm H}$  &
         $8.6\times 10^{-8} (T/300\; {\rm K})^{-0.50} $ $\cms$  
         \tablenotemark{k}\\
${\rm H_2O^+} + e \rightarrow {\rm O} + {\rm H} +{\rm H}$  &
         $3.1\times 10^{-7}  (T/300\; {\rm K})^{-0.50} $ $\cms$  
         \tablenotemark{k}\\
${\rm H_3O^+} + e \rightarrow {\rm OH} + {\rm H} + {\rm H}$ &
         $3.4\times 10^{-7} (T/300\; {\rm K})^{-0.74} $ $\cms$  
         \tablenotemark{n}\\
${\rm H_3O^+} + e \rightarrow {\rm H_2O} + {\rm H}$ &
         $1.4\times 10^{-7} (T/300\; {\rm K})^{-0.74} $ $\cms$  
         \tablenotemark{n}\\
${\rm H_3O^+} + e \rightarrow {\rm H_2} + {\rm OH}$ &
         $7.9\times 10^{-8} (T/300\; {\rm K})^{-0.74} $ $\cms$  
         \tablenotemark{n}\\
${\rm H_3O^+} + e \rightarrow {\rm H_2} + {\rm O} + {\rm H}$ &
         $7.4\times 10^{-9} (T/300\; {\rm K})^{-0.74} $ $\cms$  
         \tablenotemark{n}\\
${\rm H_3^+} + {\rm CO} \rightarrow {\rm HCO^+} + {\rm H_2}$ & 
          $1.7\times 10^{-9}$ $\cms$ \tablenotemark{k}\\
${\rm C^+} + e\rightarrow {\rm C} + h\nu$ &
         \tablenotemark{o} \\
${\rm C^+} + {\rm OH}\rightarrow {\rm CO^+} + {\rm H}$ & 
     $2.9\times 10^{-9} (T/300\; {\rm K})^{-0.33}$ $\cms$
         \tablenotemark{p}  \\
${\rm C^+} + {\rm H_2}\rightarrow {\rm CH^+} + {\rm H}$ & 
     $1.0\times 10^{-10} \exp (-T/4640\; {\rm K})$ $\cms$
         \tablenotemark{k}  \\
${\rm CO^+} + {\rm H}\rightarrow {\rm CO} + {\rm H^+}$ & 
     $7.5\times 10^{-10}$ $\cms$ 
         \tablenotemark{k}  \\
${\rm O} + {\rm H^+}\rightarrow {\rm O^+} +{\rm H}$ & 
         \tablenotemark{q} \\
${\rm O} + {\rm H}\rightarrow {\rm OH} +{h\nu}$ & $9.9\times 10^{-19} (T/300\; {\rm K})^{-0.38}$ 
         \tablenotemark{k} \\         
${\rm O^+} + {\rm H}\rightarrow {\rm H^+} +{\rm O}$ & $5.7\times 10^{-10} (T/300\; {\rm K})^{-0.36}e^{8.6\; {\rm K}/T}$ $\cms$
         \tablenotemark{q} \\
${\rm O^+} + {\rm H_2}\rightarrow {\rm OH^+} +{\rm H}$ & $1.7\times 10^{-9}$ $\cms$
         \tablenotemark{k} \\
${\rm OH^+} + {\rm H_2}\rightarrow {\rm H_2O^+} +{\rm H} $ & $1.0\times 10^{-9}$ $\cms$
         \tablenotemark{k} \\
${\rm H_2O^+} + {\rm H_2}\rightarrow {\rm H_3O^+} +{\rm H} $ & $6.4\times 10^{-10}$ $\cms$
         \tablenotemark{k} \\
${\rm H_2^+} + {\rm H_2}\rightarrow {\rm H_3^+} +{\rm H} $ & $2.1\times 10^{-9}$ $\cms$
         \tablenotemark{k} \\
${\rm H_3^+} + {\rm O}\rightarrow {\rm OH^+} +{\rm H_2} $ & $8.4\times 10^{-10}$ $\cms$
         \tablenotemark{k} \\
${\rm H_3^+} + {\rm O}\rightarrow {\rm H_2O^+} +{\rm H} $ & $3.6\times 10^{-10}$ $\cms$
         \tablenotemark{k} \\
${\rm H_2} + {\rm O}\rightarrow {\rm OH} +{\rm H} $ & 
    $3.40\times 10^{-13} (T/300\; {\rm K})^{2.67} \exp(-T/3160\; {\rm K})$ $\cms$
         \tablenotemark{k} \\

\enddata
\noindent 

${\rm H_2O^+}$ has no photodissociation channels longward of 13.61 eV and the
    rate is set to zero \citep{vandishoeck2006}.
\tablenotetext{a}{Non-photo PAH rates are calculated
    using the equations of \cite{draine1987}. Representative rates are
    given at  $T=300$ K for disk PAHs. $\Phi_{\rm PAH} = 0.5$ 
    from \cite{wolfire2008}.}
\tablenotetext{b}{Additional collisonal rates scale as $(m)^{-0.5}$
    where $m$ is the mass of the collision partner.}
\tablenotetext{c}{ $\chi$ is the FUV field measured in units of the 
    \cite{draine1978} field. Rate for ${\rm N_C = 100}$. The shape of the FUV and optical
     field from \cite{mathis1983} used for $\chi=1$.}
\tablenotetext{d}{Absorption cross section and ionization
     potential of circumovalene \citep[IP=5.7 eV;][]{malloci2007}, 
     and linear  yield function \citep{jochims1996}.}
\tablenotetext{e}{Absorption cross section and electron affinity 
     of circumovalene \citep[EA=1.9 eV;][]{malloci2007}, 
     and maximum yield ($Y=1$).}
\tablenotetext{f}{ $\chi$ is the FUV field measured in units of the 
    \cite{draine1978} field. Rate for ${\rm N_C = 100}$. FUV and optical
     field using a $T=30,000$ K blackbody used for $\chi>1$.}
\tablenotetext{g}{\cite{vandishoeck2006} except for the attenuation factors for H$_2$O and OH photodissociation. 
Here we adopt Roberge et al (1991), see footnote h, because  otherwise the models overproduce these molecules and O$_2$
at the second peak compared with observations (see H09).}
\tablenotetext{h}{{http://www.strw.leidenuniv.nl/~ewine/photo/index.php?file=pd.php, Roberge et al (1991)}}
\tablenotetext{i}{Rate for assumed cross-section of $1\times 10^{-17}$ ${\rm cm^2}$ and 
    threshold of 957 \AA.  Depth dependence from \cite{vandishoeck1988} for a molecule with a
    photoionization threshold wavelength of 950 \AA.}
\tablenotetext{j}{$\zeta_{crp}$ is the primary cosmic-ray ionization rate
    per hydrogen nucleus. Various rates are investigated in this paper.
    The total rate including secondary ionizations is from \cite{dalgarno1999}.}
\tablenotetext{k}{UDFA06}
\tablenotetext{l}{\cite{brian1990}}
\tablenotetext{m}{\cite{mccall2004}}
\tablenotetext{n}{P.\ Goldsmith 2011 private communication using mean total rate from \cite{neau2000} and
    \cite{jensen2000} of $5.7\times 10^{-7}$ $\cms$; branching ratios from \cite{jensen2000}}
\tablenotetext{o}{Dielectronic plus radiative recombination rates from 
\cite{badnell2003} and \cite{badnell2006}.} 
\tablenotetext{p}{OSU\_01\_2009 rate tables; http://www.physics.ohio-state.edu/~eric/research.html}
\tablenotetext{q}{State-specific rates from \cite{stancil1999}}
\end{deluxetable}

\begin{deluxetable}{lllc}
\tablewidth{25em}
\tablecaption{Gas Phase Abundances and Grain Properties\label{tbl:abundances}}
\tablehead{
\colhead{Species} & \colhead{Symbol} &
\colhead{Value}$^{\rm a}$  & \colhead{Ref} 
}
\startdata
Carbon & $x{\rm (C)}$ & $1.6\times 10^{-4}$ & b\\ 
Oxygen & $x{\rm (O)}$ & $3.2\times 10^{-4}$ &c\\ 
Silicon & $x{\rm (Si)}$ & $1.7\times 10^{-6}$ &d\\ 
Iron & $x{\rm (Fe)}$ & $1.7\times 10^{-7}$ & d\\ 
Sulfur & $x{\rm (S)}$ & $2.8\times 10^{-5}$ & d\\ 
Magnesium & $x{\rm (Mg)}$ & $1.1\times 10^{-6}$ &d\\ 
PAHs & $x{\rm (PAH)}$ & $2.0\times 10^{-7}$ &e\\ 
Grain Area & $\sigma_{\rm H}$ & $2 \times 10^{-21}$ &f\\ 
\enddata
\tablenotetext{a}{Gas-phase abundances per hydrogen nucleus.}
\tablenotetext{b}{\cite{sofia2004}}
\tablenotetext{c}{\cite{meyer1998}}
\tablenotetext{d}{\cite{savage1996} cool diffuse cloud towards 
   $\zeta$ Oph}
\tablenotetext{e}{Abundance from \cite{wolfire2003} modified
for ${\rm N_C=100}$ planar PAHs. This abundance gives a total number
of C in PAHs of $2\times 10^{-5}$ per hydrogen or $\sim 6\%$ of C
in PAHs.}   
\tablenotetext{f}{Units ${\rm cm^{2}}$ per hydrogen; \cite{hollenbach2009}}
\end{deluxetable}

\clearpage

\appendix{Appendix B: Simple Analytic Analysis of the Results}

The results in \S 3.1 and \S 3.2 
can be understood by a simple analytic chemical model that
incorporates the main physics.
 Such a model, though approximate, has the
advantage of allowing one to determine and understand the sensitivity to
various model parameters, and serves to validate the numerical model.

{\it  Analytic solution to $x$(\oh) and $\epsilon$ in the first peak  ($A_V < 1$)}.  
The top panel of Figure 1 describes the main chemical pathways to \oh \ for $A_V < 1$,
and Table 1 lists the rate coefficients used for each of these reactions.
Figure 1 omits the formation of \oh\ by the photoionization of OH, which is only important at very low $A_V \lta 0.01$
and never dominates the production of the column of \oh\  for clouds of higher $A_V$.
This minor route produces OH by the reactions O + H$_2$(vibrationally excited),  O + H, and the formation of
water ice on grains followed by desorption leading to OH.   The photoionization of OH  leads to a constant  (low) \oh\ abundance
at the cloud surface.   We also neglect that route in the analytic solution,
as it does not affect the peak abundance of \oh, nor the values of $\epsilon$ near the peak.
In addition, we approximate $\epsilon$ for this peak by only including the rate of formation of
\oh\ by the reaction H$_2$  + O$^+$
 $\rightarrow$ OH$^+$  + H. 
 
 In the following we use rate coefficients from Table 1 with two exceptions.   The Stancil et al (1999) rate coefficients
 for the reaction of H$^+$ with O and its reverse reaction are complicated, and we simplify them here, focussing
 on results for low density clouds.
We set  $\gamma _1= 4\times 10^{-10}$exp$(-230\  {\rm K}/T)$ cm$^3$ s$^{-1}$ as the rate coefficient for the reaction 
H$^+$  + O $\rightarrow$ O$^+$  + H, and $\gamma _2 = 4\times 10^{-10}$ 
cm$^3$ s$^{-1}$ as the rate coefficient for the reaction O$^+$  + H $\rightarrow$ H$^+$  + O.  Let 
$\gamma _3$ be the rate coefficient for the reaction  H$_2$  + O$^+$
 $\rightarrow$ OH$^+$  + H;  $\alpha _{e} $
 the rate coefficient for the recombination of H$^+$ with electrons: $\alpha _{P^-}$
 the rate coefficient for the neutralization of H$^+$ by PAH$^-$; and
  $\alpha _{P}$  the rate coefficient for the charge exchange of  H$^+$ with PAH.   Let $x({\rm H_2})$ be the abundance of H$_2$ with respect to H nuclei, $x_e$ the abundance of electrons,
 $x({\rm PAH})$ the abundance of neutral PAHs, $x({\rm PAH^-})$ the abundance of PAH$^-$, and $x_O$ the
 gas phase elemental abundance of O.   The solution to $\epsilon$, the efficiency of making \oh \ from cosmic rays,
 is then:
 
  \begin{equation}
 \epsilon= {1 \over {1+y}},
\end{equation}
where the parameter y is given:
 
 \begin{equation}
 y= {{\left[\gamma _2 + \gamma _3 x({\rm H_2})\right]\left[\alpha _{e} x_e
 +\alpha _{P^-} x({\rm PAH^-}) + \alpha _{P} x({\rm PAH})\right]}\over {\gamma _1 \gamma _3 x_O x({\rm H_2})}}.
\end{equation}
We utilized the fact that $\epsilon$ approaches unity when $y<1$ to set the condition in \S 3.1.1 (Eq. 1) on the abundances of O, H$_2$, electrons,
PAHs and PAH$^-$s which lead to $\epsilon \sim 1$.
One can immediately see that for low values of $x$(H$_2$), where  $y>1$, $\epsilon$ is proportional to  $x$(H$_2$).   Here, the dominant route of H$^+$ destruction is by recombination with
electrons, PAHs, and PAH$^-$s.   The small fraction of the H$^+$ created by cosmic rays that charge exchanges with O  to form O$^+$ that then reacts with H$_2$ to form \oh\  is proportional to the H$_2$ abundance.    
Substituting likely values of $x_e\sim 3 \times 10^{-4}$ (we focus on cases with high \crt/$n_2$ and here $H^+$ as well as C$^+$ contribute electrons-see Figure 4), $x$(PAH)$\sim 1.85 \times 10^{-7}$, $x$(PAH$^-$)$\sim
1.5\times 10^{-8}$, and $x_{\rm O}\sim 3.2\times 10^{-4}$, we obtain

\begin{equation}
\epsilon \simeq {{227e^{-2.3/T_2} x({\rm H_2})}\over {\left[1 +4.25x({\rm H_2}\right] \left[\left({x_e \over 3\times 10^{-4}}\right)T_2^{-0.75} + 4.4\left({
x({\rm PAH^-}) \over 1.5 \times 10^{-8}}\right)T_2^{-0.5}  + 2.7\left({x({\rm PAH}) \over 1.85 \times 10^{-7}}\right)\right]}},
\end{equation}
where $T_2\equiv T/100$ K.  For low values of $x$(H$_2) <<0.2$, and standard values of $x_e$, $x$(PAH) and $x$(PAH$^-$), this
becomes

\begin{equation}
\epsilon \simeq 28e^{-2.3/T_2} x({\rm H_2}).
\end{equation}
For higher values of $x({\rm H_2})>>0.2$, $\epsilon$ saturates
at values near unity, as nearly every cosmic ray ionization leads to a production of \oh.   With standard values of $x_e$, $x$(PAH) 
and $x$(PAH$^-$) Equation (3) then becomes

\begin{equation}
\epsilon \simeq {1\over {1+ 0.15e^{2.3/T_2}}}.
\end{equation}
These formulae are very good fits to the results seen in Figures 4 and 6.  Note, however, that they are only valid for
$x({\rm H_2})<<0.5$, since we have implicitly assumed $x$(H)$>>x($H$_2$). 

Equation (3) shows that PAH$^-$ in particular can
lower $\epsilon$ by neutralizing H$^+$.   We emphasize that equations  (3-5) are valid only if the PAH and PAH$^-$ rate coefficients we
have adopted in Table 1 are valid.   Equations (4-5) assume the standard PAH and PAH$^-$ abundances given above.    

 Rather than a simple saturated value of $\epsilon$ at high $x$(H$_2$),  however, we see
in the figures that the first peak in \oh\ shows a value of $\epsilon$ that peaks and then falls with increasing depth,
once most of the gas is H$_2$.   This is because once the gas is more than half molecular ($x({\rm H_2})>0.25$), 
the cosmic ray ionization is mostly of H$_2$, which we have not included in the analytic treatment.    
The route to \oh\ via the ionization of H$_2$ is not as efficient
at the surface, because of the high abundance of electrons which react quickly with H$_3^+$.   Thus, as  $x$(H) declines deeper into the cloud, the value of $\epsilon$ falls with increasing $A_V$.

The analytic equation for $x($\oh$)$ requires the addition of  reactions  of \oh\ with electrons and H$_2$ that destroy \oh.   Let $\gamma_4$
be the rate coefficient for the destruction of \oh\ by H$_2$, and $\gamma _5$
be the coefficient for destruction by electrons.  Then, we find

\begin{equation}
x({\rm OH^+})= {{(\zeta_{crt} /n) \epsilon }\over {\gamma_4 x({\rm H_2}) + \gamma_5 x_e}}.
\end{equation}
Here, $\zeta_{crt} \sim 1.5$\crt\ is the total rate of cosmic ray ionization, including secondaries.
Ignoring the photodissociation term, we can again scale to likely values of parameters to obtain

\begin{equation}
x({\rm OH^+}) \simeq \left[{{10^{-7} \epsilon}\over{\left({x_e\over 3\times 10^{-4}}\right)T_2^{-0.5} + 51x({\rm H_2})}}\right] \left({\zeta _{crt}\over 10^{-16}\ {\rm s^{-1}}}\right) \left({{100\ {\rm cm^{-3}}}\over n}\right).
\end{equation}
Again, we can break this equation into two regimes.  For $x($H$_2$)$<< 0.02 (x_e/3\times 10^{-4})T_2^{-0.5}$,  \oh\ is mainly destroyed by dissociative recombination with electrons and the second term in the denominator can be ignored.  Then, assuming standard parameters
for the abundances of $x_e$, $x$(PAH), and $x$(PAH$^-$), we obtain
\begin{equation}
x({\rm OH^+}) \simeq 2.1\times 10^{-6}e^{-2.3/T_2} T_2^{0.5} x({\rm H_2}) \left({\zeta _{crt}\over 10^{-16}\ {\rm s^{-1}}}\right) \left({{100\ {\rm cm^{-3}}}\over n}\right).
\end{equation}
However, if $0.02 (x_e/3\times 10^{-4})T_2^{-0.5} << x($H$_2$)$<0.2$, then H$_2$ mainly destroys \oh\ and, again using standard values
for the parameters, we obtain the peak and plateau value of the \oh\ abundance:

\begin{equation}
 x({\rm OH^+}) \simeq 4.1\times 10^{-8}e^{-2.3/T_2}  \left({\zeta _{crt}\over 10^{-16}\ {\rm s^{-1}}}\right) \left({{100\ {\rm cm^{-3}}}\over n}\right).
\end{equation}
For higher values of $x($H$_2) >0.2$, the abundance of \oh\ declines as $\epsilon$ saturates, and as $x$(H) declines leading to the less
efficient production of \oh\ initiated by cosmic ray ionization of H$_2$.  
The column of \oh\ in the first peak can then be estimated
\begin{equation}
N({\rm OH^+}) \simeq 8\times10^{12} e^{-2.3/T_2} \left({\zeta _{crt}\over 10^{-16}\ {\rm s^{-1}}}\right) \left({100\ {\rm cm^{-3}}\over n}\right) (\Delta A_V/0.1)\    {\rm cm^{-2}},
\end{equation}
where $\Delta A_V$ is the width of the peak region.

This shows the important result that the peak abundance (and approximately the column of \oh\ in the first
peak) is proportional to $\zeta _{crt}/n$.   In order to find regions of high \oh\ column that can be observed,
and where cosmic rays are the ultimate cause of the ions, 
we therefore must look at low density regions with high cosmic ray ionization rates.

{\it  Analytic solution to the second peak of \htho\ in the regime $A_V >1$ and with PAHs present}.  
Deeper in the cloud ($A_V > 1$) the dust and gas shield the FUV photons and carbon is no longer in the form
of C$^+$ but converts to CO.  Here, the electron abundance drops rapidly with depth, and the gas is almost
entirely molecular so that the route to \htho\ is started with the cosmic ray ionization of H$_2$.   We then define
$\epsilon$(\htho) so that the rate of formation of \htho\ per unit volume is $\zeta _{crt} n \epsilon$(\htho).\footnote{
Note that if
the total ionization rate per H is $\zeta _{crt} $, then the total rate per H$_2$ is roughly 2$\zeta _{crt}$.  The cosmic ray ionization rate per unit volume is then 2$\zeta _{crt} n({\rm H_2})= \zeta _{crt}n$ in fully molecular zones.}
Again $\epsilon$(\htho) is equivalent to the fraction of cosmic ray ionizations that lead to \htho.  

Figure 1 shows the route to \htho\ from the ionization of H$_2$.   The key competition that determines
$\epsilon$(\htho) revolves around H$_3^+$.   With all the hydrogen molecular, the routes via H$_2$ dominate
the rates via electrons.  Thus, the only interruption in the chain that leads to \htho\ is the possibility that H$_3^+$ will
react with either electrons or CO rather than with O to form either \oh\ or \hto.   We then define $\gamma _6$ as the rate coefficient for H$_3^+$ reacting with O to form either \oh\ or \hto; $\gamma _7$ is the rate coefficient for
H$_3^+$ to dissociatively recombine with electrons; $\gamma _8$ is the rate coefficient for H$_3^+$ to react with CO; and $\gamma _9$ is the rate coefficient for \htho\ to dissociatively recombine with electrons (all channels). Therefore,

\begin{equation}
\epsilon({\rm H_3O^+}) \simeq {{\gamma _6 x({\rm O})}\over {\gamma _6 x({\rm O}) + \gamma _7 x_e + \gamma _8
s({\rm CO})}}
\end{equation}

The solution for the abundance of \htho\  follows
\begin{equation}
x({\rm H_3O^+}) = {{(\zeta _{crt}/n) \epsilon ({\rm H_3O^+})}\over { \gamma _9 x_e}}
\end{equation}

We find in our model runs that electrons are formed by the cosmic ray ionization of H$_2$ and destroyed
by attachment to neutral PAHs.    Most of the PAHs deep in the cloud are neutral, and therefore using a neutral PAH
abundance of $2\times 10^{-7}$ we find

\begin{equation}
x_e \simeq 7.7\times 10^{-6} (\zeta _{crt}/10^{-16}\  {\rm s^{-1}})(100\ {\rm cm^{-3}}/n)
\end{equation}
If either O or CO dominate the destruction of H$_3^+$ over electrons (or $x_e < 8\times 10^{-3} x({\rm CO}) + 6\times 10^{-3} x({\rm O})$), which we find is usually the case 
for most of our runs at the second peak of the \htho\ abundance, then we find, substituting $x_e$ into the equation
for $x({\rm H_3O^+})$,

\begin{equation}
x({\rm H_3O^+}) \simeq {{4\times 10^{-8} (T/30\ {\rm K})^{0.74} x({\rm O})}\over {x({\rm O}) + 1.4x({\rm CO)}}}
\end{equation}
At the peak, we generally find that most of the carbon is in CO, so $x({\rm CO})\sim 1.6 \times 10^{-4}$ and that
most of the remaining O is atomic, or $x({\rm O}) \sim 1.4 \times 10^{-4}$.  Therefore, we predict a peak \htho\
abundance of  
\begin{equation}
x_p({\rm H_3O^+}) \simeq 1.6\times 10^{-8} (T/30\ {\rm K})^{0.74} 
\end{equation}
This explains why the second peak abundance of \htho\ is independent of  $\zeta _{crt}$ and $n$.   The production rate per unit volume of \htho\  depends on  $\zeta _{crt}n$ but the destruction rate per unit volume depends on
$n^2x_e$.   Since electrons are formed by cosmic ray ionization and destroyed by PAHs, $x_e \propto \zeta _{crt}/n$.
Therefore, both production and destruction of \htho\ is proportional to $\zeta _{crt}n$, and the resulting 
abundance of \htho\ is independent of both parameters.   As one goes to higher $A_V$ from the peak, the
atomic O freezes out on grains as water ice, and thus the abundance of \htho\ declines.    The thickness of
the region of peak \htho\  abundance is therefore of order $\Delta A_V \sim 1$, or a hydrogen column of
$2\times 10^{21}$ cm$^{-2}$.   Multiplying this by $x_p({\rm H_3O^+})$, we estimate columns of
\htho\ of $N({\rm H_3O^+})\sim 4\times 10^{13}$ cm$^{-2}$, as we found in our numerical runs.   There has
been some indication (Rimmer et al 2011) that the cosmic ray ionization rate may decline with depth into a cloud.
From the above result, this should have little consequence on the \htho\ column.

We should emphasize that this prediction is dependent on the presence of PAHs deep ($A_V \sim 5$) in the cloud.
If PAHs coagulate on grain surfaces, then PAHs do not remove free electrons from the gas phase.   In this case, the electron abundance
at the peak of the \htho\ can be higher than in the PAH case if gas phase metal ions are present. This suppresses the \htho abundance at the peak.   Therefore, the observation
of high columns of \htho \ deep in molecular clouds may indicate either the presence of PAHs lowering the electron
abundances there, or, in the absence of PAHs, may indicate a high depletion of gas phase metals at the second peak.

\clearpage
\appendix{Appendix C: Sensitivity of results to other parameters}

Recently, Klippenstein et al (2010) have provided  theoretical rate coefficients for the reactions of H$_3^+$
with O and with CO that are 10 to 20\% higher than the rate coefficients we adopted.   We have run our
standard cases with these new coefficients and find that the columns of \oh, \hto, and \htho \ increase by roughly
10\% with the new coefficients.    In addition, Cartledge et al (2004) suggest a gas phase oxygen abundance
of 2.84$\times 10^{-4}$, whereas Jensen et al (2005) suggest a value of 4.2 to 4.7$\times 10^{-4}$.  We have adopted
3.2$\times 10^{-4}$ but the variation in the literature suggests that we test the dependence of the ion columns
on the gas phase abundance of elemental O.  Assuming an abundance of 4.5$\times 10^{-4}$, an increase of 40\%
over our standard rate, increases the columns of \oh, \hto, and \htho\ by 10-20\%.  Therefore, the slight possible
variations in these rate coefficients and/or the gas phase elemental O abundance have a negligible effect on our conclusions.
 
We have also tested our results for their sensitivity to the H$_2$ formation rate coefficient for formation on grain
surfaces.   We find that lowering this rate by a factor of 3 hardly affects the results for high density gas, and that the
effect is most pronounced at low density.    At $n= 100$ cm$^{-3}$, lowering the H$_2$ formation rate by 3 lowers the columns of \oh\ and \hto\  by factors of less than
2 in both peaks.   The \htho\  column in both peaks decreases by a bit less than a factor of 3. The main effect of lowering the H$_2$ rate coefficient is to drive the H/H$_2$ transition deeper into
the cloud,  which moves the first peaks of the ions to greater depths, but their columns and peak abundances do not change
appreciably.   

In the text we discussed the effect of lowering the gas phase abundance of low ionization potential metal atoms such as S, Fe, Mg, and Si if PAHs are not present.    Here, we examine the effect of raising their abundances.  Our standard gas phase abundances for these species are  given in Table 2 of 
Appendix A.   In our test cases, we raise the gas phase abundances of these species to 10$^{-5}$, a factor of 10-100
times their standard values.   For our standard case of $n= 10^4$ cm$^{-3}$ and \crt$=2\times 10^{-16}$ s$^{-1}$, but
with no PAHs, this raises the electron abundance at high $A_V$ in the gas from about 10$^{-6}$ to about $10^{-5}$.  This rise
depresses the (second) peak abundance of \htho\ from about $10^{-9}$ to about $10^{-10}$ if the photodesorption yields of Fe, Mg, and
Si are $10^{-3}$, a likely upper limit (H09).   Smaller yields lower the electron abundance, thereby changing the
\htho\ less.   However, if PAHs are
present,  metal ions
recombine with PAH$^-$ and PAH, which is a much faster process than with electrons. Hence, the electron abundance
does not depend much on the metal abundances.
Consequently, the (second) peak \htho\ abundance does not change significantly.   The first peaks of the ions are not affected, because the
electrons are supplied by C$^+$, whose gas phase abundance is more than 10 times the (high) abundance of the
metal ions.

Section 4  discusses observations that indicate  $N$(\htho)$\gta$ $
10^{14}$ cm$^{-2}$ in some clouds, possibly even as high as $10^{16}$ cm$^{-2}$.    The only way our models can
accommodate very large columns of \htho\ in a single cloud is if there is incomplete freeze-out of water ice on grains.   Note that our
models are steady state, so that time dependent effects might leave more elemental oxygen in the gas phase than
steady state results would indicate.   Another possibility is that the grains are sufficiently warm, $\gta 100$ K, to thermally
evaporate water ice off the grains, or that cosmic ray rates are sufficiently high to desorb the ice mantles.   We have
run our code fixing the grain temperature to be $> 110$ K, so that water ice cannot form and elemental oxygen is
plentiful in the gas phase at all $A_V$.   In the case of PAHs, and for our standard case of $n= 10^4$ cm$^{-3}$ and
\crt$= 2\times 10^{-16}$ s$^{-1}$, we find that $x$(\htho) plateaus at a value of about $2\times 10^{-8}$ at about $A_V \sim 6$,
and stays constant at that value for all higher $A_V$.  (Note that given the discussion in \S 3.1, this result
is independent of \crt\ and $n$).  Therefore, a cloud with a hydrogen nucleus column
of $10^{23}$ cm$^{-2}$ would have $N$(\htho)$\sim 2\times 10^{15}$ cm$^{-2}$, for example.   Without PAHs, and 
assuming the low ionization potential metal atoms strongly deplete on grains, we would predict similar columns.

\clearpage
\begin{figure}[ht!]
\label{fig:gasphase}
\plotone{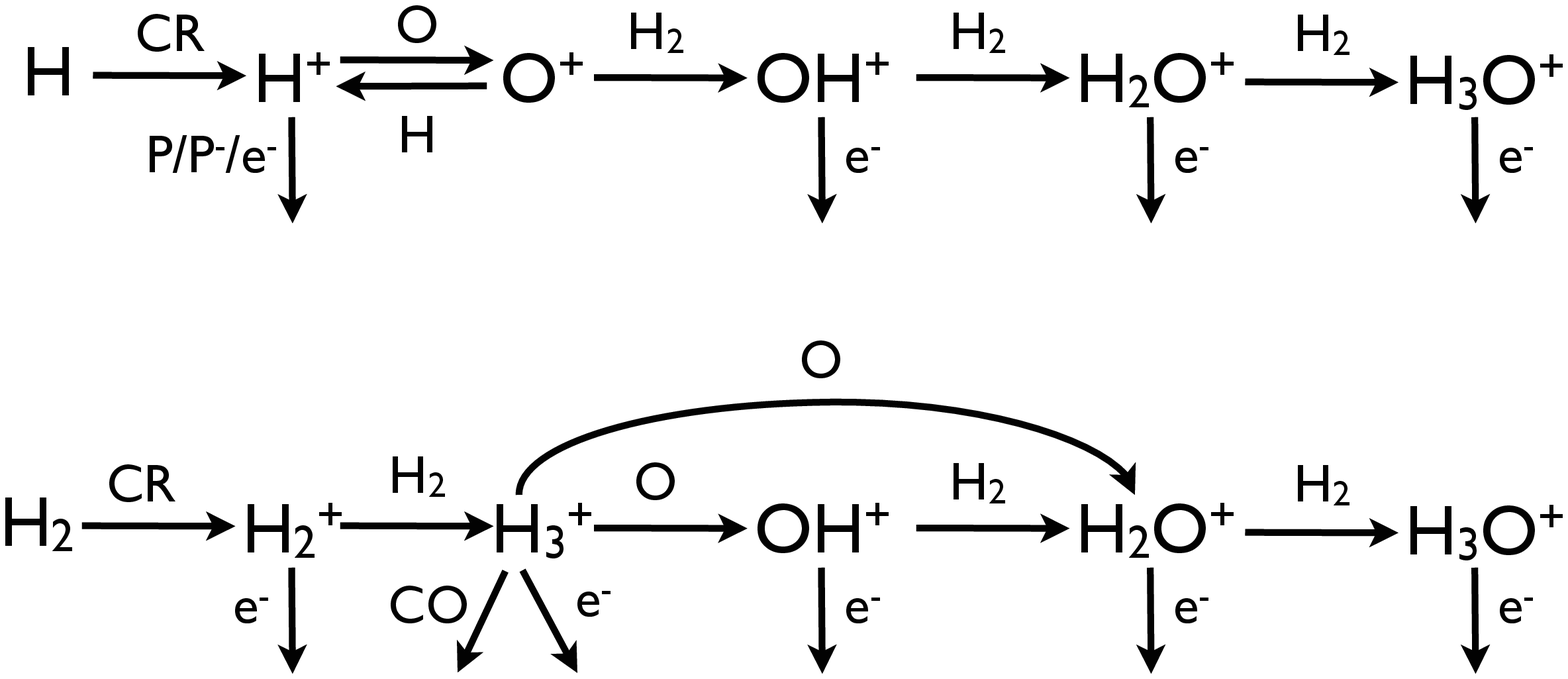}
\caption{Standard ion-neutral chemistry leading to the formation of \oh,
\hto, and \htho\  in clouds via ionization of atomic H (top panel), and
via ionization of molecular H$_2$ (bottom panel).  In the top panel,
the destruction of H$^+$ labelled "P/P$^-$/e$^-$" means the neutralization
of H$^+$ by reacting with PAHs, PAH$^-$s, or electrons. ``CR"  means cosmic ray ionization.}
\end{figure}
\clearpage

\begin{figure}[ht!]
\plotone{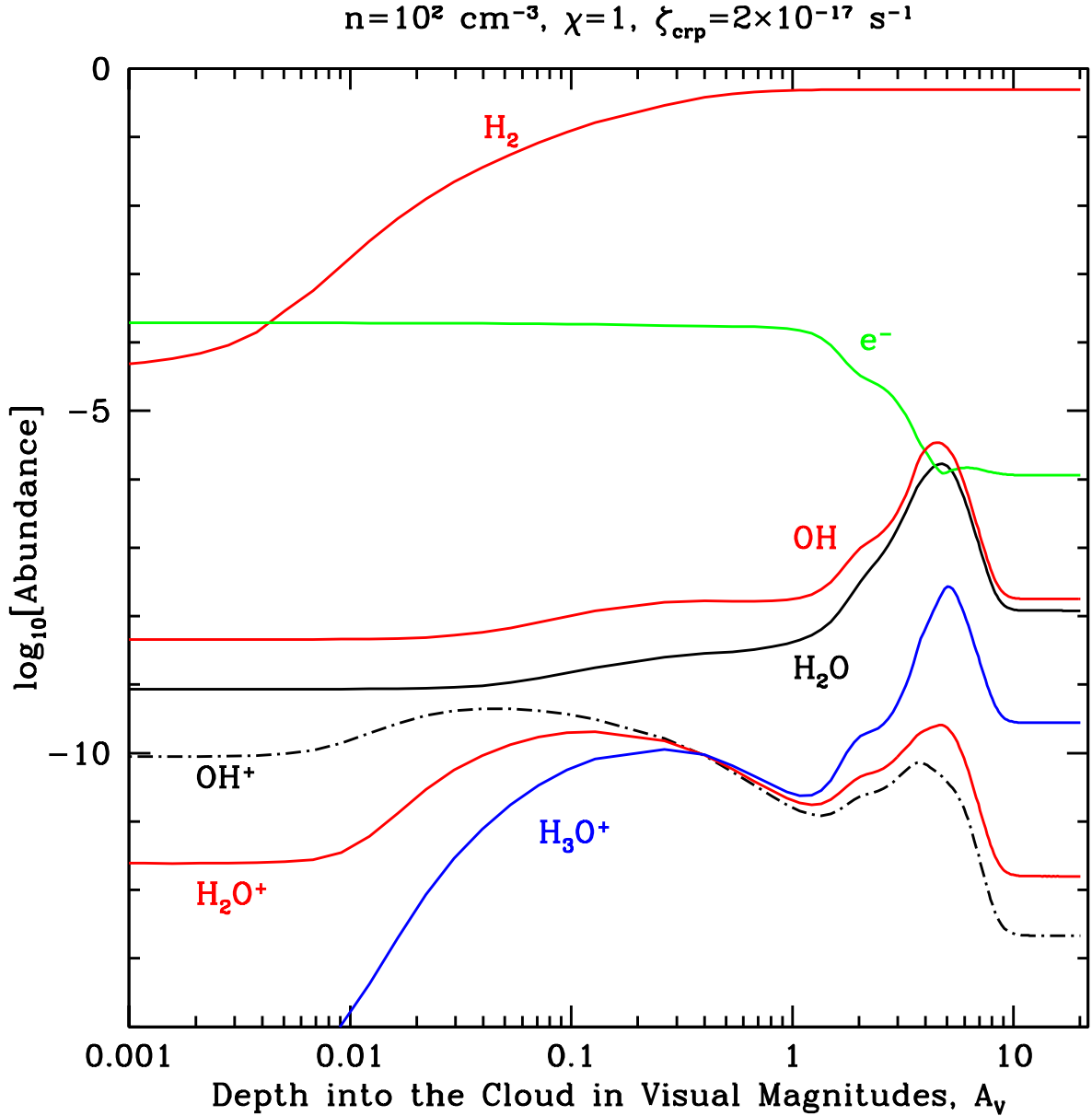}

\caption{The variation of gas phase abundances of species as a function of depth  
$A_V$ into a cloud for the standard case $n= 100$ cm$^{-3}$, $\chi = 1$, and \crt = 2$\times 10^{-17}$
s$^{-1}$.   To convert $A_V$ to hydrogen nucleus column $N$, use
$N= 2\times 10^{21}A_V$ cm$^{-2}$. This case probes diffuse cloud-like condition to high $A_V$, or could apply
to low density surfaces of GMCs experiencing the local interstellar radiation field. Note these are columns when
the cloud is viewed face-on. Color versions of the figures available in the on-line manuscript }
\label{fig:std1}
\end{figure}

\clearpage
\begin{figure}[ht!]
\plotone{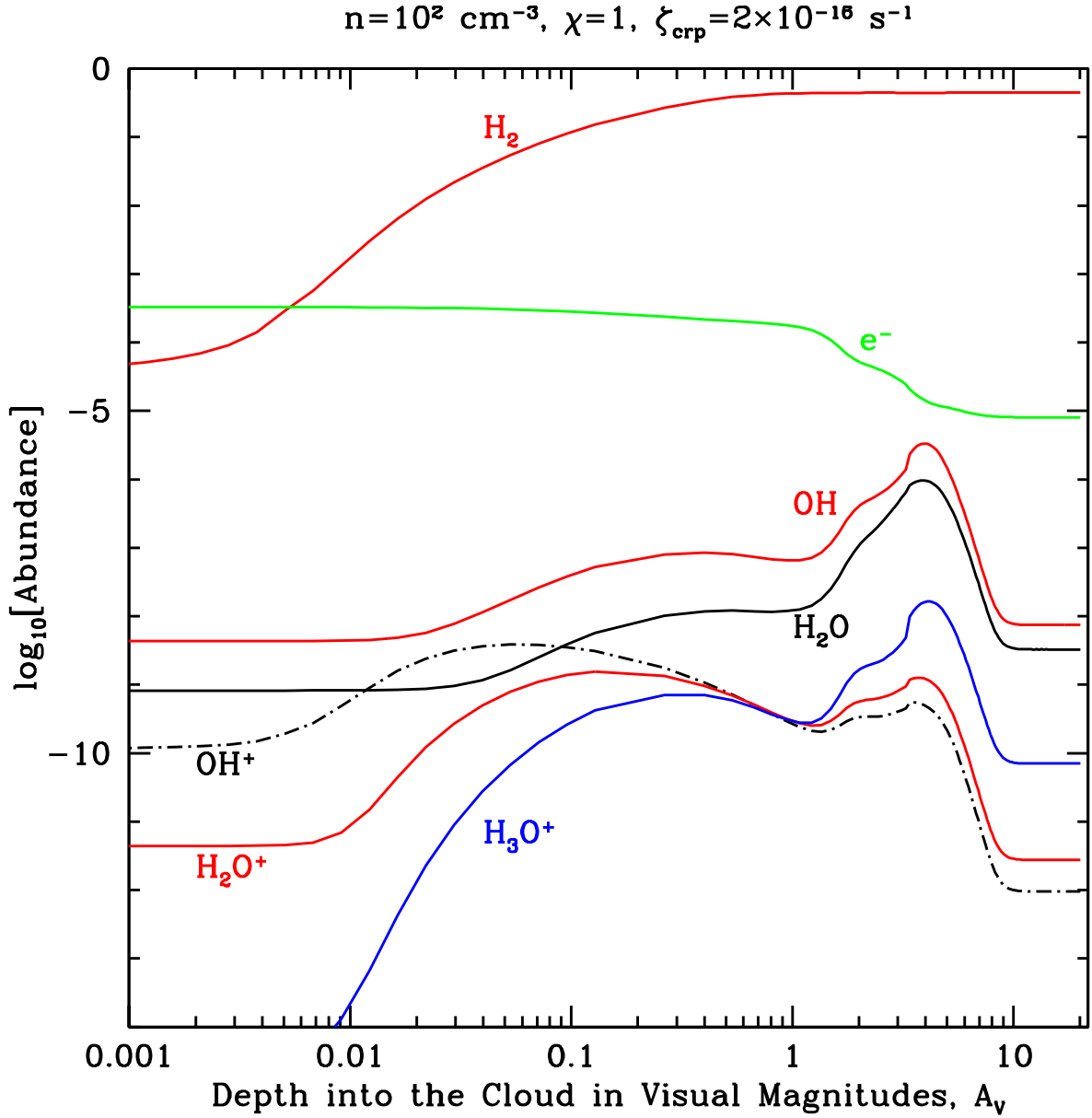}

\caption{The variation of gas phase abundances of species as a function of depth 
$A_V$ for the standard case $n= 100$ cm$^{-3}$, $\chi = 1$, and \crt = $2\times 10^{-16}$
s$^{-1}$.   This is the same case as Figure 2 but with 10 times the cosmic ray ionization rate.
This case probes diffuse cloud-like condition to high $A_V$, or could apply
to low density surfaces of GMCs experiencing the local interstellar radiation field.}
\label{fig:std3}
\end{figure} 

\clearpage
\begin{figure}[ht!]
\plotone{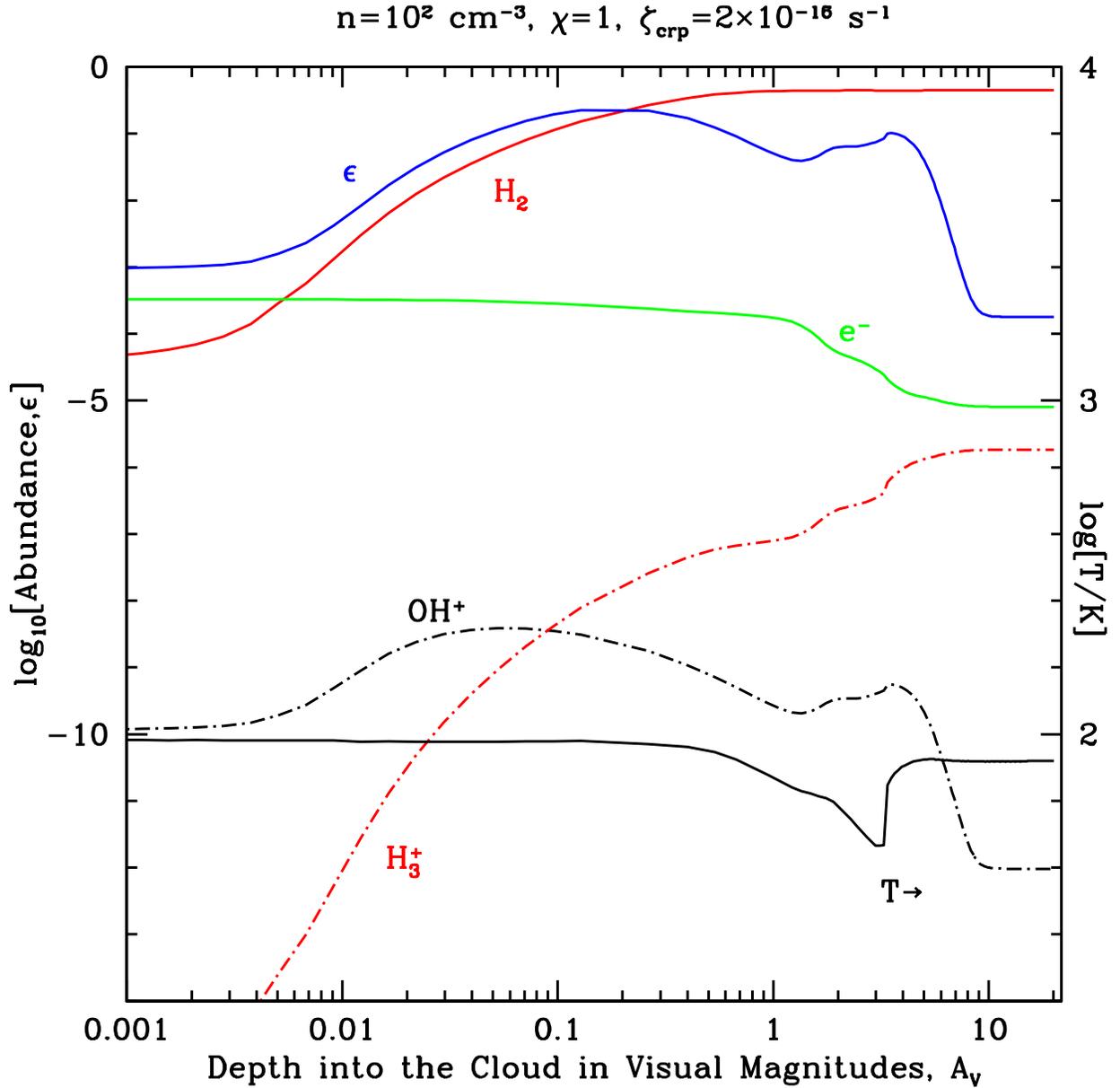}

\caption{The variation of the parameter $\epsilon$, the ratio of  the rate of \oh\ formation to the 
cosmic ray ionization rate of H and H$_2$, the gas temperature T (labelled on right), and the gas phase abundances of electrons, H$_2$, H$_3^+$,
and \oh\ as a function of depth 
$A_V$ for the case $n= 100$ cm$^{-3}$, $\chi = 1$, and \crt = $2\times 10^{-16}$
s$^{-1}$.   This case probes diffuse cloud-like condition to high $A_V$, or could apply
to low density surfaces of GMCs experiencing the local interstellar radiation field.  This is the same case as the previous figure.}
\label{fig:std4}
\end{figure}

\clearpage
\begin{figure}[ht!]
\plotone{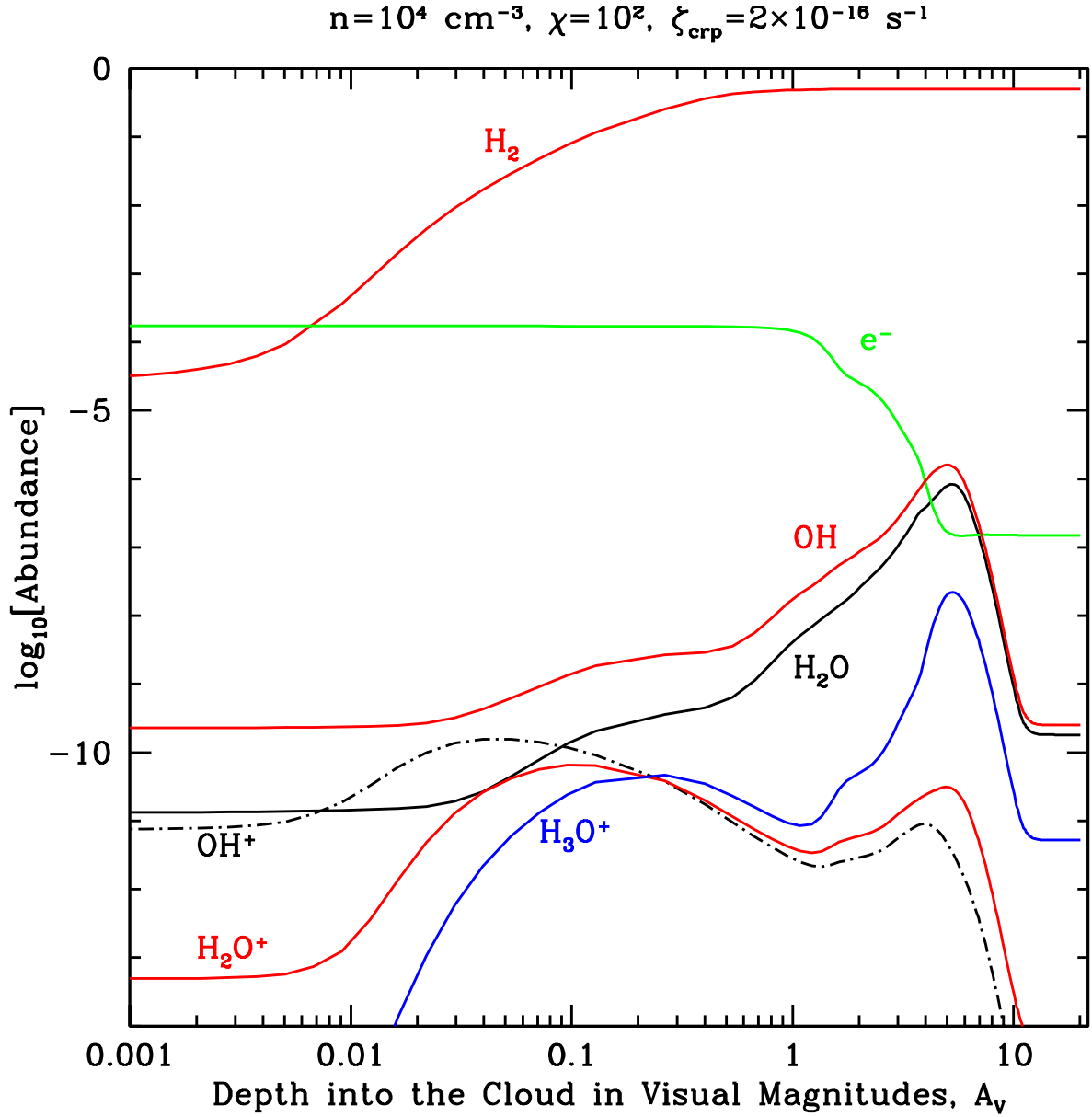}

\caption{The variation of gas phase abundances of species as a function of
depth $A_V$ into the cloud for the standard case $n= 10^4$ cm$^{-3}$, $\chi = 100$, and
\crt= $2\times10^{-16}$ s$^{-1}$.   This case may be appropriate to GMCs with elevated
FUV fluxes incident due to nearby O and B stars.  }
\label{fig:std5}
\end{figure}

\clearpage
\begin{figure}[ht!]
\plotone{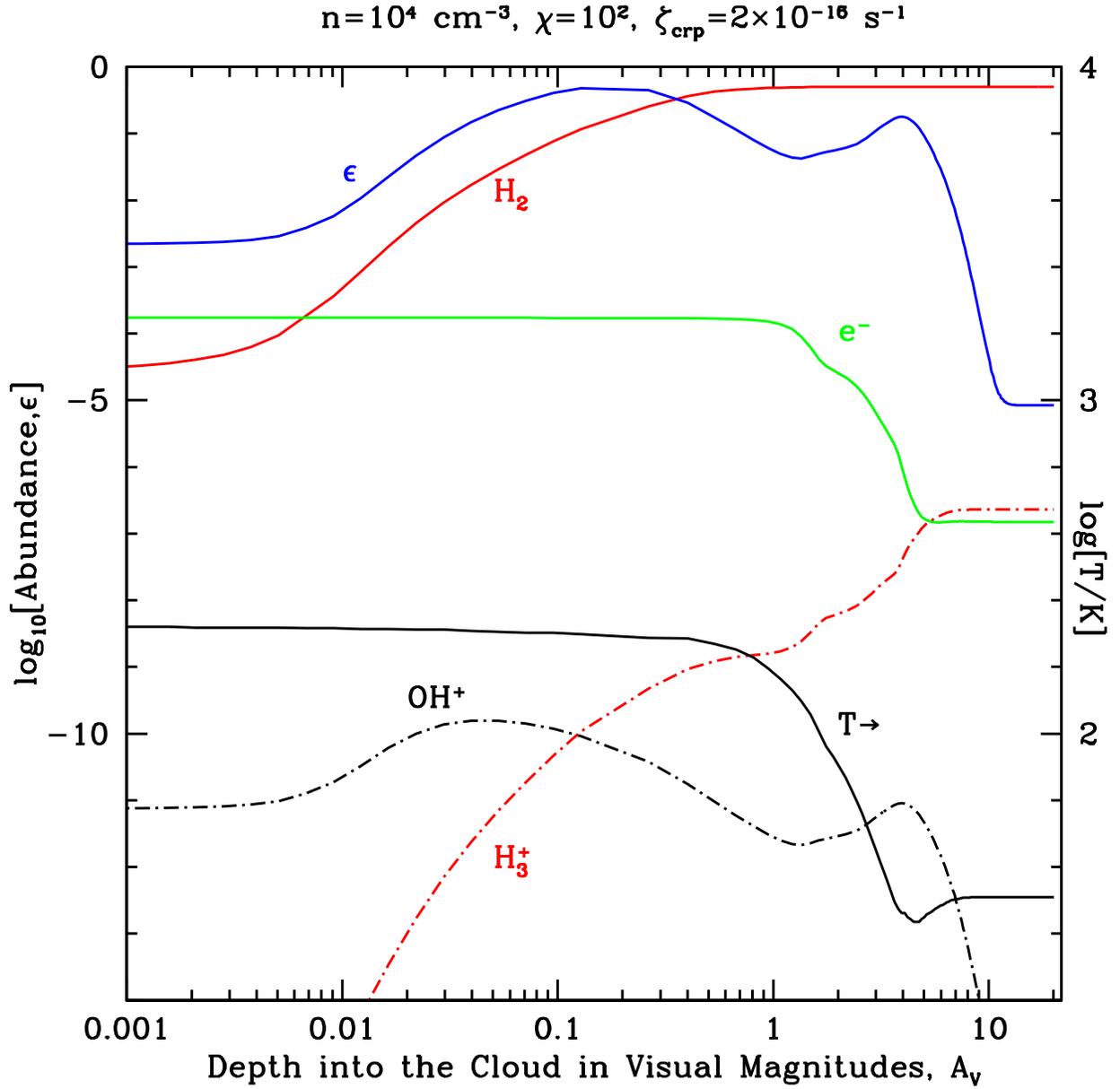}

\caption{The variation of the parameter $\epsilon$, the ratio of  the rate of \oh\ formation to the 
cosmic ray ionization rate of H and H$_2$, the gas temperature T (labelled on right), and the gas phase abundances of electrons, H$_2$, 
H$_3^+$, and \oh\ as a function of depth 
$A_V$ for the case  $n= 10^4$ cm$^{-3}$, $\chi = 100$, and
\crt= $2\times10^{-16}$ s$^{-1}$.  This is the same case as the previous figure.}
 \label{fig:std6}
\end{figure}

\clearpage

\begin{figure}[ht!]
\plotone{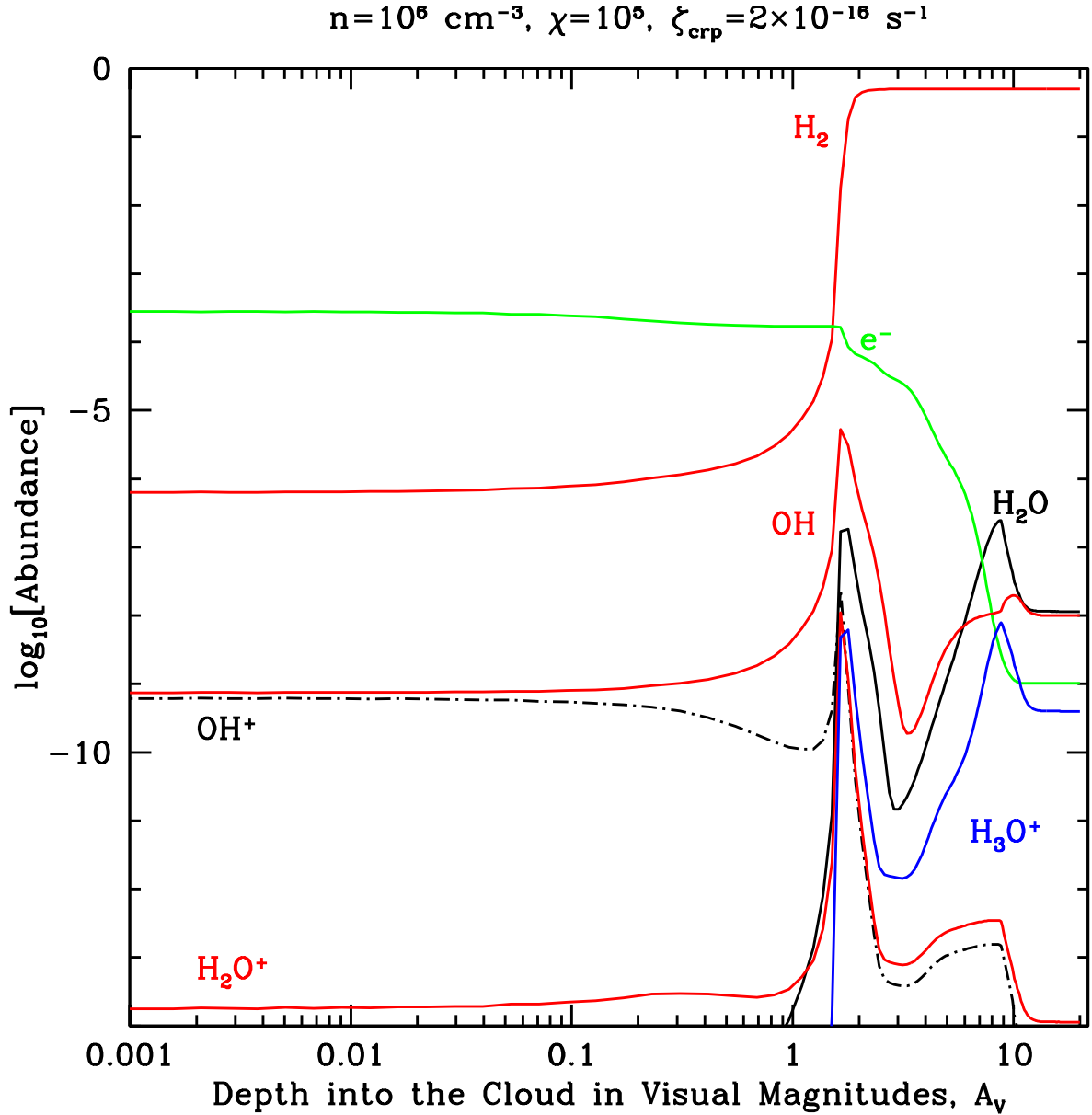}

\caption{The variation of gas phase abundances of species as a function of depth 
$A_V$ for the case $n= 10^6$ cm$^{-3}$, $\chi = 10^5$, and \crt = 2$\times 10^{-16}$
s$^{-1}$.  This case demonstrates the structure of PDRs with both high density and high
FUV fluxes, where elevated ($> 300$ K) temperatures in the region with significant H$_2$ leads
to the enhanced production of H$^+$ by chemical routes not initiated by cosmic ray ionization
(see text).  One mark of this is the enhanced OH abundance, produced by the neutral-neutral reaction
of H$_2$ with O that is seen at $A_V \sim 1$, where the gas temperature is $T \sim 1000$ K.
The enhanced OH reacts with C$^+$ to produce CO$^+$, which then reacts with H to form H$^+$.
The rest of the chemistry leading to \oh, \hto, and \htho\ is seen in Figure 1.}
\label{fig:std7}
\end{figure} 

\clearpage

\begin{figure}[ht!]
\plotone{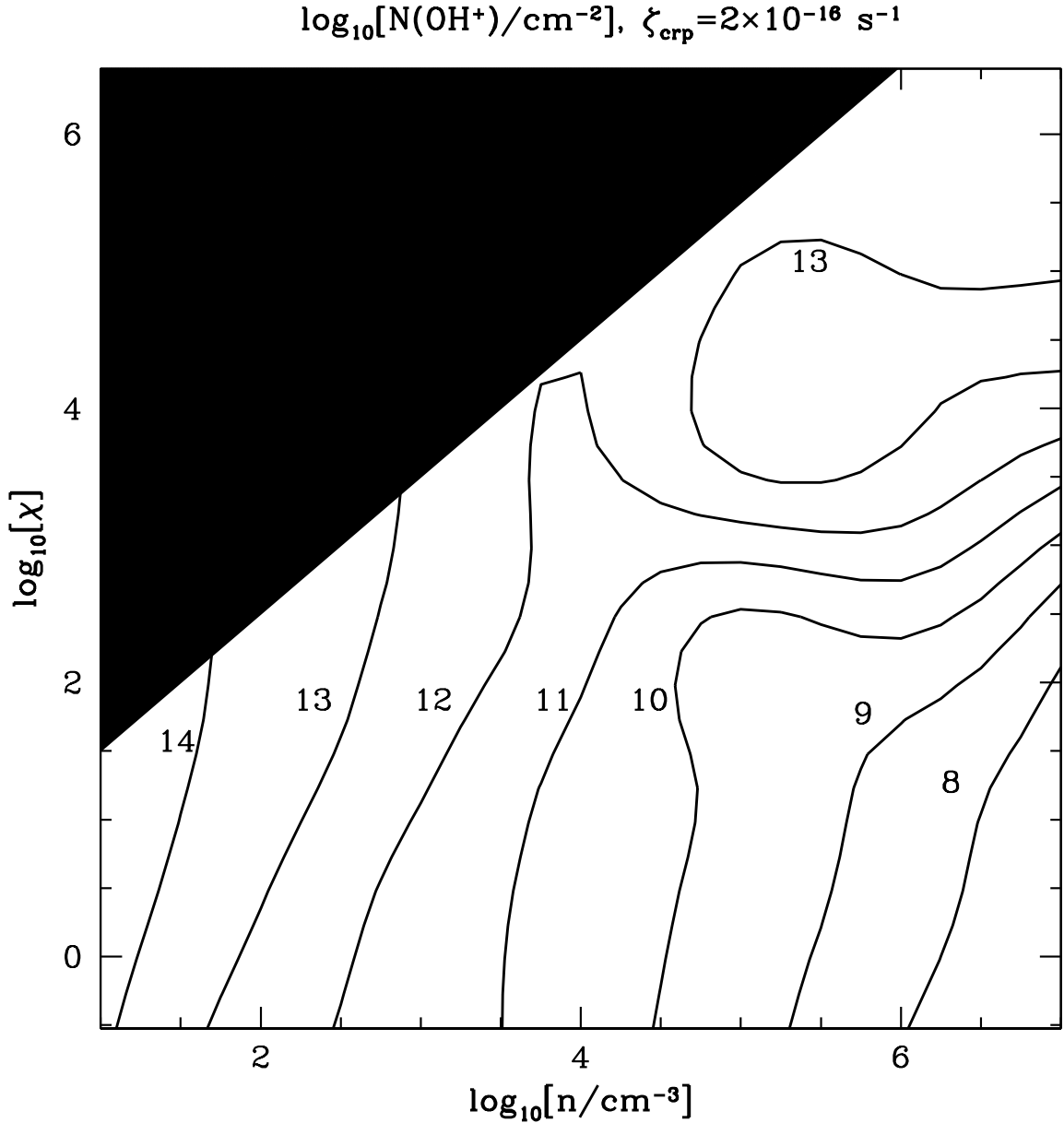}

\caption{Contours of the column of \oh, labelled in log units, as a function of $n$ and $\chi$ for
a fixed primary cosmic ray ionization rate of \crt=$2\times 10^{-16}$ s$^{-1}$ per H atom.   The upper left portion
of the figure is blacked out because radiation pressure on dust drives dust quickly through the PDR
in this region, invalidating the physics assumed in the model.   This combination of $\chi$ and $n$
are rarely observed in any case. 
Except in the upper 
right hand corner of this figure (high $n$ and high $\chi$), we see that for fixed cosmic ray
ionization rate, the column is roughly proportional to $n^{-1}$, and independent of $\chi$.   The
upper right hand corner shows a secondary peak in the \oh\ column, caused by the
alternate chemical routes described in text and shown in more detail in Figure 7.  
 Low densities $n\lta100$ cm$^{-3}$ are required to obtain columns greater than about $10^{13}$ cm$^{-2}$ created by cosmic rays.}
\label{fig:OH+high}
\end{figure} 
\clearpage

\begin{figure}[ht!]
\plotone{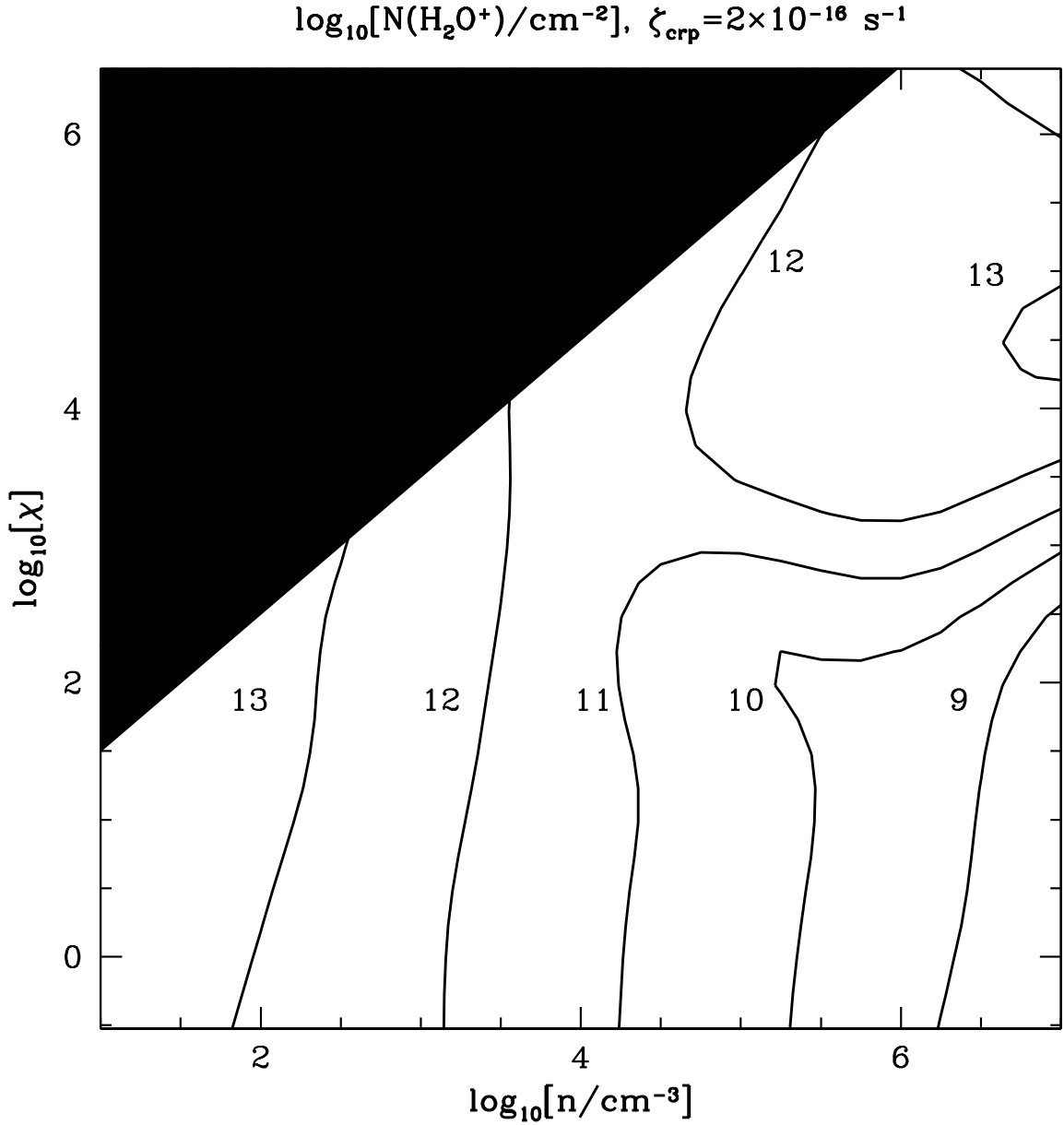}

\caption{Contours of the column of \hto, labelled in log units, as a function of $n$ and $\chi$ for
a fixed primary cosmic ray ionization rate of \crt=$2\times 10^{-16}$ s$^{-1}$ per H atom.  The same discussion
as in the previous figure applies here. To obtain 
columns greater than about $10^{13}$ cm$^{-2}$ created by cosmic rays requires low densities $n\lta 100$ cm$^{-3}$.}
\label{fig:H2O+high}
\end{figure} 
\clearpage

\begin{figure}[ht!]
\plotone{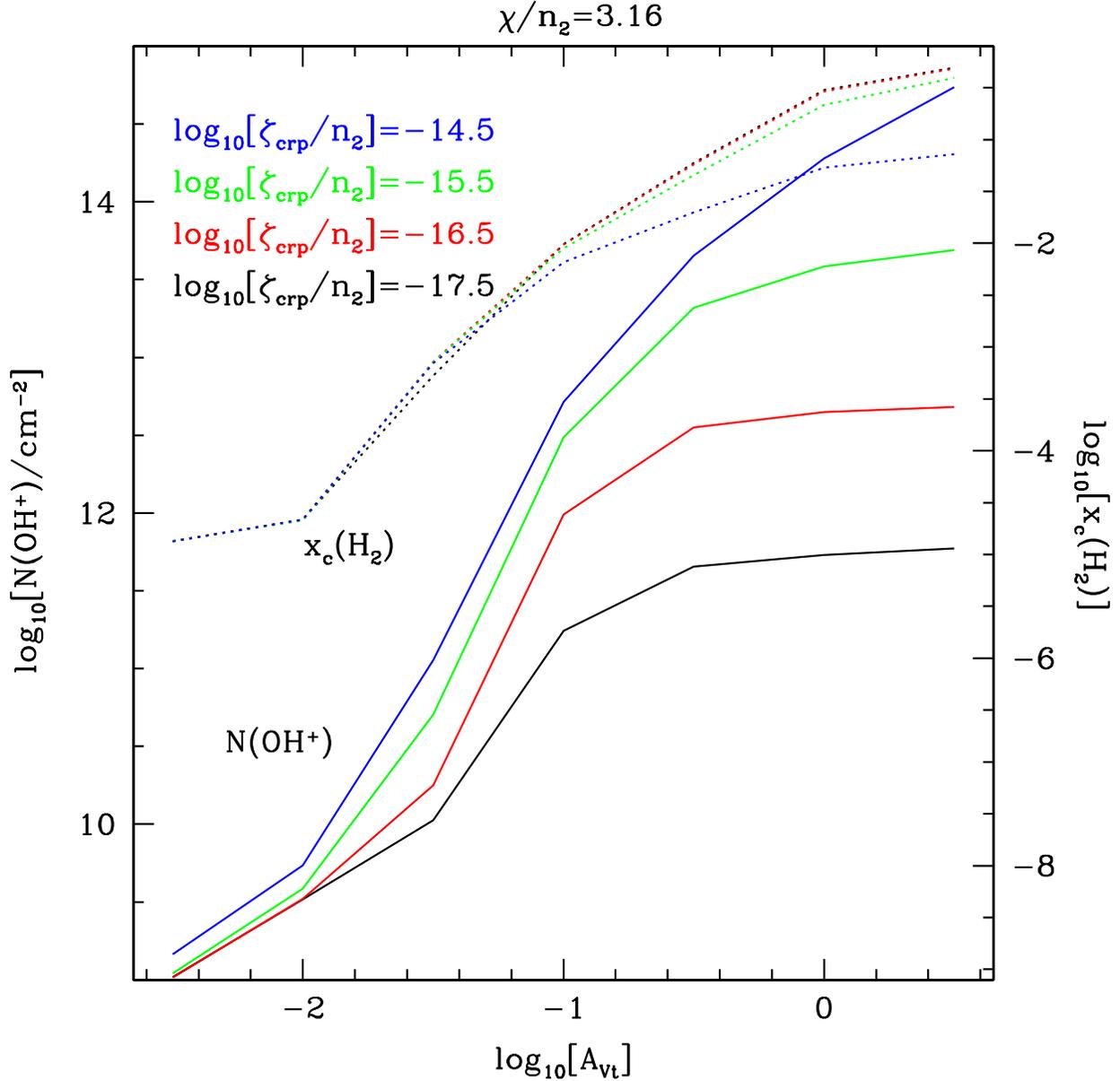}

\caption{The columns of \oh, $N$(\oh), are plotted as a function of $A_{Vt}$ for four values of the primary cosmic ray ionization
   rate per H atom divided by $n_2 \equiv n/ 100$ cm$^{-3}$ and for $\chi/n_2= 3.16$.   The  dotted line plots 
   the abundance of H$_2$ at cloud center, and these values appear on the right of the figure.  Recall $A_{Vt}$ is the total
   $A_V$ through the diffuse or translucent cloud. The H$_2$ abundance at cloud center, $x$(H$_2$), is also plotted (dotted lines) and its
   values noted on the right. }
\label{fig:nohpav3}
\end{figure} 
\clearpage

   \begin{figure}[ht!]
\plotone{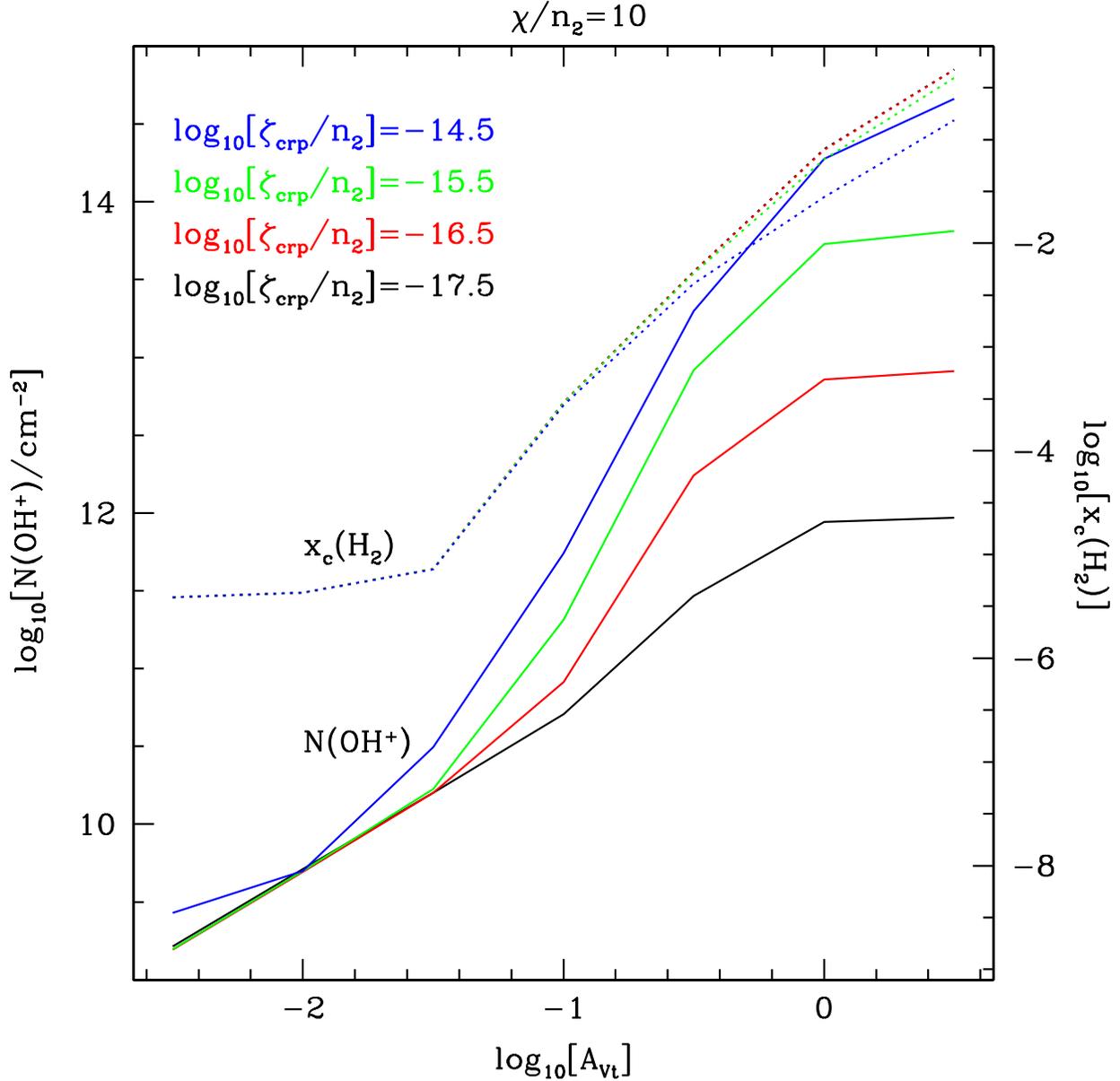}

\caption{ The columns of \oh, $N$(\oh),  are plotted as a function of $A_{Vt}$ for four values of the primary cosmic ray ionization
   rate per H atom divided by $n_2 \equiv n/ 100$ cm$^{-3}$ and for $\chi/n_2= 10$.   The  dotted line plots 
   the abundance of H$_2$ at cloud center, and these values appear on the right of the figure.  This figure is the same as
   Figure 10, only  with $\chi/n_2$ raised by 3.16. }
\label{fig:nohpav1}
\end{figure} 
\clearpage

    \begin{figure}[ht!]
\plotone{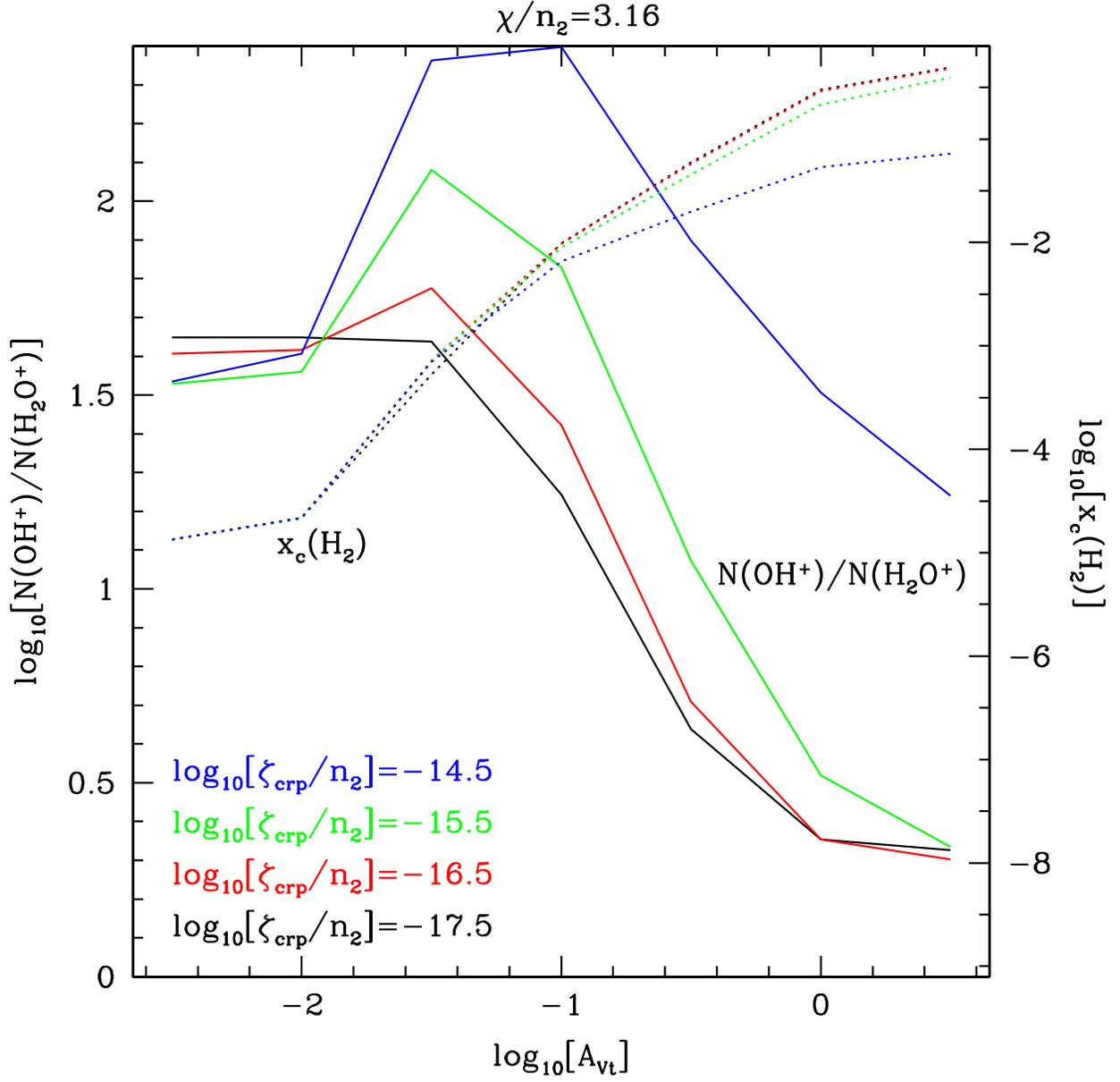}

\caption{The ratio  $N$(\oh)/$N$(\hto) is plotted as a function of $A_{Vt}$ for four values of the primary cosmic ray ionization
   rate per H atom divided by $n_2 \equiv n/ 100$ cm$^{-3}$ and for $\chi/n_2= 3.16$.   The dotted line plots 
   the abundance of H$_2$ at cloud center, and these values appear on the right of the figure. }
\label{fig:r8}
\end{figure} 
\clearpage

    \begin{figure}[ht!]
\plotone{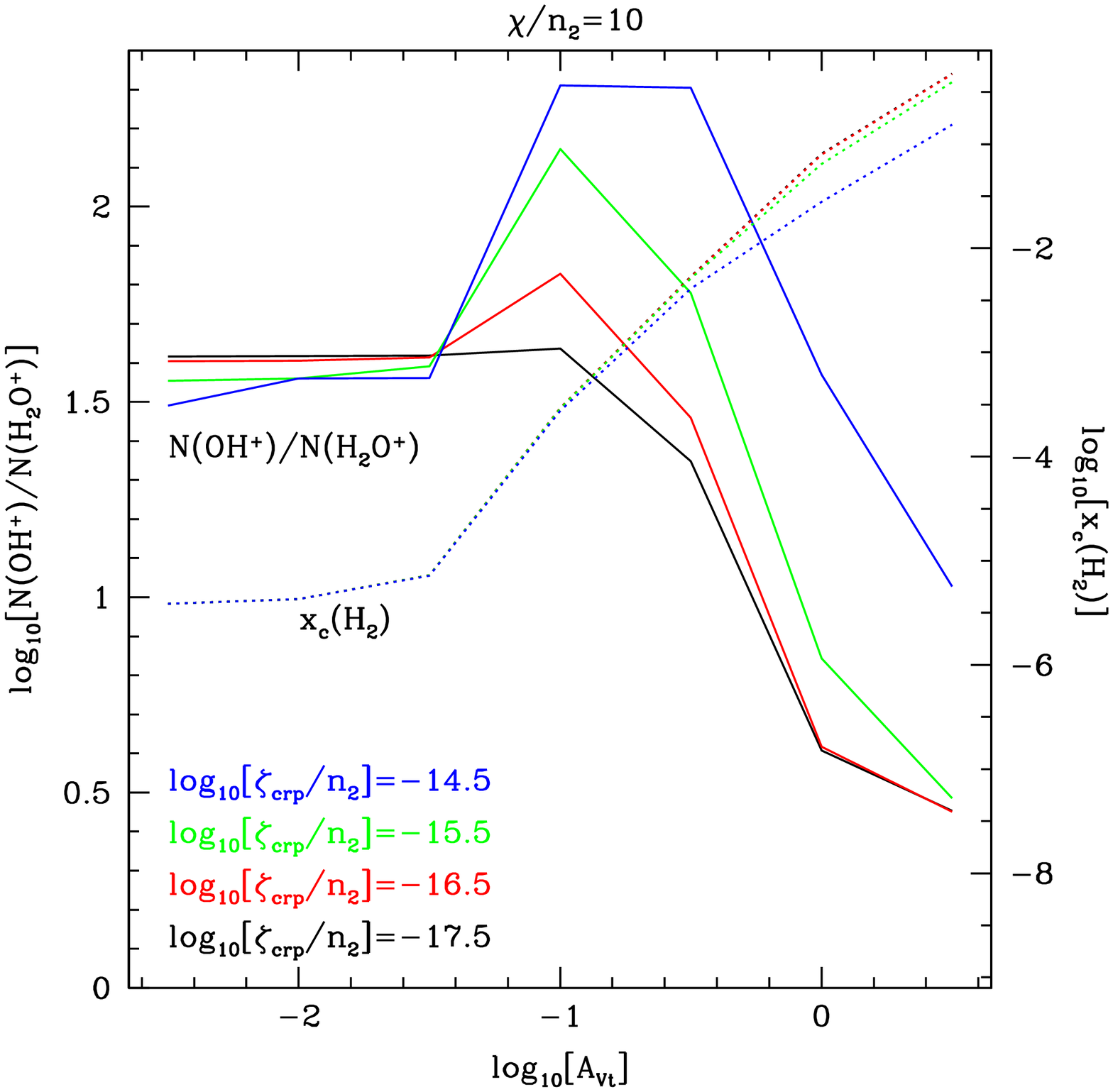}

\caption{The ratio $N$(\oh)/$N$(\hto) is plotted as a function of $A_{Vt}$ for four values of the primary cosmic ray ionization
   rate per H atom divided by $n_2 \equiv n/ 100$ cm$^{-3}$ and for $\chi/n_2=10$.   The  dotted line plots 
   the fraction of H$_2$ in the cloud, and these values appear on the right of the figure. This figure is the same as
   Figure 12, only  with $\chi/n_2$ raised by 3.16.}
\label{fig:r6}
\end{figure} 
\clearpage

\begin{figure}[ht!]
\plotone{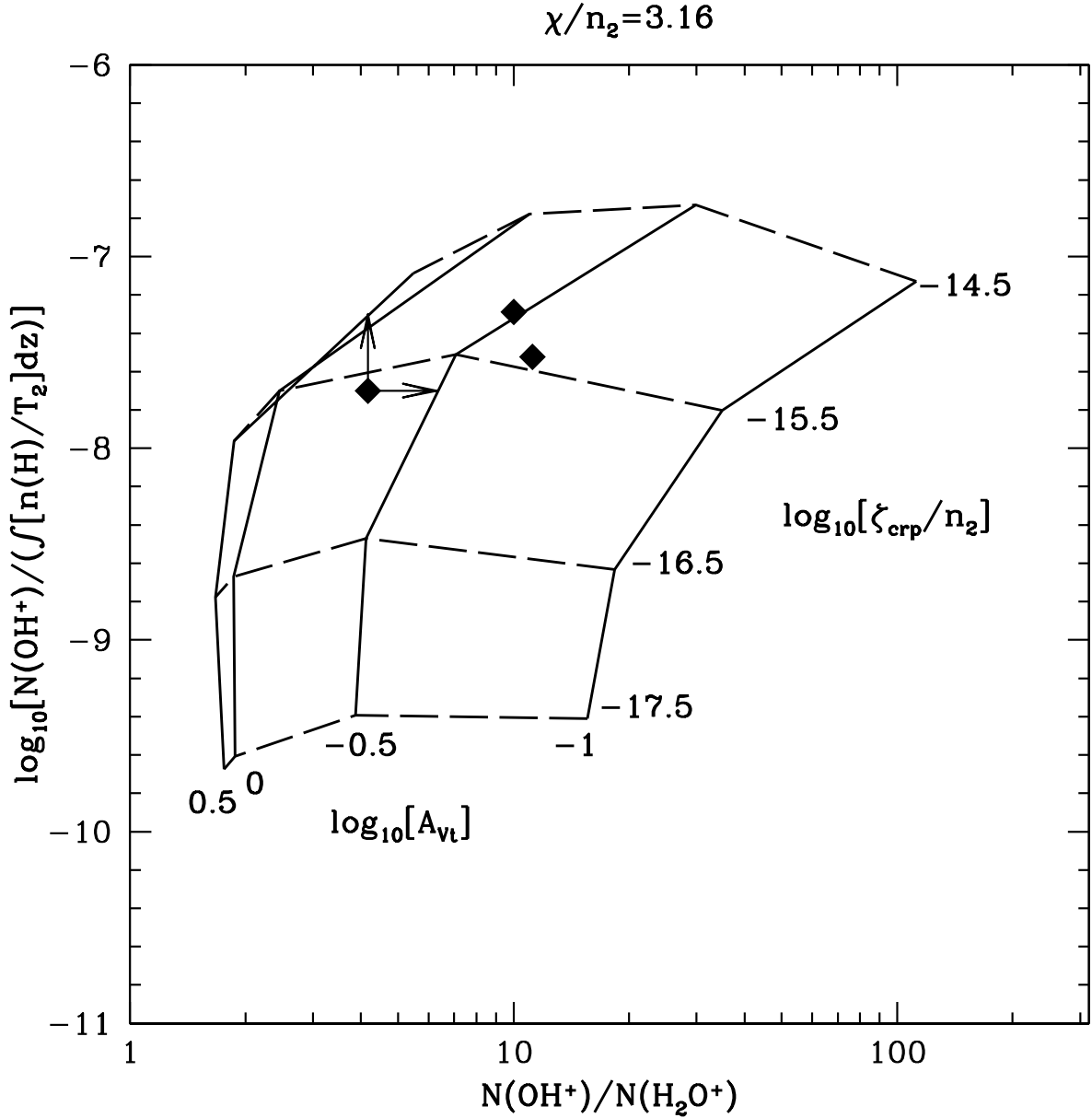}

\caption{  Log$_{10}$ $[N$(\oh)$/[\int (n$(H)/$T_2)\  dz$] is plotted on the vertical axis and $N$(\oh)/$N$(\hto)
on the horizontal axis for the case $\chi/n_2=3.16$.    Plotted
   as solid lines are constant values of log$_{10}[A_{Vt}]$, labelled on the bottom of these lines.  Plotted
   as dashed lines are contours of constant log$_{10} $[\crt/$n_2]$, labelled on the left, and in units of s$^{-1}$.  
   The two data points at right are two velocity components
    (diffuse clouds) toward W49N.  The lower limit data point is toward W31C (see text). }
\label{fig:mesh2}
\end{figure} 
\clearpage

\begin{figure}[ht!]
\plotone{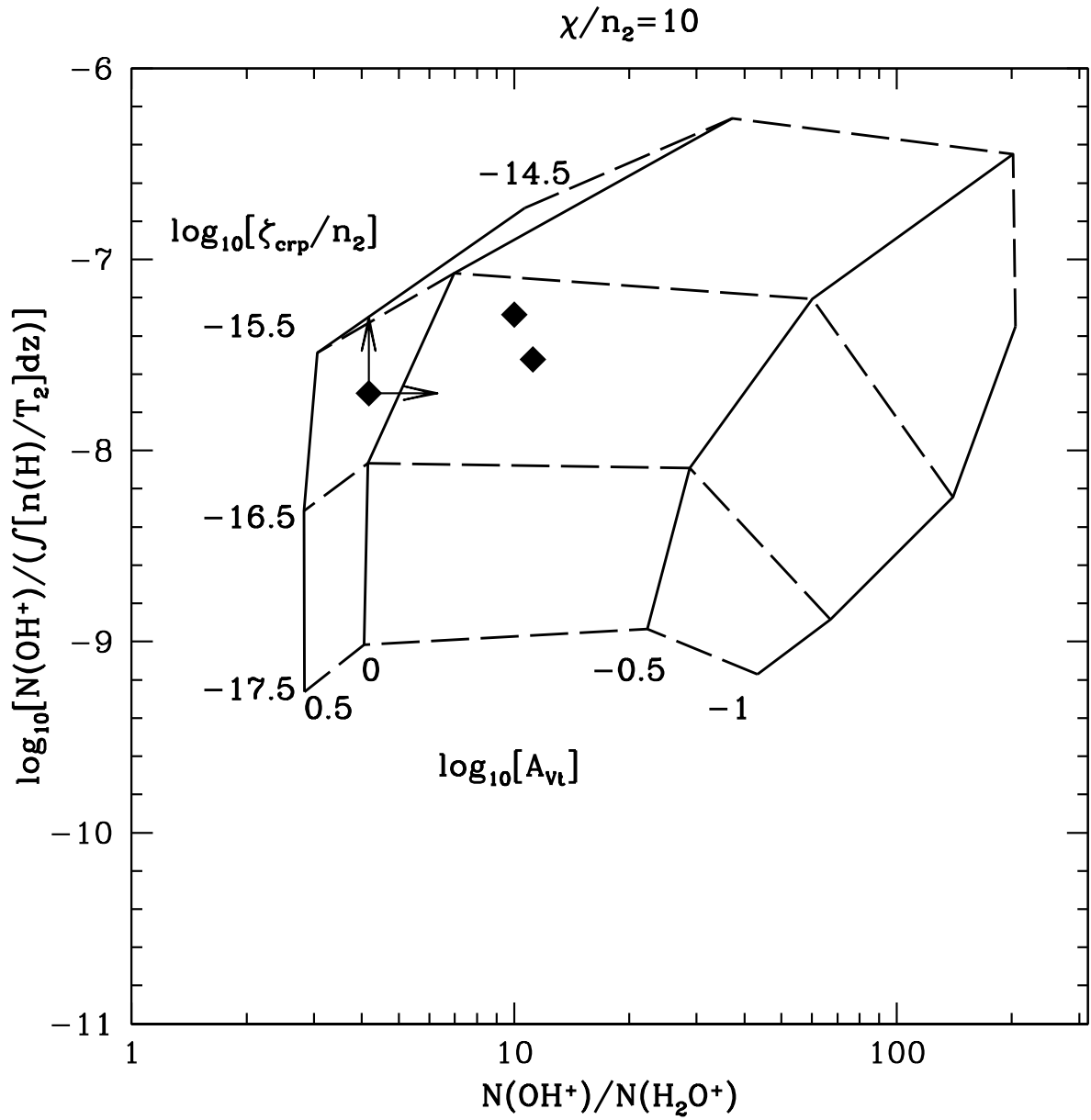}

\caption{ This figure is identical to Figure 14 except that $\chi/n_2=10$. }
\label{fig:mesh3}
\end{figure} 
\clearpage

\begin{figure}[ht!]
\plotone{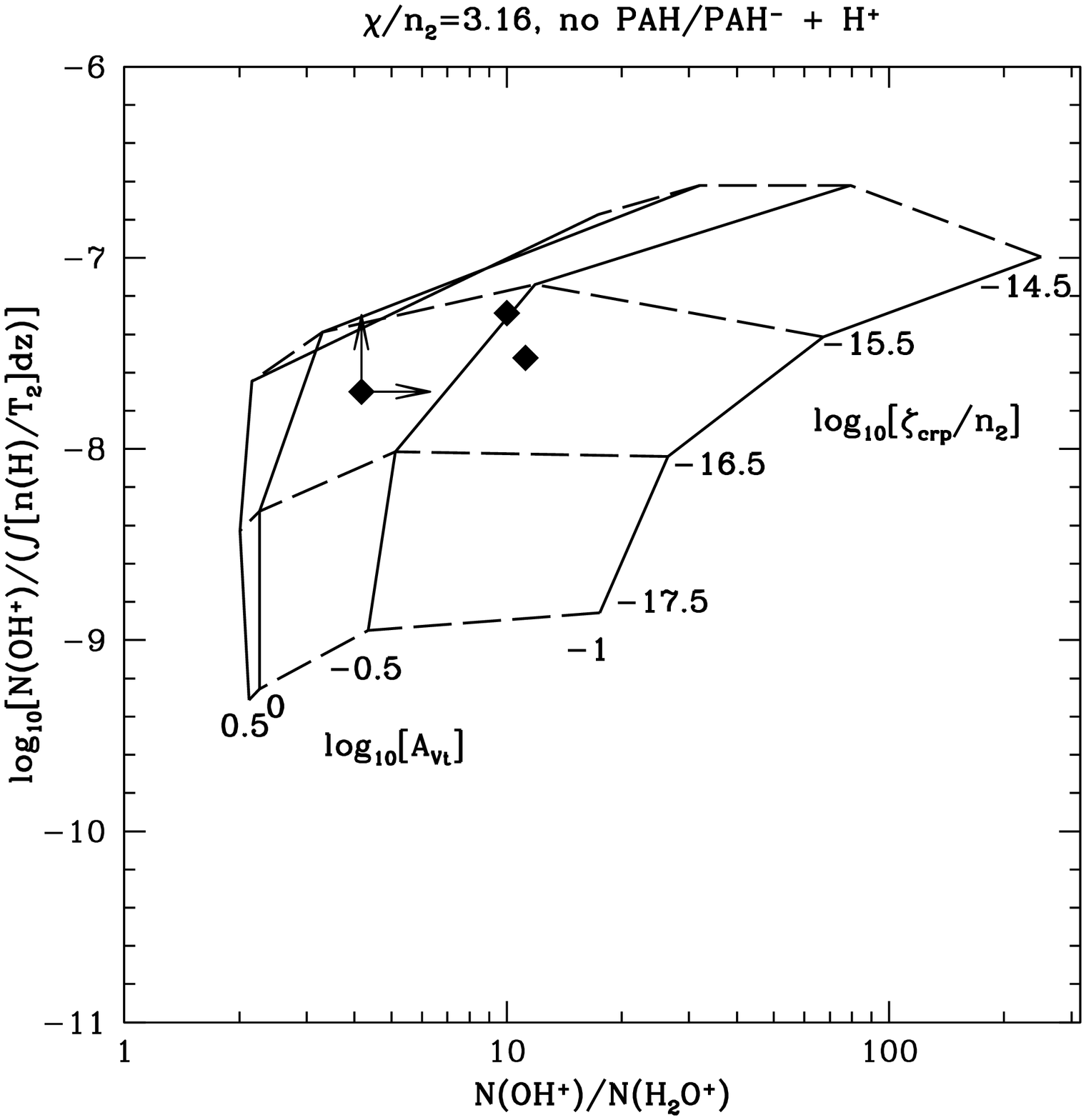}

\caption{ This figure is identical to Figure 14 ($\chi/n_2=3.16$)  except that the PAH and PAH$-$ rates with H$^+$ have
been reduced by $\gta 4$ so that H$^+$ is mainly destroyed by electrons or by forming O$^+$ which
then reacts with H$_2$ to form \oh. Note that the inferred \crt/$n_2$ values decrease compared to the case with PAHs. } 
\label{fig:mesh4}
\end{figure} 
\clearpage

\begin{figure}[ht!]
\plotone{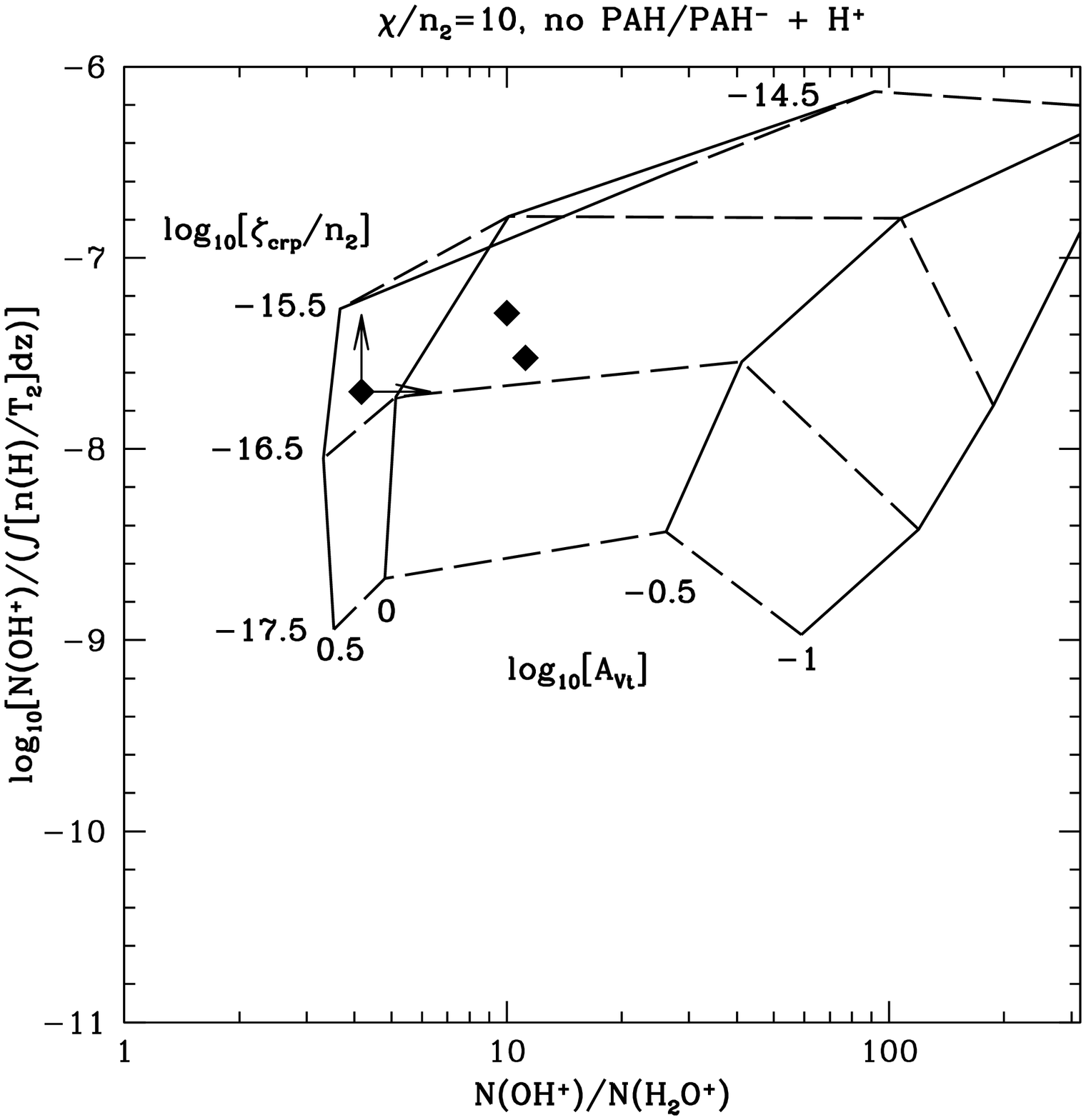}

\caption{This figure is identical to Figure 15 ($\chi/n_2=10$) except that the PAH and PAH$-$ rates with H$^+$ have
been reduced by $\gta 4$ so that H$^+$ is mainly destroyed by electrons or by forming O$^+$ which
then reacts with H$_2$ to form \oh.  Note that the inferred \crt/$n_2$ values decrease compared to the case with PAHs. }
\label{fig:mesh5}
\end{figure} 
\clearpage

\end{document}